\documentclass[aps, twocolumn, tightenlines]{revtex4}
\usepackage{graphicx,subfigure}
\usepackage{amsmath,amssymb}
\usepackage{multirow}
\usepackage{bm}

\newcommand{\be}{\begin{eqnarray}}
\newcommand{\ee}{\end{eqnarray}}
\newcommand{\nn}{\nonumber}

\def\slashchar#1{\setbox0=\hbox{$#1$}           
   \dimen0=\wd0                                 
   \setbox1=\hbox{/} \dimen1=\wd1               
  \ifdim\dimen0>\dimen1                        
 \rlap{\hbox to \dimen0{\hfil/\hfil}}      
  #1                                        
 \else                                        
    \rlap{\hbox to \dimen1{\hfil$#1$\hfil}}   
    /                                         
 \fi}                                         %

\begin{document}

\title{Baryon  clustering at the critical line \\
and near the hypothetical critical point \\
in heavy-ion collisions}

\author{Edward Shuryak and Juan M. Torres-Rincon}

\affiliation{Department of Physics and Astronomy, Stony Brook University,
Stony Brook NY 11794-3800, USA}

\begin{abstract}
We study clustering of baryons at the freeze-out point of relativistic heavy-ion collisions. Using a Walecka-Serot model for the nucleon-nucleon (NN) interaction we analyze how the
modified critical $\sigma$ mode---responsible for the NN attraction---allows for clustering of nucleons when the system is close to a possible critical point of QCD. We investigate 
clusters of few nucleons, and also the internal cluster configuration when the system is long lived. For realistic heavy-ion collisions we study to how extend such clusters can be formed
in a finite time, and perform the statistical analysis of cumulants and higher-order moments (skewness and kurtosis) for collisions at the Beam Energy Scan of RHIC. 
\end{abstract}

\maketitle
\tableofcontents
\section{Introduction}

We start by contrasting known facts about high- and low-energy heavy-ion collisions,
after which we will define the phenomena to be discussed in this work. 

By high-energy collisions we mean those at Relativistic Heavy-Ion Collider (RHIC) full energy $\sqrt{s_{NN}}\approx 200$ GeV  and at the
Large Hadron Collider (LHC) $\sqrt{s_{NN}}\approx 2-8$ TeV. In these cases the particle yields are very accurately
described by the so-called ``resonance gas model", assuming that all interactions between hadrons can be effectively treated 
as an ideal gas of all known resonances, as suggested by the Beth-Uhlenbeck formula.
As shown e.g. in Ref.~\cite{Andronic:2017pug}, the yields per degree of freedom are on the same thermal exponent, from 
the lightest species---pions, kaons, etc---up to the baryons, hyperons and their antiparticles, all the way up to light nuclei such as $^4$He. 
This trend is observed to hold over about 9 decades. The lesson is that, 
at chemical freeze-out temperatures $T_{ch} \simeq 150$ MeV and 
at near-zero baryonic chemical potential $\mu_B \simeq 0$ the
fireball  is very well thermally equilibrated. Such high degree of equilibration
undoubtedly is related with kinetic properties of the strongly coupled quark-gluon plasma (QGP),
preceding the freeze-out stage of these collisions.
 
At low non-relativistic collision energy, $\sqrt{s_{NN}}< 1$ GeV, creating nuclear matter with the temperature $T\simeq 10$ MeV,
one observes the so-called ``multifragmentation'' phenomena, production of 
large variety of  nuclear fragments, with a wide power-like distributions. It is attributed to a nearby presence
of a critical point, separating liquid nuclear matter from a gas-like phase. For a review see e.g.~\cite{Bondorf:1995ua}.
The production of various nuclear clusters is not in equilibrium, and is very sensitive to
the relation between the temperature and time available for cluster formation.    
This regime can be compared to that in atomic physics studying various out-of-equilibrium
situations, for example ``snow production" machines, operating in between water and ice
phases. 

Heavy-ion collisions at intermediate energies have been studied in 1980s,
both at CERN Super Proton Synchrotron and the Brookhaven National Laboratory Alternating Gradient Synchrotron,
but not in sufficient detail. Many models predict that baryon-rich 
matter will also have the first order transition line, ending in certain critical point. 
Its search using enhanced fluctuations was proposed in Refs.~\cite{Stephanov:1998dy,Stephanov:1999zu}. The Beam Energy Scan (BES) towards the lowest energies
possible at RHIC is currently under way.  Significant modification of the baryon number distributions, such as its large {\it kurtosis}  is indeed observed
at the low energy end~\cite{Luo:2015ewa}, perhaps indicating out-of equilibrium fluctuations related with criticality. Using STAR detector at RHIC in a fixed-target mode is in the plans.  

The topic of this paper is the baryon clustering phenomenon happening at the so-called {\it freeze-out}
stages of heavy-ion collisions, in this intermediate baryon-rich domain. We 
will show how relatively small modification of nuclear forces at distances $r=1-2$ fm
can dramatically change the binding of baryonic clusters, as well as the kinetics of
their production.  
  
The paper is structured as follows. We start in Sec.~\ref{sec_motivation} with a motivation for the modified inter-nucleon potential, which
are quantified in Sec.~\ref{sec_forces}. Then we present preliminary studies of how such modifications change the binding of
clusters in Sec.~\ref{sec_binding}, using two opposite limits: the uncorrelated Gaussian-shape clusters in Sec.~\ref{sec_Gaussian}, and 
fully correlated clusters forming certain classical shapes in Sec.~\ref{sec_correlated}. Afterwards we form mean-field clusters consisting of bound
nucleons only: the corresponding approximation follows the theory of globular clusters
in galaxies which is briefly described in App.~\ref{sec_globular}. Connection to globular clusters is a new element
of this work, which very instructively shows how a dynamical system can be in
out-of-equilibrium partially-clustered stage, for billions of years. 
We use a similar approximation to evaluate the properties of mean-field
self-consistent clusters of bound nucleons in Sec.~\ref{sec_meanfield}. Before we describe our main body of simulations, we classify the observables
to be used in Sec.~\ref{sec_observables}. The bulk of our studies introduced in Sec.~\ref{sec_simulations} is done using classical dynamical
approaches, such as molecular dynamics (MD) complemented by Langevin (MD+L) 
forces representing the effects of the mesonic heat bath. We  proceed from
small number of nucleons and finite clusters to rather large ones, with $N\sim 100$
particles, see Sec.~\ref{sec_clustering_at_fo}. Finally, in Sec.~\ref{sec_exp} we calculate the resulting skewness and kurtosis
in a setting modeling experimental conditions of the RHIC BES program,
and indeed find that its growth can be caused by the modified inter-baryon forces. 


The remainder of this introduction contains a summary of the ideas motivating this work. 

One important notion is the very {\it high  sensitivity} of the  dynamical clustering
to the details of the inter-nucleon effective potential. Since the time of Yukawa's suggestion, nuclear forces are traditionally described 
in terms of certain meson exchanges. Furthermore, as all nuclear physicists know, any model of nuclear forces needs special tuning,
needed to reproduce two delicate phenomena:
(i) strong cancellation between repulsion and attraction in the mean potential energy; and (ii)
partial cancellation of the remainder in the mean potential energy by quantum kinetic
energy. The final result should be that neutron systems, and in fact
many species of light nuclei,  are not bound. Even infinite nuclear matter, 
with equal number of protons and neutrons (and QED effects switched off) is only slightly bound.  

Because of these cancellations, a small modification of the inter-nucleon potential can induce
quite significant changes in binding, even up to an order of magnitude. This is of crucial importance, because the temperatures of the hadronic phase
we discuss is ranging from the critical temperature 
$T_c \approx 120-155$ MeV down to the kinetic freeze-out temperature of baryons $T_{kin} \approx 80-100$ MeV.
Such temperatures may appear large compared to the usual nuclear potential depth $\sim 50$ MeV
and binding-per-nucleon $\sim 10$ MeV. And yet, even with such conditions we do 
find significant clusters of trapped baryons. We therefore suggest
to look not only at higher-order moments of the net-baryon distribution, but also out-of-equilibrium production of  
light nuclei.

Why do we think that inter-nucleon effective potentials might be modified in the conditions 
discussed, from well-known forces in cold nuclear matter?

One generic reason---suggested  many times before---is that in the baryon-rich end of  the phase diagram 
certain modification of meson masses and couplings should be much larger than
in the (well studied) small-$\mu_B$ meson-dominated regime.  
In the spirit of the resonance gas model, one may argue that there are much more
baryonic resonances than mesonic ones. Studies of dilepton spectral density~\cite{Arnaldi:2006jq} and
related $\rho$-meson modifications~\cite{Rapp:1999ej} have indeed shown such baryonic dominance. 
It is furthermore quite reasonable to think that what happens with $\rho$ should happen with other wide
resonances, the $\sigma$ in particular.  

Another generic reason, emphasized in Ref.~\cite{Stephanov:1998dy} and also widely known, is the possible existence of  
the (hypothetical) {\it QCD critical point}, as the endpoint of the first-order phase transition line. On general theoretical grounds
we know that second-order phase transitions have massless modes, which lead to the phenomenon of critical opalescence at scales much larger than
the microscopic scales of matter. If exchanges of such long-range critical modes do appear in the inter-nucleon potential---even with relatively small coupling---we
will find a significant enhancement of both the binding of certain nuclear clusters, and the kinetic clustering rates.

%
%
%

Finally, as multiple studies on the kinetics near the phase transitions indicate, the so-called ``critical slowing down" phenomenon prevents
complete equilibration, and opens the door to multiple out-of-equilibrium scenarios, some with significant cluster production.

\section{Baryonic densities and forces at freeze-out} \label{sec_motivation}

We already mentioned the ``resonance gas model'', which is so successful
for predicting hadronic yields for high-energy heavy-ion collisions. It is based on
the standard statistical expression for the equilibrium particle densities at
 number of baryons of the type $i$ 
\be  \label{eqn_fermi_gas} N_i=\gamma_i V_{tot} \int \frac{d^3p}{(2\pi)^3} \frac{1}{\exp[(m_i -\bar\mu+p^2/2m_i)/T_{ch}]+1} \ , 
\ee
where $ \gamma_i , V_{tot},  T_{ch}$ are the statistical weight, total effective volume
of the  chemical freeze-out surface, and the corresponding chemical freeze-out temperature. We put a
bar on the chemical potential indicating that we include the mean value of the
inter-baryon potential in it, $\bar\mu =\mu-\bar V$.

There are certain important distinctions between high-energy collisions and the 
conditions we are going to study. First of all, in the former case $\mu_B \approx 0$ 
and baryons/antibaryons are both very much suppressed by the Boltzmann
factor, since $m_i/T_{ch} \sim 10$. Second, at $T_{ch}\approx T_c\approx 155$ MeV,
excitation of baryonic resonances $N^*,\Delta^*$ and their strange counterparts is
very significant. For example, the population of the $S=3/2,I=3/2$ $\Delta$ resonance,
relative to that of the nucleon, is about  $4 \exp \left[ (m_\Delta-m_N)/T_{ch} \right]\approx 0.7$. 
On the other hand, in the time between the chemical and kinetic freeze-out,
with $T_{kin}\approx 80-100$ MeV, most of them decay into a baryon and one (or more) mesons,
providing large ``feed-down corrections" to nucleon yields. 

For the conditions of the BES, on the other hand, the chemical potential is in the range 
$\bar \mu=500-700$ MeV, and the Boltzmann factors  $ \exp[ -(m_i -\bar\mu)/T_{ch}]$ 
are not so punishingly small. Furthermore, the number of antibaryons is negligible,
and we will not discuss them in the following. The chemical freeze-out temperature is lower,
and thus a fraction of excited baryonic resonances is much smaller. In the following
we will (maybe crudely) ignore their existence and feed-down. 
Another way to explain this approach is to assume that effectively the baryonic
resonances have the same effective potentials as the nucleons.  We thus normalize our calculations to the total  
final nucleon number observed experimentally.

  Let us finally comment on the distinctions between our molecular dynamics computations and
those for low-energy heavy-ion collisions. If the temperature $T\sim 10$ MeV,
the thermal kinetic energy is comparable with the Fermi energy of matter at nuclear densities, and
therefore quantum effects play a significant role and needs to be taken care of, by some kind of approximation. 
The freeze-out temperatures we deal with are significantly higher, many states are
excited and the role of Fermi repulsion is significantly reduced. In essence, the
baryon component of the system can be approximated by a classical gas. Nevertheless, we will study some 
effective quantum corrections in App.~\ref{sec_quantum}.

\section{Modified baryonic potentials} \label{sec_forces}

Let us now proceed to the discussion of in-matter forces between the baryons,
starting with the so called ``mass shifts" issue, which is somewhat controversial.
On one hand, a significant part of the nucleon mass is believed to be due to ``constituent quark masses" induced by chiral symmetry breaking.
If so, in view that the freeze-out is not far from the restoration line of the chiral symmetry,
it was predicted by many phenomenological and theoretical models that there
should be a significant downward shift of such contributions to the effective quark mass. 
On the other hand, as we mentioned already, the successful thermodynamical description of the particle yields at chemical freeze-out
uses the``resonance gas model" without any modifications of the particle masses.

Furthermore, the range of  the inter-nucleon potentials is defined by masses of the corresponding mesons.  
For one of them, the vector meson $\rho$, we have direct access to its
spectral density via the dilepton production, and its significant widening has indeed been observed~\cite{Arnaldi:2006jq}. For the $\omega$ meson no
changes are observed, which is expected, since due to its longer lifetime most of
them decay outside of the fireball. The $\sigma$ meson, wide even in vacuum~\cite{Pelaez:2015qba}, is often represented as a correlated  $\pi\pi$ pair, and is
perhaps getting even wider in matter.
The effective potential, convoluting Yukawa potential with its spectral density, is expected to
become longer-range or even infinite-range at the critical point.  

Unfortunately, lattice QCD at the moment can only extrapolate to $\mu_B/T<2$ or so,
which is far from the regime we are interested in. Some hints can perhaps be
gained from the lattice study by the Graz group~\cite{Glozman:2012fj}, which performed  
restoration  of  chiral symmetry ``by surgery" i.e. simply removing the lowest Dirac 
eigenstates from the hadronic mass evaluation. What is observed is that the chiral partners
(such as the nucleon $P=+1$ and $N^*$ $P=-1$, $\rho$ and $a_1$, etc.)
modify their masses in the opposite directions, meeting somewhere in between.
Perhaps such effects cancel each other in the calculations of the total baryon and meson yields.  
If so, note that the chiral partner of the $\sigma$ is the pion. Moving towards it means
reducing its mass, maybe to a half of it, or even all the way to zero (close to the second-order phase transition).

Completing the motivation, we now explain the reader the simplified form of nuclear forces
we will be using. It follows from the popular relativistic model by
Serot and Walecka~\cite{Serot:1984ey}. One important simplifying characteristic is that it only includes the isoscalar mesons,
scalar $\sigma$ and vector $\omega$, so there is no difference between the interaction of protons and neutrons.
We will also ignore electromagnetism, as the clusters studied are not so large as to make it important.

 The Lagrangian density of their model is
\begin{eqnarray}
 {\cal L} &=& \frac{1}{2} \left( \partial_\mu \phi \partial^\mu \phi -m_\sigma^2 \phi^2\right) - \frac{1}{4} F_{\mu\nu} F^{\mu\nu} +\frac{1}{2} m_\omega^2 V_\mu V^\mu  \nonumber \\
          &+& \bar \psi \left[ \gamma_\mu (i\partial^\mu -g_\omega V^\mu)-(m_N-g_\sigma \phi)  \right] \psi \ , \label{eq:lagrangian}
 \end{eqnarray}
where the Abelian field strength of the vector field $F_{\mu\nu}\equiv \partial_\mu V_\nu -\partial_\nu V_\mu$
is the same as in electrodynamics. There are thus three fields, Dirac nucleons $\psi$, vector $\omega$-mesons $V_\mu$ and scalar $\sigma$-mesons $\phi$, interacting with each other
in relativistically invariant way. Their masses are considered to be an input. For definiteness we use $m_\sigma=500$ MeV, $m_\omega=782$ MeV and $m_N=938$ MeV.

The resulting static potential between nucleons is 
\be \label{eq:WalPot} V_A(r)=- \frac{g_\sigma^2}{4\pi r}e^{-m_\sigma r}+ \frac{g_\omega^2}{4\pi r}e^{-m_\omega r} \ , \ee
where the coupling values selected by Serot and Walecka~\cite{Serot:1984ey} are
\be \label{eq:WalPotPar} g_\sigma^2 = 267.1 \left( \frac{m_\sigma^2}{m_N^2} \right),  \qquad
g_\omega^2 = 195.9  \left( \frac{m_\omega^2}{m_N^2}  \right) \ .
\ee

\begin{figure}[ht]
\begin{center}
\includegraphics[width=6.5cm]{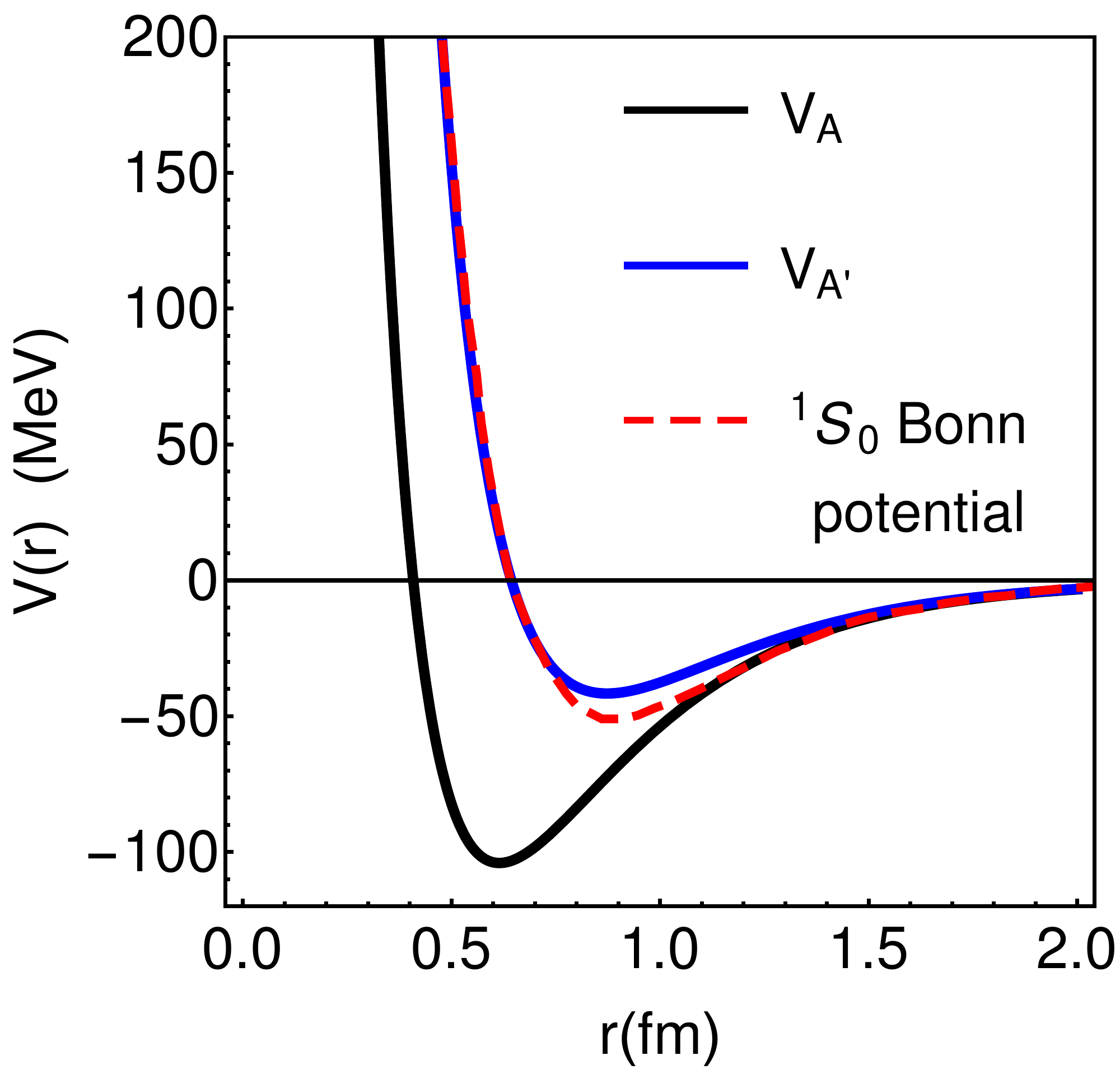}
\caption{Phenomenological nucleon-nucleon potentials. Solid line: Serot-Walecka potential (\ref{eq:WalPot}) with parameters in (\ref{eq:WalPotPar}).
Dotted line: Same as before but with the repulsive strength increased by a factor 1.4. Dashed line: Bonn $NN$ potential~\cite{Machleidt:2000ge} in channel $^1S_0$, taken from~\cite{Ishii:2006ec}.}
\label{fig_Walecka_potA}
\end{center}
\end{figure}

The $\omega$ coupling is stronger, thus dominating at small distances. Note further that these two terms nearly cancel each other, leaving us with a relatively shallow
potential, $|V_A|<100$ MeV $\sim m_N/10$, see Figs.~\ref{fig_Walecka_potA} and~\ref{fig_Walecka_potAbis}. It is also important to notice that the couplings are selected not to
fit the binary scattering phase-shifts and deuteron binding, as done for all other phenomenological
potentials, but from the fit to nuclear matter in the {\it mean-field approximation}. The details of that are further delegated to Sec.~\ref{sec_meanfield}.

For our studies of the baryonic clustering in this work
we will use the Serot-Walecka model in four different versions of the mesonic masses:
\begin{itemize}
\item [($A$)] The unmodified Walecka potential (\ref{eq:WalPot}) with the parameters computed at mean field quoted in (\ref{eq:WalPotPar}).
\item [($A'$)] Walecka potential with increased repulsion $g^2_\omega \rightarrow 1.4g^2_\omega$ to make it closer to the phenomenological Bonn potential.
\item [($B1$)] One in which the $\sigma$ mass squared decreases ``half way" (that is $m_\sigma^2\rightarrow m_\sigma^2/2$), presumed to hold at the critical line
for $\mu_B<\mu_c$. The ``minimal modification" version changes the coupling as well $g_s^2 \rightarrow g_s^2 /2$, keeping the mean potential energy constant. 
\item [($B2$)] This version is the same as (B1) except that the scalar coupling is $not$ modified.
The mean potential from $\sigma$ thus is a factor 2 larger than in B1.
\item [($C$)]  An admixture of the (B2) potential
 with the one with very light critical mode $\sigma$, $m_\sigma^2\rightarrow m_\sigma^2/6$ (denoted as $V_{crit}$),
\be V_C(r;x)= (1-x) V_{B2}(r)+x V_{crit}(r)\ . \ee  
\end{itemize}

\begin{figure}[ht]
\begin{center}
\includegraphics[width=6.5cm]{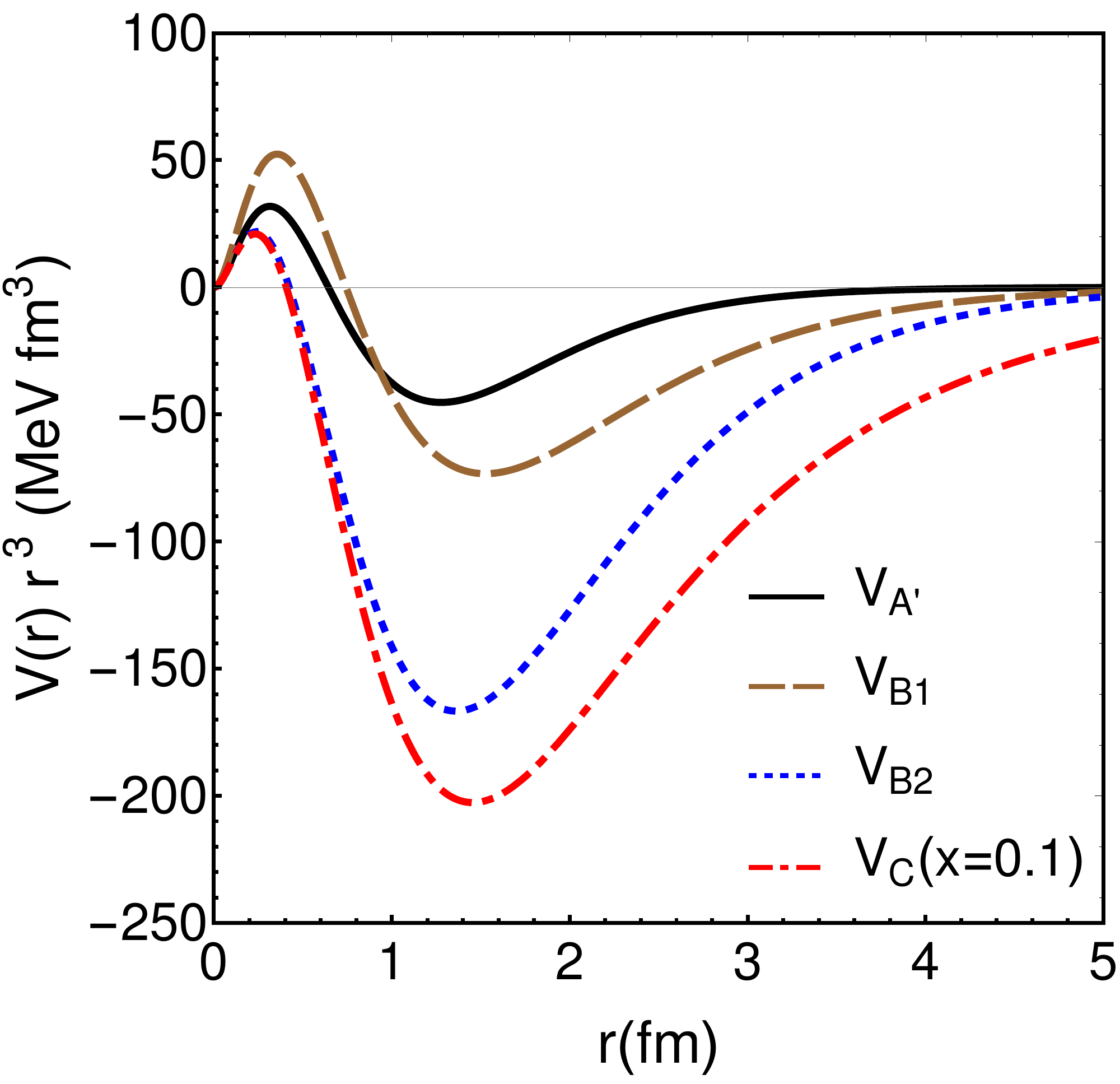}
\caption{$V(r) r^3$ (MeV fm$^3$) versus $r$ (fm) for the four models used in this work. The original Walecka potential (with increased repulsion as compared to the parameters in (\ref{eq:WalPotPar})) is shown by the
black solid line, the versions $B1$ and $B2$ correspond to the dashed brown and dotted blue dashed lines, respectively.
The version $C$ potential, with $x=0.1$ is represented by the red dash-dotted line.}
\label{fig_pots_times_r3}
\end{center}
\end{figure}

In Fig.~\ref{fig_pots_times_r3} we show the corresponding potentials, multiplied
for convenience  by $r^3$ (note that $4\pi r^3/3$ times the density of other baryons
tell us effectively how many ``partners" a given baryon has). 
As one can see, these four models show progressively increasing depth and range of the attractive potential.


\section{Preliminary studies of cluster binding} \label{sec_binding}

Before we discuss our dynamical out-of-equilibrium studies of multi-baryon systems,
it is instructive to report some simplified approaches. We considered either $N=4-13$ nucleons, or clusters of certain
fixed size, and  use all versions of the modified potentials described above. In Sec.~\ref{sec_Gaussian} we consider a limit in which there are no correlations
between locations of the nucleons, so that the $N$-body distribution is simply factorizable
into a product
\be n(\vec r_1,\vec r_2,...,\vec r_N)=\prod_i n(\vec r_i) \ , \ee
with the same Gaussian-like spatial distribution. In Sec.~\ref{sec_correlated} we turn to the opposite limit, in which the nucleons are 
set to specific locations, defined by symmetry considerations, which in turn depend
on the particle number, and study the dependence of the total energy on the scale
parameter. Finally, in  subsection~\ref{sec_meanfield} we calculate properties of self-consistent mean
field clusters, formed of {\it only bound} nucleons.

\subsection{Clusters made of uncorrelated nucleons} \label{sec_Gaussian}

Before we study clustering rates (correlation growth), it is instructive to illustrate the effect of different potentials defined above in a
simple model. Let us consider a Gaussian-shaped cluster, with the nuclear matter density 
$n_0=0.16 \textrm{ fm}^{-3}$ at its core,
\be n(r)=n_0 \exp \left( - \frac{r^2}{2 R^2} \right) \ , \ee 
and the r.m.s. size $R=2$ fm. The integral $N=\int d^3r\ n(r) \approx  20$,
so this is a crude model of a medium-size nucleus.

Using the Thomas-Fermi expression for local Fermi momentum with $\gamma$ degrees of freedom $p_F(r)=[6\pi^2 n(r)/\gamma]^{1/3} $ one can calculate the kinetic energy per nucleon.
For $\gamma=4$ it is
\be K/N= \left\langle  \frac{p_F(r)^2}{2m_N} \right\rangle \approx 17.1 \, \rm{MeV} \ , \ee
independent or $R$. For pure neutron matter, with $\gamma=2$, it is $K/N \approx 27.1$ MeV. 

\begin{figure}[ht]
\begin{center}
\includegraphics[width=6.5cm]{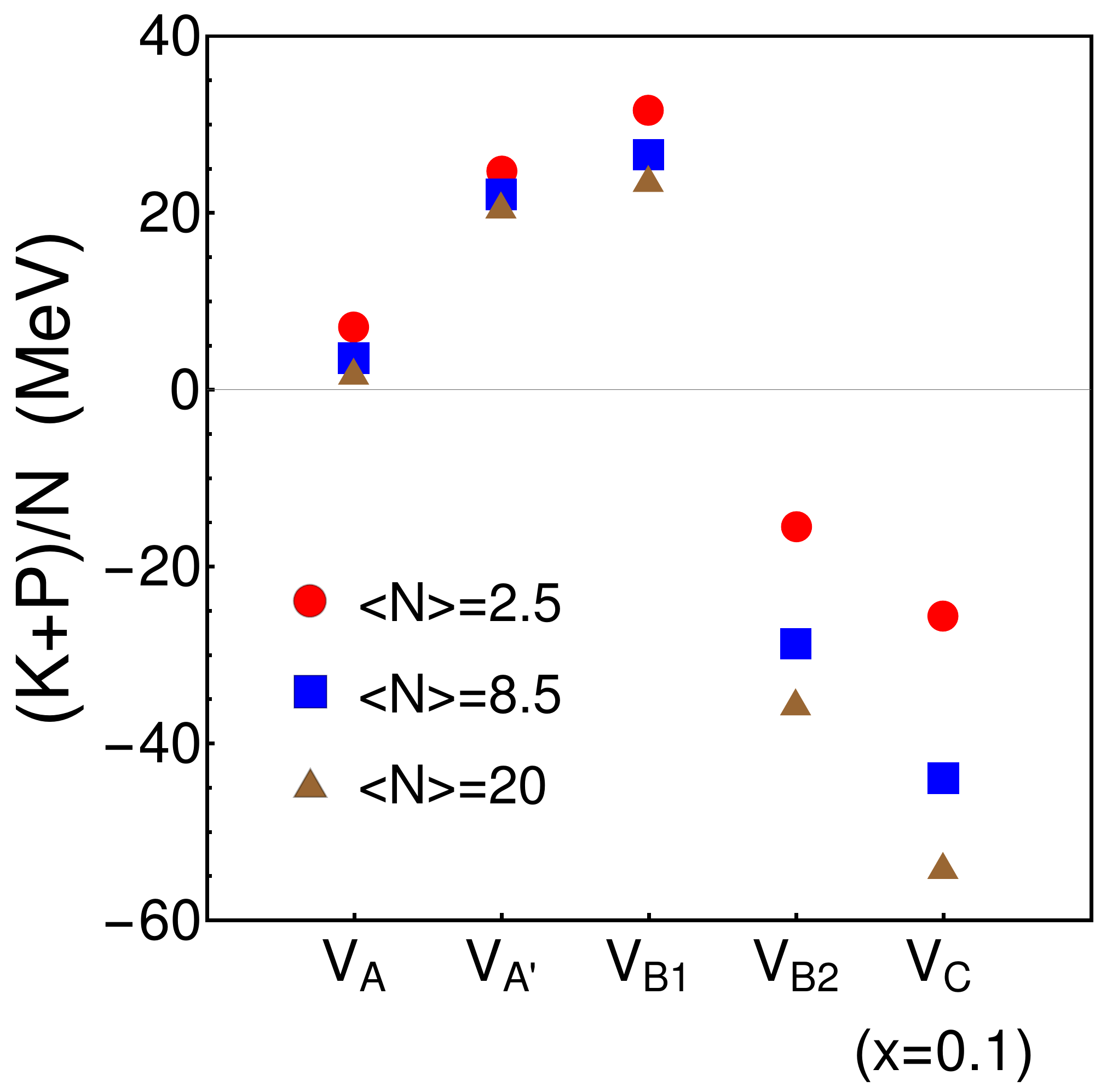}
\caption{The total energy per nucleon in MeV/nucleon, for all versions of the binary potentials and the number of nucleons $N=2.5, 8.5, 20$;
 corresponding to Gaussian density with r.m.s radii $R=1,1.5,2$ fm, respectively.}
\label{fig:energyMF}
\end{center}
\end{figure}

Now, ignoring pair correlations $n(\vec r_1, \vec r_2)\rightarrow n(\vec r_1)n( \vec r_2)$ (mean-field approximation),
one can calculate the potential energy
\be P=\frac{1}{2} \int d^3r_1 d^3r_2 \ n(r_1) V(\vec r_1 -\vec r_2) n(r_2) \ , \ee 
corresponding  to forces defined in the preceding section. The results are $P=-15.1, 3.7, 6.7, -52.5$ and $-70.9$ MeV per nucleon,
for models $A, A', B1, B2$ and $C$, respectively. In Fig.~\ref{fig:energyMF} we summarize our results for the total energy per nucleon, for the different 
potentials used in this work. We also consider different r.m.s. of the nuclear density, with a total number of nucleons of 2.5 (circles), 8.5 (squares) and 20 (triangles).

%

The total energy per nucleon $(K+P)/N$ for Walecka model $A$ (whose parameters are chosen at mean field) is, within the accuracy of about 2 MeV,
equal to zero for nuclear matter ($p,n$ equal mixture), and $+10$ MeV for pure neutrons.

The model $A'$ is chosen to be much more shallow than the previous Walecka potential, and it is not able to bind this kind of clusters in the mean-field approximation. Similarly, the potential $B_1$ has comparable
effect, with a total energy per nucleon of several dozens of MeV. However models $B2$ and $C$ lead to an large binding. As we will detail later, the addition of binary correlation
function can increase this binding even more. The main lesson from this initial calculation is that significant binding (non-negligible
compared to the temperature $T \approx 100$ MeV) can be produced at mean field only for significantly modified potentials (models $B2$ and $C$).

\subsection{Clusters made of strongly correlated nucleons}
\label{sec_correlated}

For vanishingly small temperatures and small values of the particle number $N$, the geometry of the classical lowest energy states is suggested by symmetry. 
In this section we present some expectations as functions of $N$. 
Later, we will study not only the near-freeze-out $T\sim 100$ MeV cases, but also cool the systems
down to $T\approx 1$ MeV and even $T\approx 10^{-3}$ MeV and test that
the symmetric configurations considered in this section are indeed obtained from the MD+L simulations.

For definiteness, the potential used in this section is $V(r)=V_{A'}(r)$,
which is enough to bind the nucleons as we will not account for thermal effects ($T=0$).

The smallest number of particles we consider is four, $N=4$, which form a {\it tetrahedron}.
As it is known from studies of few-body nuclei, such correlation between four nucleons is indeed
rather strong inside the $^4$He, and persists in ``alpha-particle nuclei" such as $^{12}$C,$^{16}$O. All 6 pair distances between the 4
nucleons are in this case the same, denoted by $a$. In general, $a$ is defined as the minimum distance between 2 nucleons in equilibrium---which is not necessarily the minimum of the inter-nucleon potential. 
The energy per nucleon $\langle V \rangle$ in this simplest case is just
 \be \langle V \rangle_4=\frac{3}{2} V(a) \ . \ee
  
{\it Octahedron} has $N=6$ particles and 15 pairs: 12 of them of distance $a$ and 3 of distance
$\sqrt{2}a$ .
The energy per nucleon is in this case
\be \langle V \rangle_6=2  V(a)+\frac{1}{2}  V(\sqrt{2}a) \ . \ee

 The next cluster to consider, of $N=8$ particles is the {\it hexahedron} (or cube). It has 12 distances 
 $a$, 12 distances  $\sqrt{2}a$ and 4 of distances  $\sqrt{3}a$, 28 in total, 
\be \langle V \rangle_{8h}=\frac{3}{2} V(a) + \frac{3}{2} V(\sqrt{2}a) + \frac{1}{2} V(\sqrt{3} a) \ . \ee  
 
  The largest particular cluster we discuss is the {\it icosahedron} with 12 vertices, to which we added one particle at the center, making $N=13$. 
It has 78 distances: 12 distances at $a$, 30 distances at $\sqrt{2-2/\sqrt{5}} a$, 30 at $\sqrt{2+2/\sqrt{5}} a$,
and 6 at $2a$,
\begin{eqnarray}
 \langle V \rangle_{12+1} & = & \frac{12}{13} V(a) + \frac{30}{13} V(\sqrt{2- 2/\sqrt{5}}a) \nn \\
&+& \frac{30}{13} V(\sqrt{2+ 2/\sqrt{5}}a) + \frac{6}{13} V(2 a) \ . 
\end{eqnarray}

\begin{figure}[ht]
\begin{center}
\includegraphics[width=7cm]{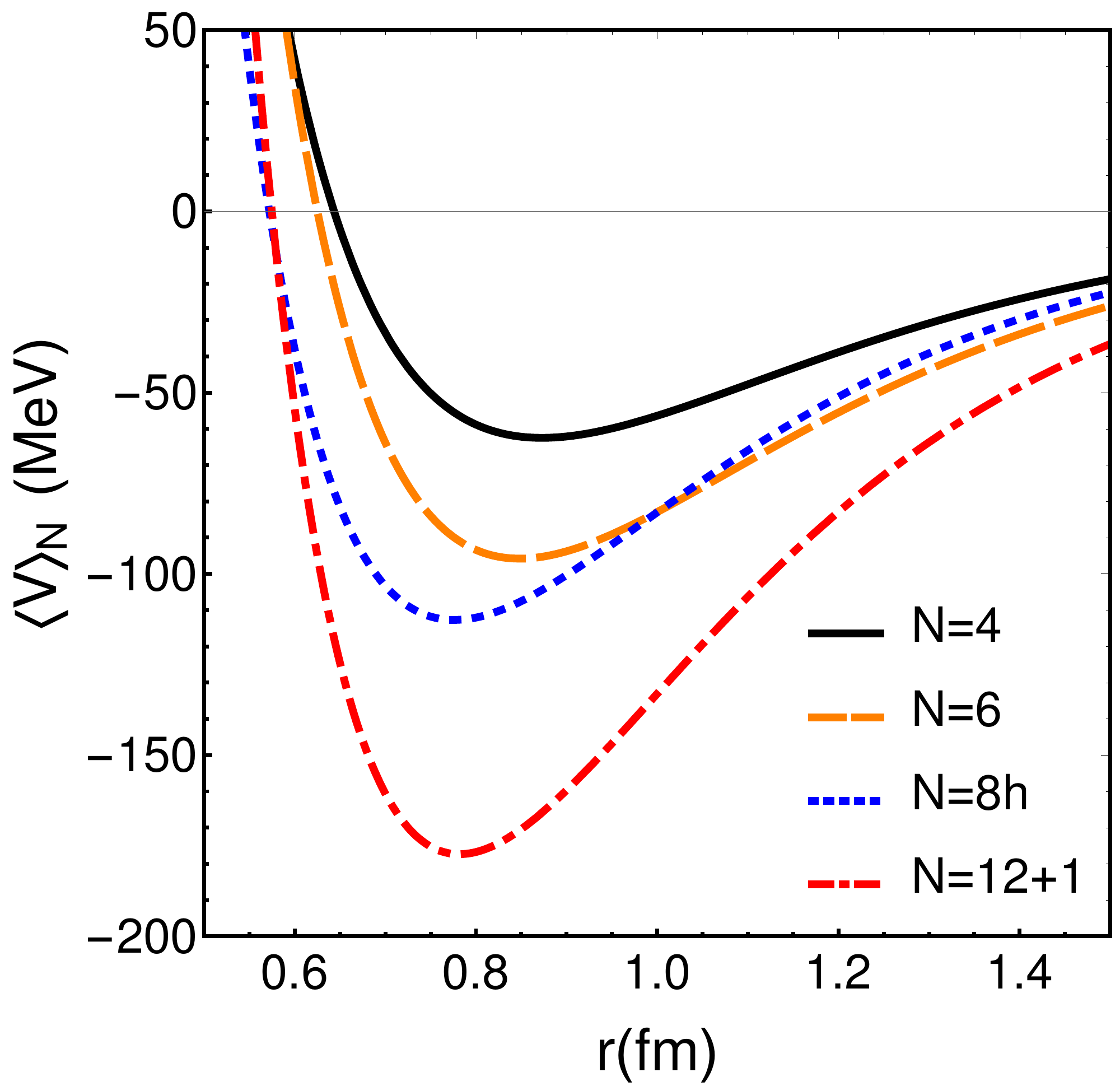}
\caption{The potential energy per particle  $\langle V\rangle_N$ (MeV/nucleon) 
as a function of distance $r$ (fm) for four clusters: $N=4$ tetrahedron
(the upper thin solid line), $N=6$ octahedron (the dashed line), $N=8$ cube (the dotted line)
and $N=12+1$ icosahedron+one particle (the lower dot-dashed line).
This calculation is done with the Walecka potential $V_{A'}$.
}
\label{fig_four_clusters}
\end{center}
\end{figure}

 The energy per particle $\langle V\rangle_N (r)$ for all four clusters, as a function of the distance $r$,
is shown in Fig.~\ref{fig_four_clusters}.  One can see that augmenting $N$ this potential energy increases, eventually exceeding the range of temperatures in
 the problem $T=(100-150)$ MeV by a significant factor (even assuming that the potential is not modified by the temperature).
 Previous experience of working with strongly coupled Coulomb plasmas, see Ref.~\cite{Gelman:2006sr}
 and references therein,  tells us that for such range of $ \langle V\rangle_N/T$
 the factorized mean field theory is completely inadequate, and the correlations
 are significant. At the same time, this range of the ratio is also too small to 
 cause solidification of the system, keeping the system in the 
 strongly correlated but still liquid phase. 

  The value of the minimal distance between 2 nucleons in equilibrium was denoted by $a$, and it can be obtained by minimizing
the potential energy per particle in Fig.~\ref{fig_four_clusters} for each $N$. For future reference, we summarize these distances and the associated 
potential energy in Table~\ref{tab_minima}. Notice that $a$ coincides with the minimum of the potential $V_{A'}(r)$ only
for the case $N=4$.

  Finally note that suggested by a totally different minimization problem (Thomson problem in electrostatics~\cite{thomson}), we have tried a different configuration for $N=8$,
the {\it square antiprism}, whose energy per particle is denoted as $\langle V\rangle_{8s}$. We indeed find a lower potential energy than the cubic configuration, providing an example where the expectation based on 
symmetry (provided in this case by the Platonic solids) does not provide the optimal configuration. We will come back to these geometries when applying our MD+L simulations to cold systems.  

\begin{table}[ht]
\begin{center}
\begin{tabular}{|c|c|c|c|}
\hline
$N$ & Polyhedron & $a$ (fm) &  $\langle V\rangle_N$  (MeV/nucleon) \\ 
\hline
\hline
$4$ & Tetrahedron & 0.8727 & -62.47 \\
$6$ & Octahedron & 0.8481 & -95.78 \\
$8h$ & Hexahedron (cube) &0.7761 & -112.70 \\
$8s$ & Square antiprism &0.8096 & -117.03\\
$12+1$ & Icosahedron +1 & 0.7816 & -177.32 \\
\hline
\end{tabular}
\caption{Minimal distance between nucleons and potential energy per nucleon for several configurations with $N$ nucleons.\label{tab_minima}}
\end{center}

\end{table}

\subsection{Mean-field baryon clusters at freeze-out} \label{sec_meanfield}

Before we study the clustering phenomenon dynamically, it is important to 
see what kind of clusters can {\em in principle} be 
self-consistent, in analogy to globular clusters in galaxies.

Let us assume homogeneous matter at rest, with certain mean density (\ref{eqn_fermi_gas}) 
and the mean potential $\bar V$, and on top of it a cluster, as a {\it deviation} from the mean. It is cause by a
deviation of the mean potential $\delta V(r)=V(r)-\bar V$.
In thermal equilibrium it will add an {\it extra} density of baryons,
\begin{eqnarray} \label{eqn_extra_density}  
  \delta n_i(r) &=& \gamma_i \int \frac{d^3p}{(2\pi)^3} \left[ \frac{1}{\exp \left( \frac{m_i -\bar\mu+\frac{p^2}{2m}+\delta V(r)}{T_{ch}} \right)+1}  \right. \nn \\ 
  &-& \left.  \frac{1}{\exp \left( \frac{m_i -\bar\mu+\frac{p^2}{2m}}{T_{ch}} \right)+1 } \right] \ . 
\end{eqnarray}

Furthermore, following the setting of the globular clusters in the galaxies
described in App.~\ref{sec_globular}, we will consider times at which  
all unbound particles has already left the cluster, and in the phase space
integral we include {\it only bound particles}. This means in the momentum integral we only integrate over the region where
\be \frac{p^2}{2m}+\delta V(r) <0 \label{eqn_bound} 
\ee

To make cluster self-consistent, this extra potential $\delta V(x)$ should be created by
the extra density {\it itself}. We write this condition in the integral form
\be \delta V(\vec r_1)=\int d^3 r_2 V(\vec r_1-\vec r_2) \delta  n_i(\vec r_2) \label{eqn_selfconsist}
\ee
equivalent to the Poisson Eq.~(\ref{eqn_Poisson}) for the Newtonian potential in App.~\ref{sec_globular}. The two
equations (\ref{eqn_extra_density},\ref{eqn_selfconsist}) together make a system
of equations which needs to be solved.

One simplification is to ignore $+1$ in (\ref{eqn_extra_density}), that
is proceed from Fermi to Boltzmann statistics. Note further, that when $\delta V/T$ is
small, one can expand the bracket to the first order in it, and then take the momentum integral
using the binding condition (\ref{eqn_bound}). The resulting contribution is
$\delta n \sim \delta V^{5/2}$. The exact integral without expansion can also be done
analytically, leading to the following function of $z \equiv \delta V/T$, given  with its (rather well convergent) series
\begin{eqnarray} 
N(z) & = & e^z \textrm{ Erf}(\sqrt{z})-\frac{2 \sqrt{z}(3+2z)}{3 \sqrt{\pi}}  \\
&+& \frac{8 z^{5/2}}{15 \sqrt{\pi}}\left(1+ \frac{2z}{7}+\frac{4z^2}{63} + \frac{8 z^3}{693}+ \frac{16 z^4}{9009}+... \right) \ . \nn
\end{eqnarray}
(see also  Eq.~(\ref{dens_function_of_pot})).

In practice we adopt the following procedure: start with a certain ansatz for $\delta V(r)$,
e.g. Gaussian with two parameters, the amplitude and the radius. Then, via $N(z)$
function, calculate numerically the r.h.s. of Eq.~(\ref{eqn_selfconsist}), and tune the parameters
to minimize the difference between the l.h.s., the obtained $\delta V$, and the input one.
Of course, inside a given variational ansatz one cannot get a very good match of the shape,
but the overall difference was kept at a reasonable level, of the order of 15-20 \%.   

We found it instructive to keep the radius of the cluster fixed, say r.m.s. radius $R=2.2$ fm,
and modify only the potential depth. For different potentials defined above, we find
the best depth of the potential:  the resulting number of nucleons in the cluster and the mean potential
per nucleon in it, see Table~\ref{tab_selfconsist}. One can see that while the
original Walecka potential require quite deep potential and large number of baryons,
the modified potentials $B2, C$ expected near the critical point can, due to its longer range, bind a smaller
number of nucleons.
 
\begin{table}[ht]
\begin{center}
\begin{tabular}{|c|c|c|c|}
\hline
 Potential & $N$ & $\langle V\rangle/N$ (MeV) & $V(r=0)$  (MeV) \\ 
 \hline \hline
$A$       & 25.4 & -180 & -295\\
$B2$      & 10.5  & -113  & -207\\
$C$       & 7.8  & -119  & -187\\
\hline
\end{tabular}
\caption{The parameters of the self-consistent clusters for various input potentials, 
all with the same r.m.s. radius $R=2.2$ fm.
$N$ is the integrated number of baryons in the cluster,  $\langle V\rangle/N$ and $V(r=0)$
are the mean potential per baryon, and the potential depth at the center.\label{tab_selfconsist}}
\end{center}

\end{table}

\section{Observables} \label{sec_observables} 

In this section we include some generic discussion of the observables involved.

The thermodynamical susceptibilities in equilibrium ---derivatives of  
$\log Z$ over various chemical potentials of three light quarks--- are usually 
recombined into
\be c(N_B,N_Q,N_S)=\frac{\partial^{N_B+N_Q+N_S}}{\partial^{N_B} \mu_B \partial^{N_Q} \mu_Q \partial^{N_S} \mu_S} \big( \log Z  \big). \ee
We would call those {\it global} observables, because they correspond to 
mean correlation functions of 
{\em fully integrated}  quark densities. Many of these quantities,
up to $N= N_B+N_Q+N_s=6$, are currently calculated on the lattice, see Ref.~\cite{Bazavov:2017dus}.
For their comparison to the heavy-ion data on event-by-event fluctuations  
see e.g. Ref.~\cite{Bazavov:2017tot}.
 
At the opposite end are what we will call {\em local} observables, related to unintegrated local densities. 
For example,  {\it bi-local} distribution function $n(\vec r_1, \vec r_2)$,
which is usually defined in a ``uncorrelated plus a correlation" form. In homogeneous matter it is defined as
\be  \langle n(\vec r_1, \vec r_2) \rangle=  \langle n(\vec r_1)\rangle  \langle n( \vec r_2) \rangle+ C_2(\vec r_1- \vec r_2) \ . \ee
Similar definitions can be given for the $N$-point correlators. Obviously, the local
observables include the full information about the correlations in the system.
However, for $N>2$ they are multi-dimensional functions, which is difficult to work with.
Say, for $N=4$ and homogeneous matter, there are 6 relative distances,
and it is not practical to calculate 6-dimensional histograms. 
 
 Furthermore, as we will see, in bound clusters there are strong velocity-position
correlations, so that in classical approaches we adopt below one  has to
work with the {\em phase space} distributions, e.g. 6-dimensional one body
distribution $f(\vec r,\vec v)$. Their local-in-phase-space correlators
obviously are even of higher dimensions. 
 
 As a result, one needs to invent/use certain observables in between global and local ones. 
Experimentalists naturally use what we would call {\em semi-global} observables, in which
integral is done over the detector acceptance. For example, it can be a certain range
of longitudinal rapidities $y\in [-Y,Y]$ and transverse momenta  $p_\perp\in [p_{\perp,min},p_{\perp,max}]$. Typically, the included
kinematic range is comparable to the excluded one, colloquially known as 50-50 percent setting, maximizing the fluctuations.
One can measure distributions in the number of net protons $P(N_p)$, or electric net charge $P(Q)$, or net strangeness $P(S)$, deduce the
corresponding moments, cumulants, etc., or correlations between these charges. 

 As will be discussed later, for the net-proton case, the kinematical cuts imposed in experimental analyses reduce the
measured multiplicities by a factor around $5-15 \%$ of the total multiplicity (not really following 50-50 setting). Such a reduced
multiplicity allows to reach Poissonian fluctuations of protons and antiprotons, thus observing, for high energies, the Skellam expectations.
 
 Another natural set of observables, which we would call {\em semi-local} ones, in which the densities
(in coordinate space or the phase space) are integrated, but over the same small volume $V$
 \be C(V,N)=\int_V \ \prod_i^N d^3 r_i\ \langle  n(\vec r_1)...  n(\vec r_N)  \rangle \ , \ee
(or analogous small region in the phase space). In studies of clustering we will do,
the effect is of course maximal when $V$ is of the order of the volume of the clusters
produced. 

The last set of observables can in fact be directly observed in experiment, via physical clusters
in the final state. One well known indicator of the baryon clustering is the deuterium d production. The so-called coalescence
models assume that d yield is proportional to
\[  \int d^3r_1 d^3r_2 d^3p_1 d^3p_2 W_d(\vec r_1-\vec r_2, \vec p_1-\vec p_2  )\langle  f(\vec r_1, \vec r_2,  \vec p_1, \vec p_2 )  \rangle  \ , \]
where $W_d$ is the so-called Wigner function related to the deuteron wave function. 
In this case the microscopic volume $V$ is that of the deuterons
or other light nuclei, such as t,$^3$He,$^4$He (and hypernuclei, and their antiparticles), currently observed.  

With out-of-equilibrium production of $^4$He in mind, we will use below a
4-particle observable, a normalized sum of 6 inter-particle distances.

\section{Molecular dynamics+Langevin simulations} \label{sec_simulations}

We use molecular dynamics (MD) simulations to study the agglomeration of nucleons and the time scales required for cluster formation.
Previous applications of MD in nuclear matter to study clustering can be seen in Refs.~\cite{Aichelin:1986wa,Peilert:1991sm}.
We start by testing our code checking total energy and momentum conservation for finite systems, and then proceed to relatively large number of particles.
These are contained in a cubic box with periodic boundary conditions and ``reflections" on all sides, simulating infinite homogeneous matter. We reproduced a number of correlation functions
for gas and liquid argon in a comparable regime, see App.~\ref{app:atomic}. We also apply 
the same approach to a (modified) Walecka potential to access the average properties
of cold nuclear matter, introducing  an effective 
quantum localization potential, see App.~\ref{sec_quantum}. We relegate that study to an appendix because it turns out 
not to be important for systems at temperatures around the freeze-out one.

  Furthermore, we modify the MD code for a nuclear system at fixed temperature, using
Langevin dynamics. The corresponding stochastic forces can be thought of as interactions of ambient heat bath made of multiple mesons.

Presenting the results, we begin with systems with a small number of nucleons,  
starting  with $N=4$ nucleons. Using different temperatures, we check that they group into an average tetrahedral shape minimizing their energy per nucleon by sitting 
at mutual distances close to the minimum of the potential. Then, we will consider a larger number of nucleons and analyze their clustering rate. We will study the nuclear
density profile of these clusters and their higher-order correlation functions.

\subsection{Setting}

 A system with a small number of nucleons is useful to check and validate our MD+L code. Equilibrium configurations can be easily found for such systems. In finite systems we find no extra 
complications due to periodic boundary conditions, such as breaking of the periodicity of the pair-wise potential. Nevertheless, to avoid the particles to escape from the region of interest we
sometimes 	implement a confining potential,
\be \label{eq:WS} U_{c} ( |\vec x|)= V_{WS}(|\vec x|) - V_{WS} (0) \ , \ee
which is written in terms of the Woods-Saxon potential
\be V_{WS}(|\vec x|)= - \frac{V_0}{1+\exp \left( \frac{|\vec x|-R}{a} \right)} \ , \ee
where $V_0$ is the strength of the potential, $R$ is the radius of the volume, and $a$ is the skin depth. Such potential does not appreciable modify the dynamics in the region $|\vec x| \simeq 0$.
  
  The temperature of the system is fixed by the light degrees of freedom (pions and kaons), which we encode in the nucleon Langevin dynamics. Therefore, we introduce a stochastic force to the
nucleons as well as a drag force proportional and opposed to the nucleon momentum, 
\be \left\{
\begin{array}{rcl}
 \dfrac{ d \vec x_i}{dt}   & = & \dfrac{\vec p_i}{m_N} \\
  \\
 \dfrac{ d \vec p_i}{dt} & = &  -  \vec \nabla U_c ( |x_i|) - \sum\limits_{j\neq i} \dfrac{ \partial V (|\vec x_i- \vec x_j|)}{\partial \vec x_i} -\lambda \vec p_i +  \vec \xi_i \ , 
\end{array}
\right.
\ee
with $\lambda$ is the drag coefficient and $\vec \xi$ is the random noise following a white Gaussian distribution,	
\begin{eqnarray}
  \langle \vec \xi_i (t) \rangle  &=& 0 \ , \\
 \langle \xi^a_i (t) \xi^b_j (t') \rangle  &=& 2T\lambda m_N \delta^{ab} \delta_{ij} \delta(t-t') \ , 
 \end{eqnarray}
with $a,b=1,2,3$ and we made use of the fluctuation-dissipation theorem to relate the drag coefficient with the variance of the fluctuation noise. 
A reasonable value for $\lambda$ is taken from the baryon diffusion coefficient
\be \lambda = \frac{T}{m_ND_B} \ , \ee
where the latter is extracted from URASiMA simulations for similar conditions of density and temperature as those used here for the hadronic evolution until freeze-out~\cite{Sasaki:2000fk}, which
is found to be around $D_B \simeq 0.5$ fm. Incidentally this number is not to far to the often quoted estimate using strongly-coupled QGP from holography~\cite{Kovtun:2003wp} $D_B \simeq (2\pi T)^{-1}$ for temperatures around $T_{ch}=120$ MeV.

The final value used in our simulations will be $\lambda=0.256$ fm$^{-1}$. The precise number is not important as long as it allows for a rapid thermalization of the system.

\subsection{Few-nucleon configurations}

 It is instructive to remind the different distribution of distances for the first Platonic polyhedra discussed in Sec.~\ref{sec_correlated}. We summarize them in Table~\ref{tab:polyhedra}.
  
\begin{table}[ht]
\begin{center}
\begin{tabular}{|c|c| c|c|}
\hline
$N$ & Polyhedron & Distances of edges &Proportion \\
\hline
\hline
4 & Tetrahedron & $a$ & 6 \\
6 & Octahedron & $a$,\ $a\sqrt{2}$ & 12:3 \\
8h & Hexahedron (cube) & $a$, \ $a\sqrt{2}$, \ $a\sqrt{3}$ & 12:12:4 \\
8s & Square antiprism & $a,a\sqrt{2}$, \ $a\sqrt{1+\sqrt{2}}$ & 16:4:8 \\
13 & Icosahedron+1 & $a$, \ $a\phi_-$ , \ $a\phi_+$, \ $2a$ & 12:30:30:6 \\
\hline
\end{tabular}
\caption{Summary of the distances of edges of some polyhedra. $a$ denotes the length of the minimal edge.
Included for completeness the cube configuration does not appear as an equilibrium configuration, rather 
the square antiprism. We denote $\phi_\pm \equiv \sqrt{2 \pm2/\sqrt{5}}$.\label{tab:polyhedra}}
\end{center}

\end{table}

\subsubsection{$N=4$: Tetrahedron}

 \begin{figure}[htb]
\begin{center}
\includegraphics[width=7.5cm]{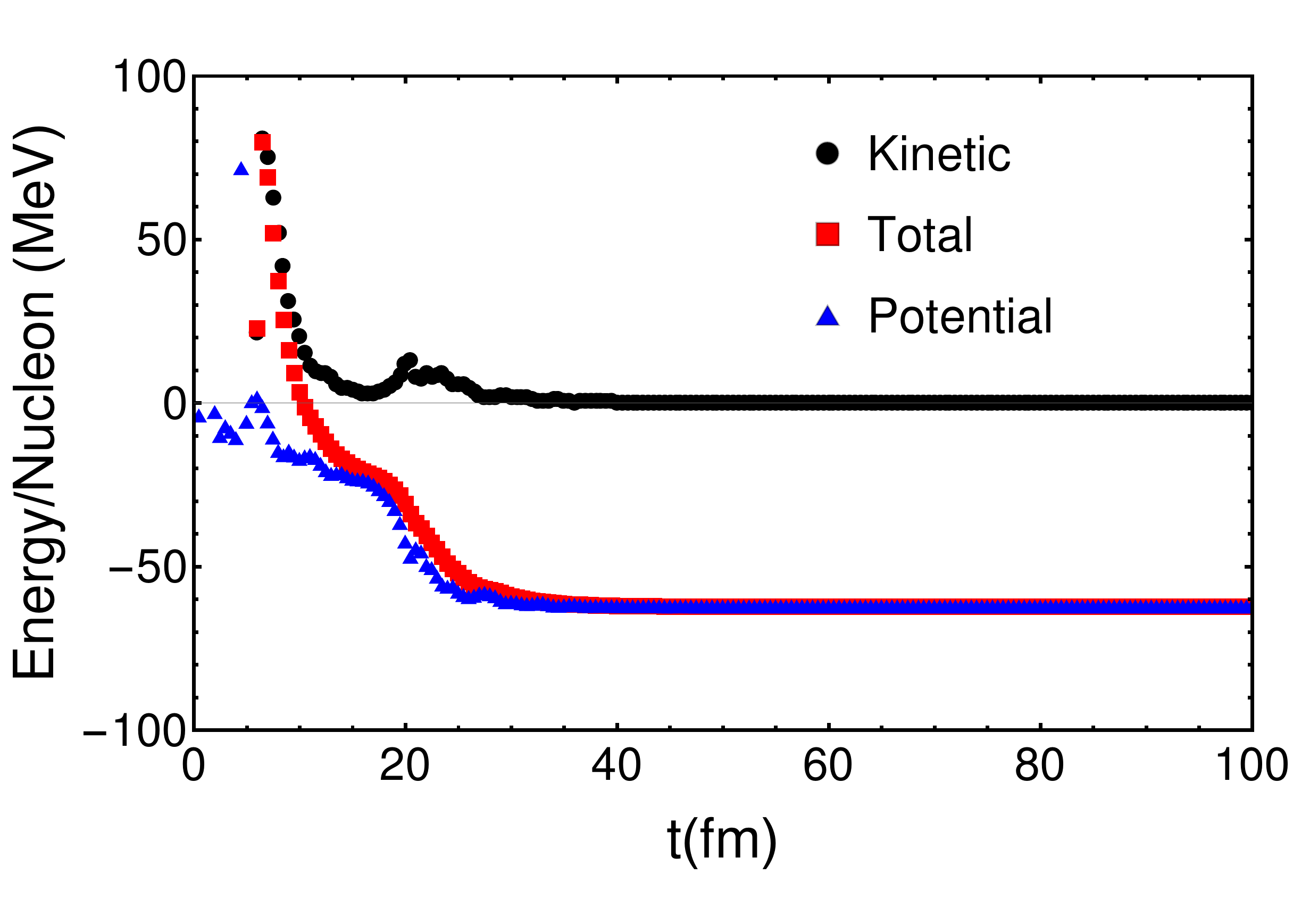}
\includegraphics[width=6.8cm]{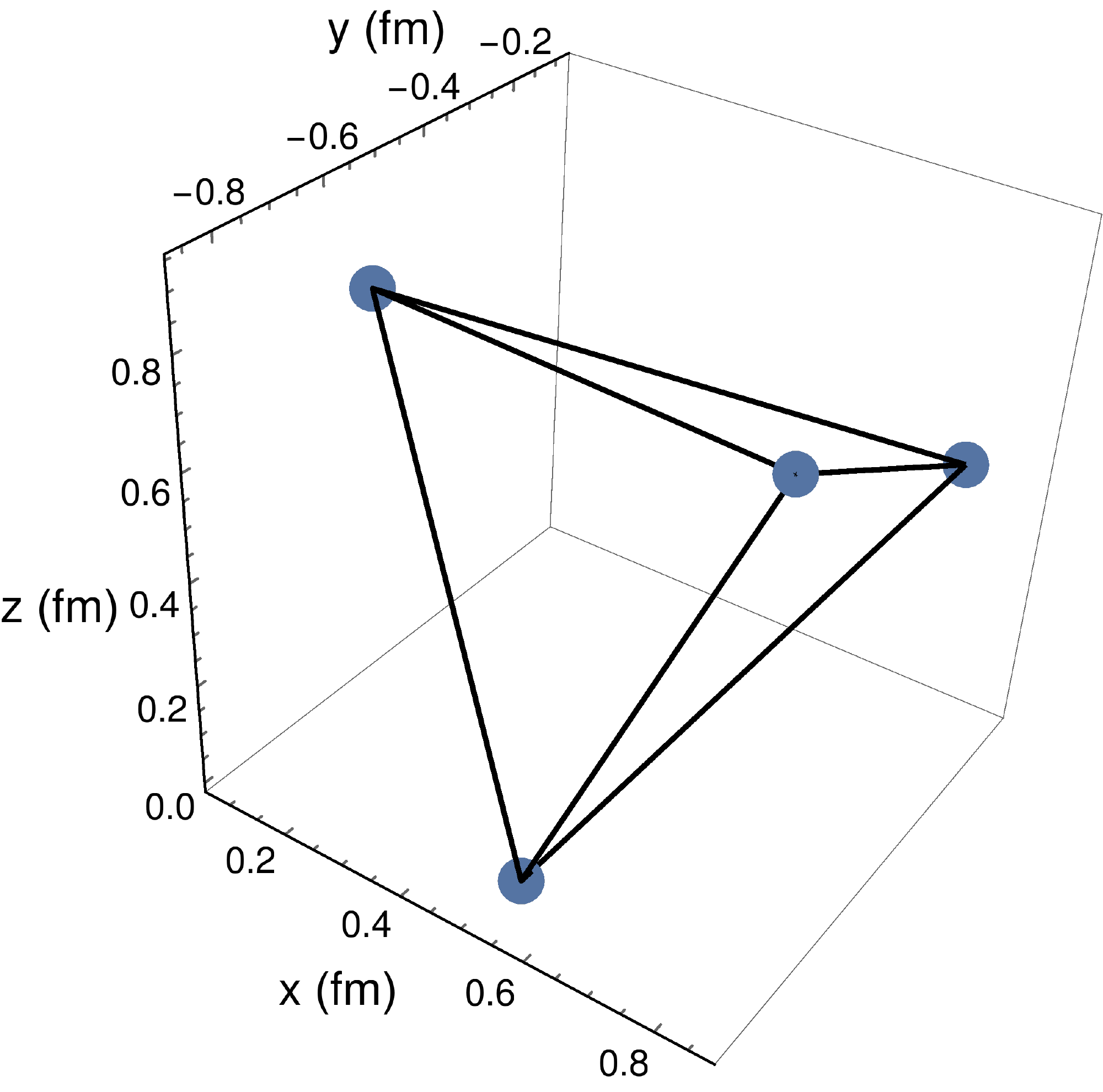}
\caption{Top panel: Kinetic (black), potential (blue) and total (red) energies per nucleon (in MeV) for $N=4$ calculation versus time (fm) at $T=10^{-3}$ MeV.
Bottom panel: Snapshot of the coordinate configuration at some time after equilibration.\label{fig:elN4}}
\end{center}
\end{figure}

 We first apply this MD scheme to a system of $N=4$ particles and $V (x_{ij})=V_A(x_{ij})$ i.e. the unmodified Walecka potential. We first run a simulation at very low temperatures $T=10^{-3}$ MeV to match the
analysis in Sec.~\ref{sec_correlated}. The initial sampling of velocities is done at higher temperatures so that the particles are given some time to acquire their equilibrium configuration. The
evolution of the potential, kinetic and total energies is seen in the top panel of Fig.~\ref{fig:elN4}. While the kinetic energy is negligible, the potential energy per nucleon takes the 
equilibrium value $\langle V\rangle_N=-62.47$ MeV (as predicted from Table~\ref{tab_minima}).

\begin{figure}[htb]
\begin{center}
\includegraphics[width=6.5cm]{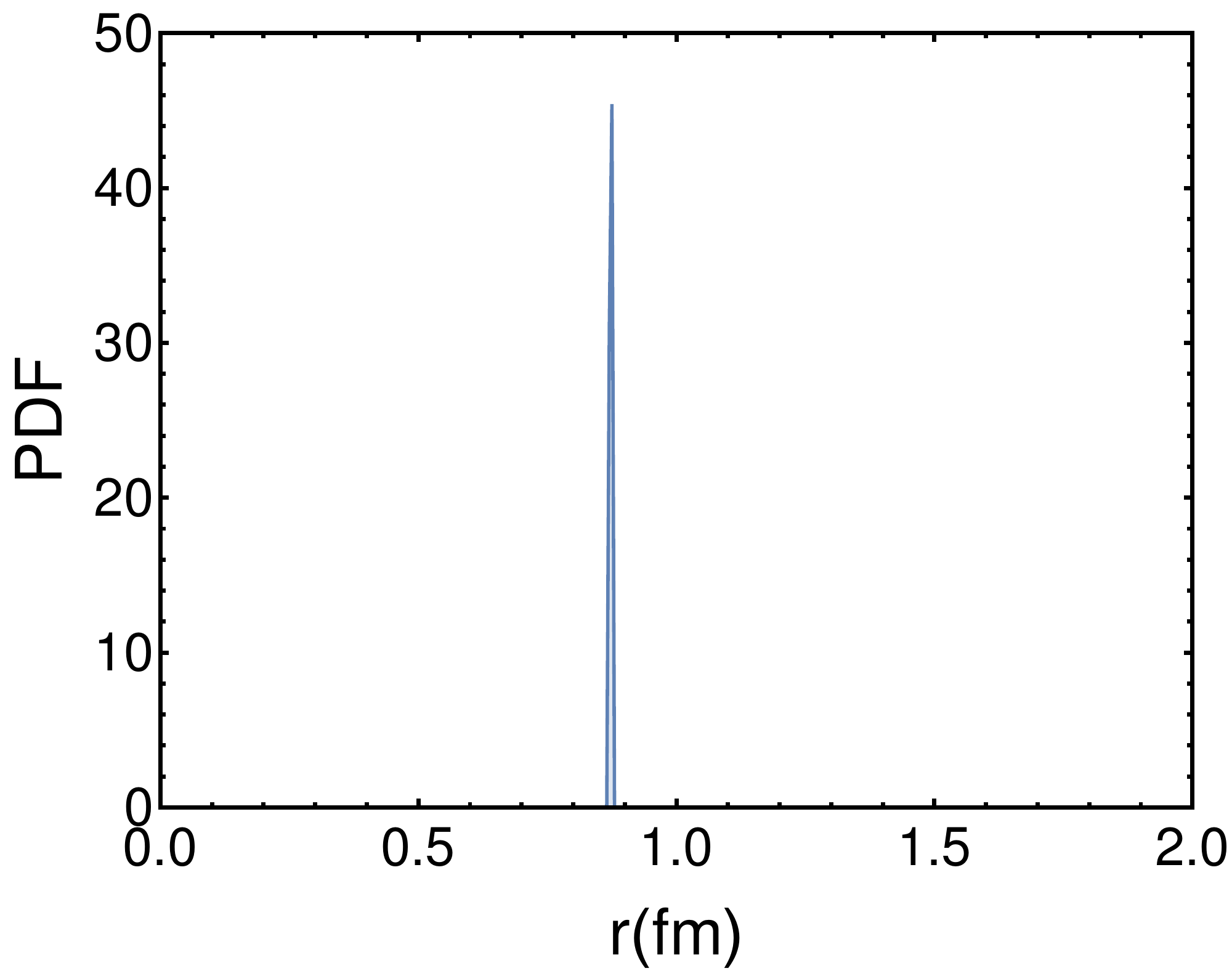}
\includegraphics[width=6.5cm]{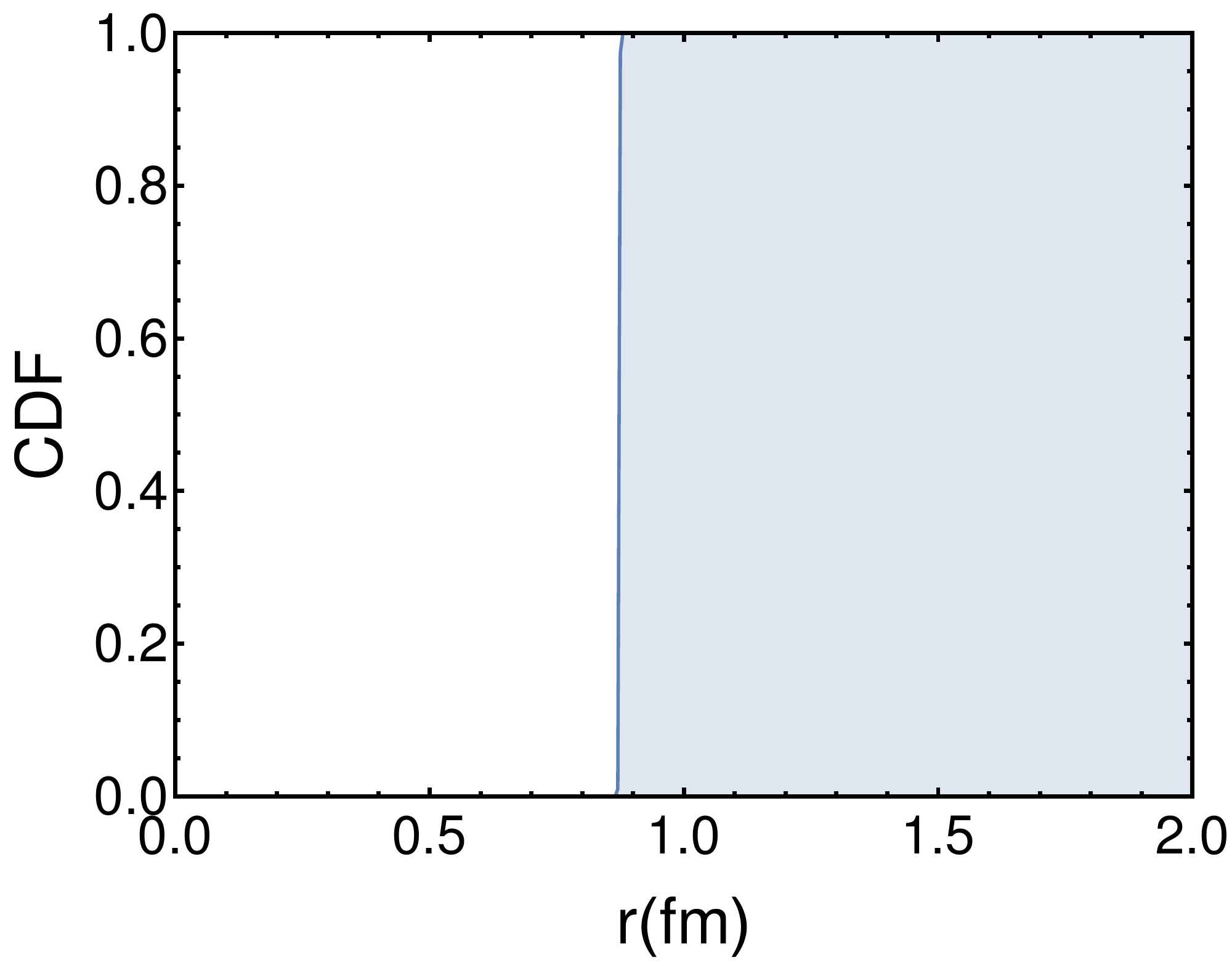}
\caption{Top panel: Histogram of the distance between nucleon pairs for $N=4$ simulation at $T=10^{-3}$ MeV. Bottom panel: Cumulative distribution function.}
\label{fig:distN4T0001}
\end{center}
\end{figure}

  It is easy to see that the geometrical configuration is the expected tetrahedron shape (Fig.~\ref{fig:elN4}, bottom), whose center of mass is evolving with time but the relative distances are preserved.
We perform a (time) distribution of the distances between pairs of nucleons in the top panel of Fig.~\ref{fig:distN4T0001}. The probability distribution function (PDF) shows a single peak at 0.873 fm, which
is the expected value quoted in Table~\ref{tab_minima}, and corresponds to the minimum of the Walecka potential $V_A$. The cumulative distribution function (CDF)
jumps from 0 to 1 precisely at this distance (bottom panel of the same figure). These distribution functions---not particularly informative in this case---will become useful for the cases with larger $N$.

\begin{figure}[htb]
\begin{center}
\includegraphics[width=6.5cm]{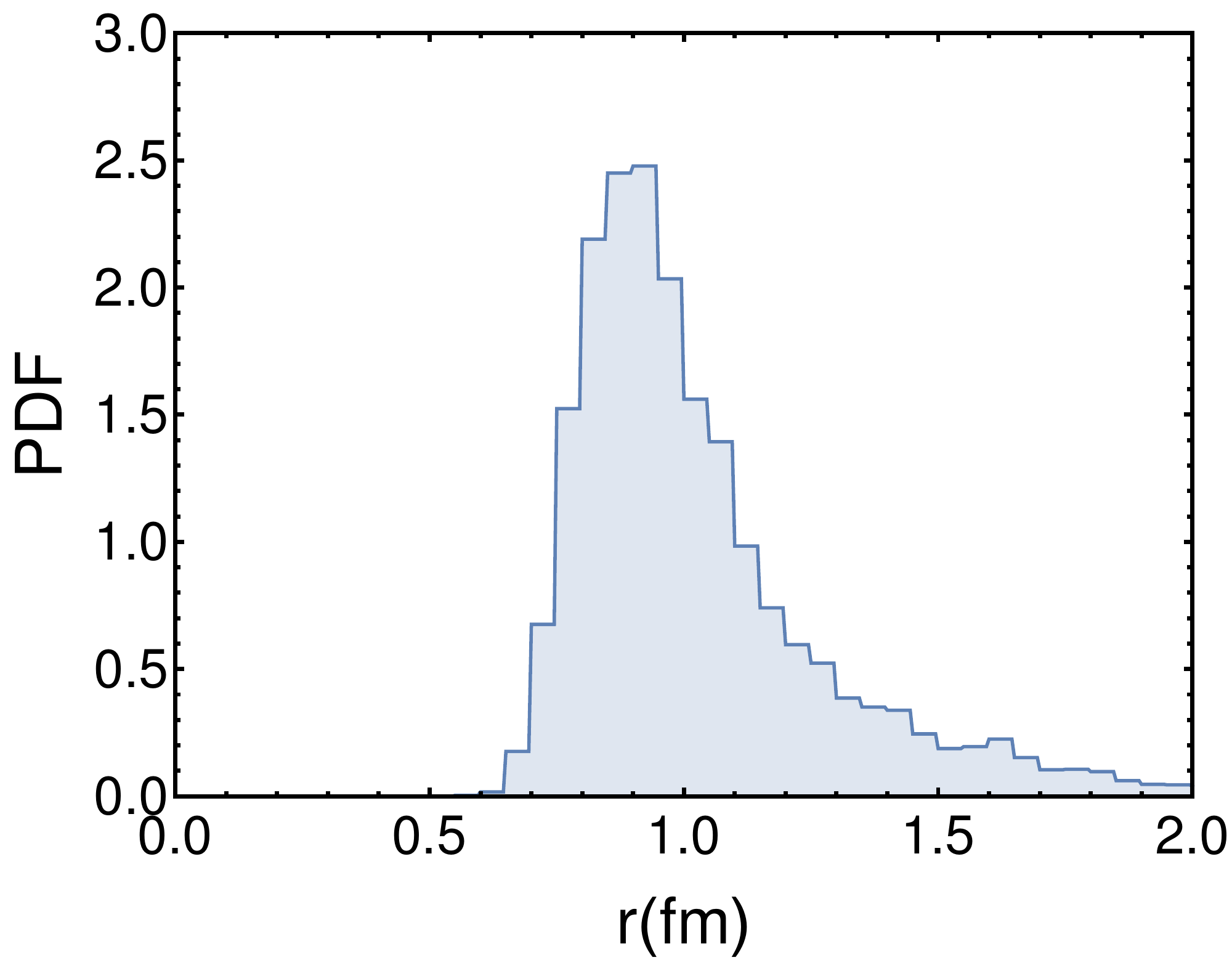}
\includegraphics[width=6.5cm]{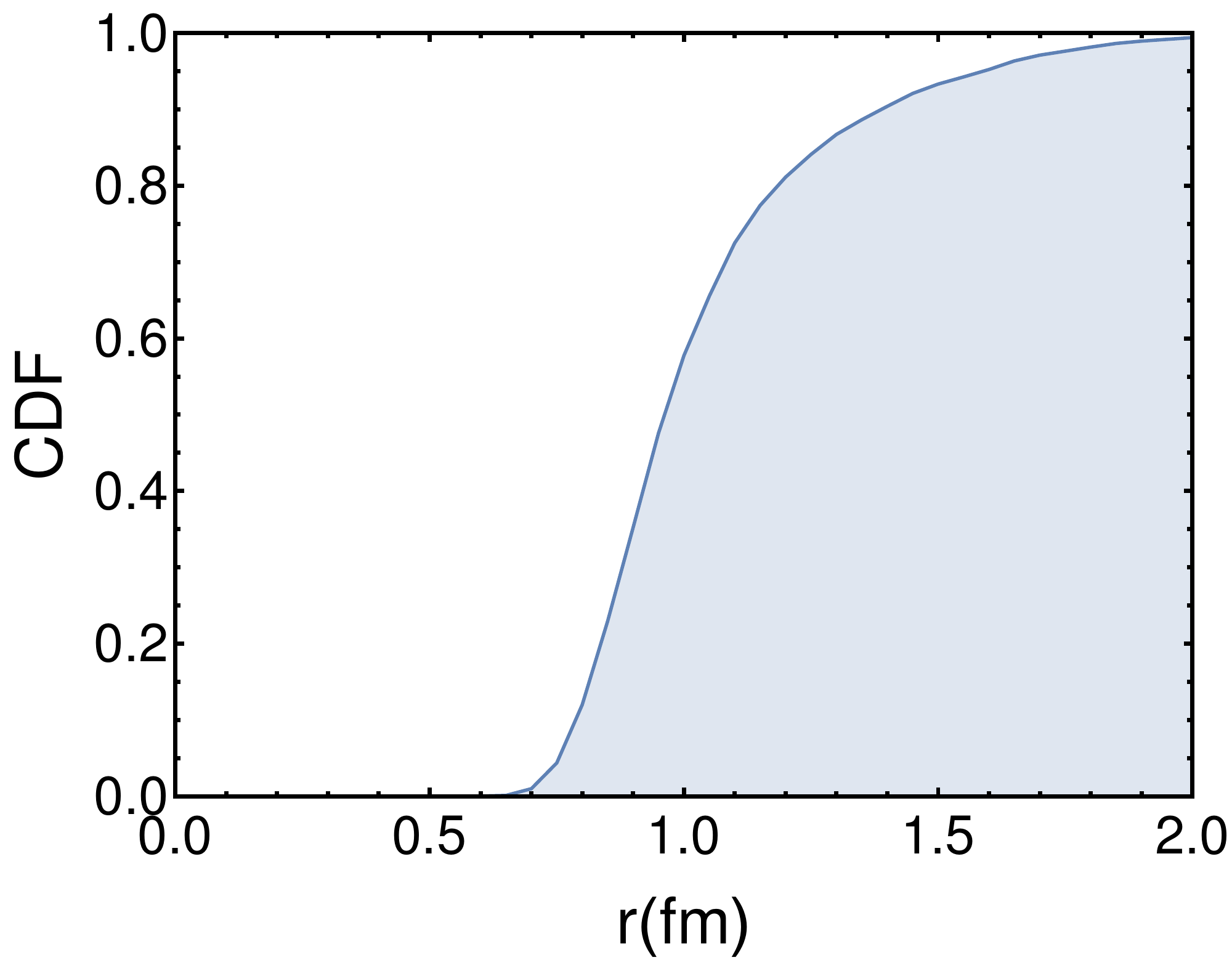}
\caption{Top panel: Histogram of the distance between nucleon pairs for $N=4$ simulation at $T=10$ MeV. Bottom panel: Cumulative distribution function.\label{fig:distN4T10}}
\end{center}
\end{figure}

  An increase of the temperature produces a broadening of the PDF (although the tetrahedral shape is still preserved for small $T$). We present the same distributions for $T=10$ MeV in
Fig.~\ref{fig:distN4T10}. The kinetic thermal energy is the responsible of making the average distances increase with temperature (in this case the average distance is computed as 1.03 fm), eventually
preventing any kind of clustering among nucleons when the temperature dominates over the attractive $NN$ potential.

We make notice that {\it at temperatures of $T=120$ MeV we obtain no bound system for $N=4$ with the 
original Walecka potential $V_A$}. The clustering of four nucleons (and the eventual formation of $^{4}$He) requires a deeper potential for the freeze-out temperatures of baryon-rich HICs.

\subsubsection{$N=6$: Octahedron}

\begin{figure}[htb]
\begin{center}
\includegraphics[width=7.5cm]{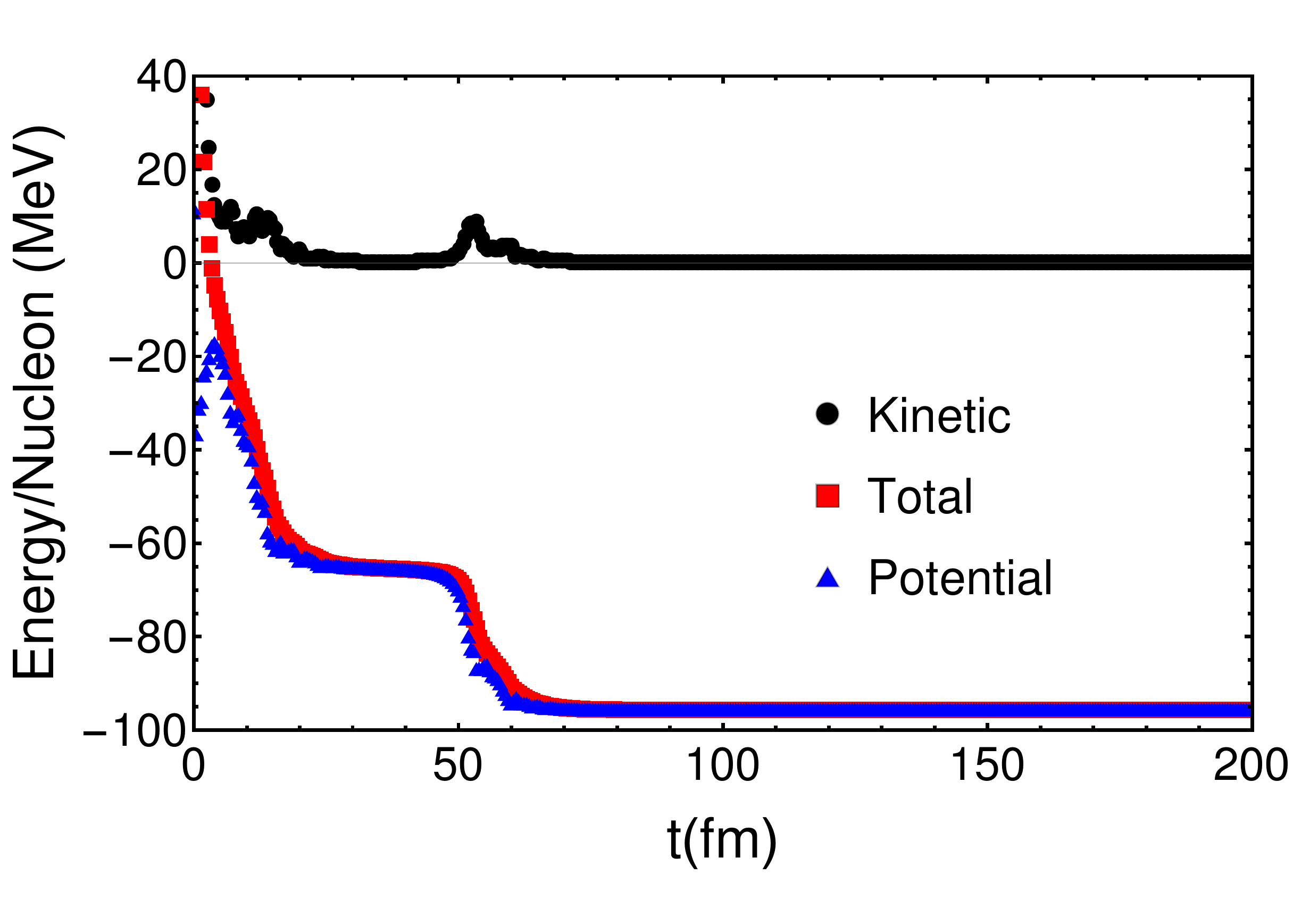}
\includegraphics[width=6.8cm]{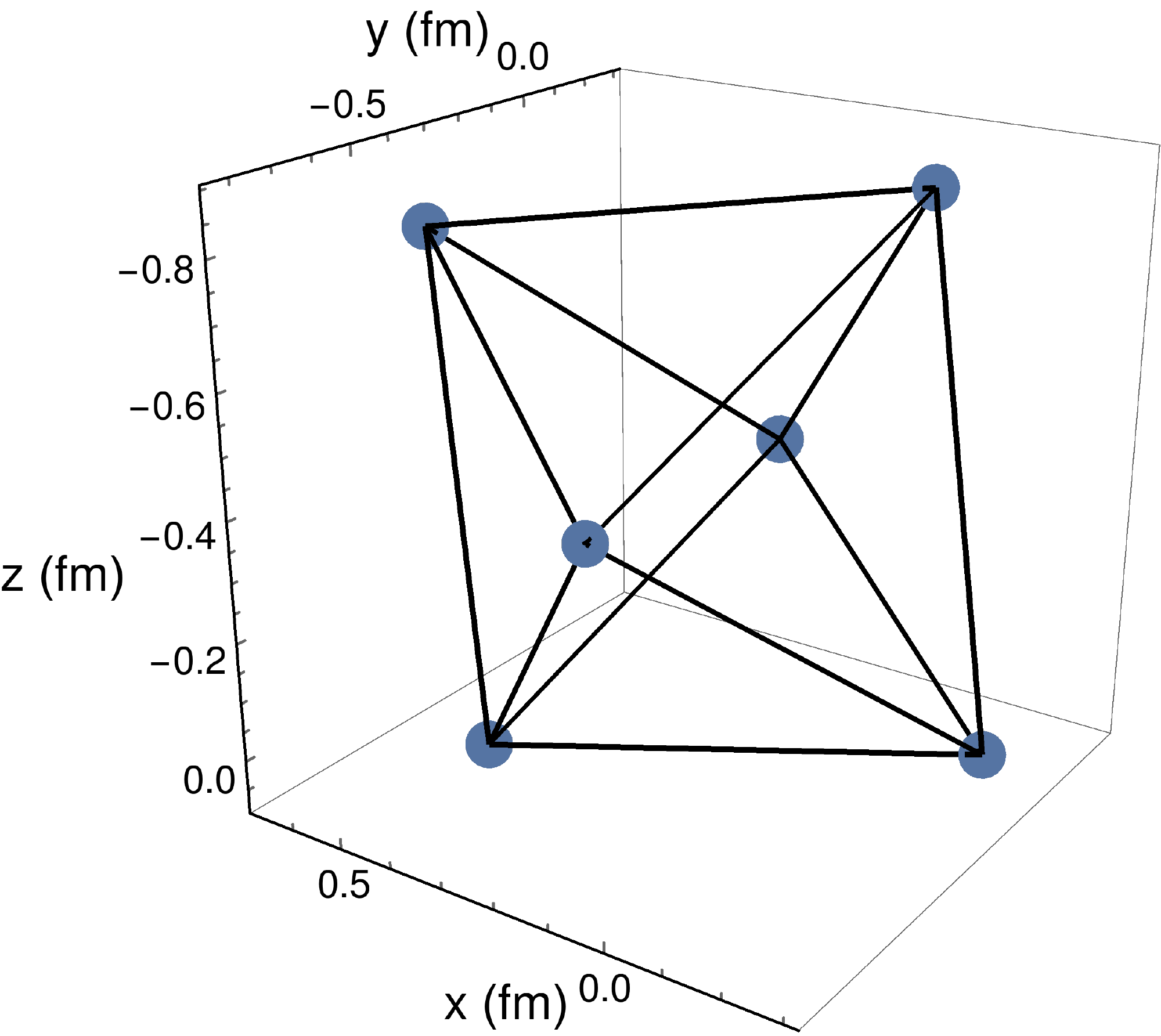}
\caption{Top panel: Kinetic (black), potential (blue) and total (red) energies per nucleon (in MeV) for $N=6$ calculation versus time (fm) at $T=10^{-3}$ MeV.
Bottom panel: Snapshot of the coordinate configuration at some time after equilibration.\label{fig:elN6}}
\end{center}
\end{figure}

The case with $N=6$ nucleons is still relatively simple to predict that the octahedron configuration will be the equilibrium shape. For
$T=10^{-3}$ MeV a fast equilibration is reached (see top panel of Fig.~\ref{fig:elN6}), sitting 
until $t=50$ fm in a metastable minimum of the potential (until the last particle is finally captured by the cluster). 
The final potential energy per nucleon is equal to $\langle V\rangle_N=-95.78$ MeV, in agreement with Table~\ref{tab_minima}.
A snapshot of the spatial configuration is presented in the lower panel of Fig.~\ref{fig:elN6}.

\begin{figure}[htb]
\begin{center}
\includegraphics[width=6.5cm]{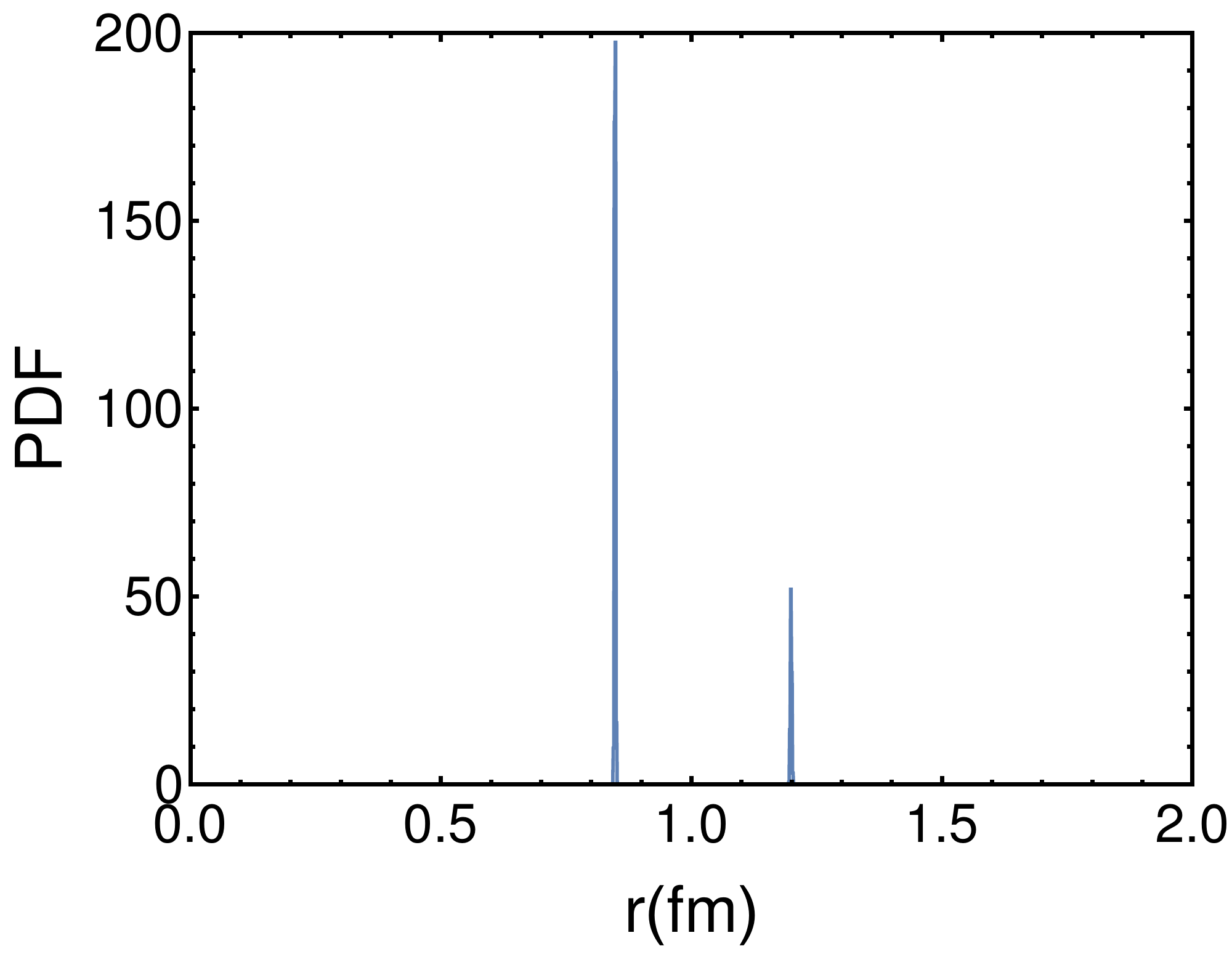}
\includegraphics[width=6.5cm]{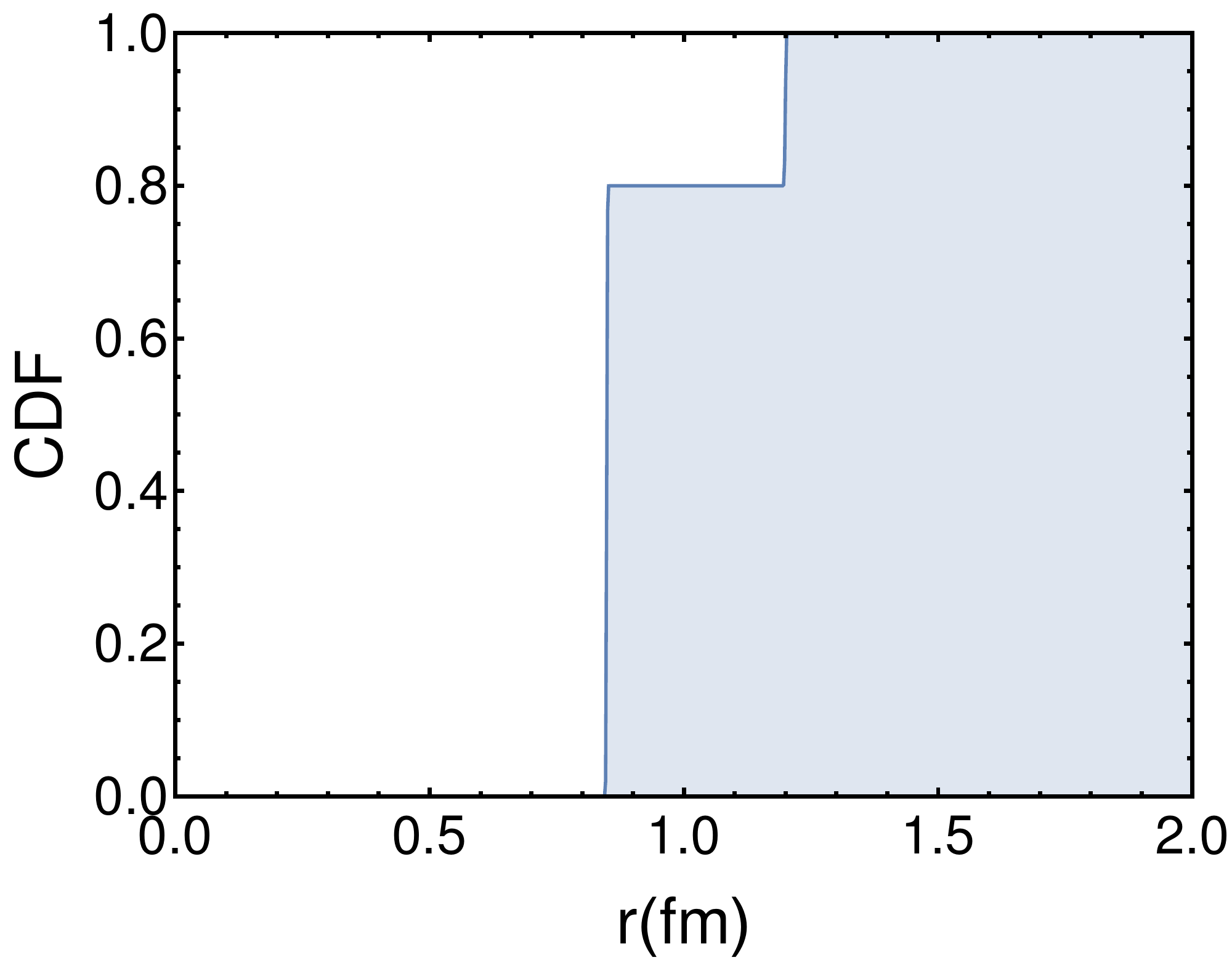}
\caption{Top panel: Histogram of the distance between nucleon pairs for $N=6$ simulation at $T=10^{-3}$ MeV. Bottom panel: Cumulative distribution function.\label{fig:distN6T0001}}
\end{center}
\end{figure}

  The distribution function of mutual distances is presented in the top panel of Fig.~\ref{fig:distN6T0001}. It is possible to verify that the geometry is consistent with the expectations of an octahedron.
This polyhedron has 2 different sets of relative distances, one at some distance $a$ and another at $\sqrt{2} a$ with relative strength 12-to-3. This is precisely what we observe in the histogram, where the ratio 
between the area under the peaks is exactly 4. This can alternatively be checked in the cumulative distribution function of the bottom panel of the same figure. The steps in this function are
located at 0.848 fm, and 1.199 fm, which correspond to the 2 distances between nucleons in the octahedron configuration. The 
minimum distance coincides with the expectations in Table~\ref{tab_minima}, and the second one is a factor $\sqrt{2}$ larger.
  
\begin{figure}[htb]
\begin{center}
\includegraphics[width=6.5cm]{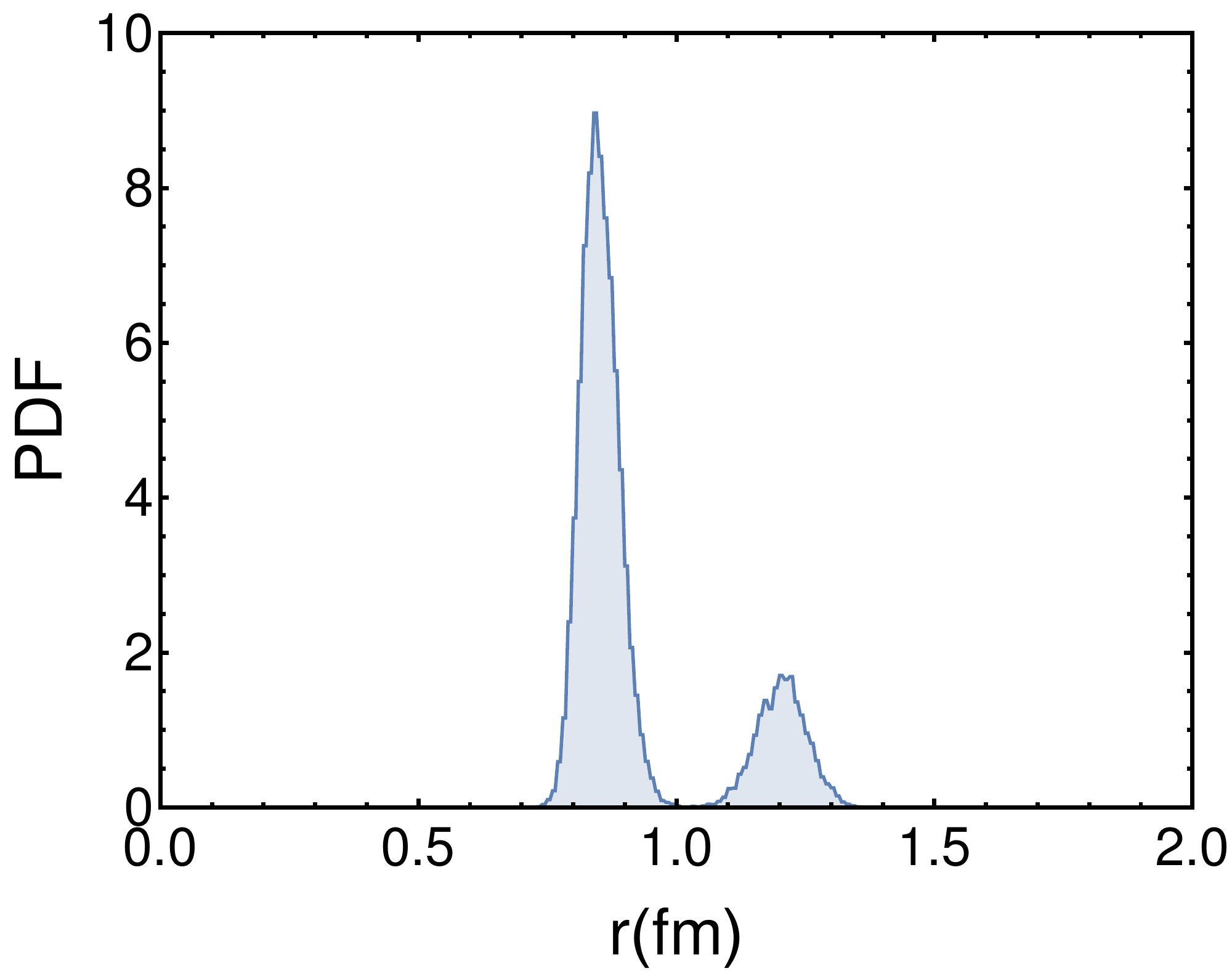}
\includegraphics[width=6.5cm]{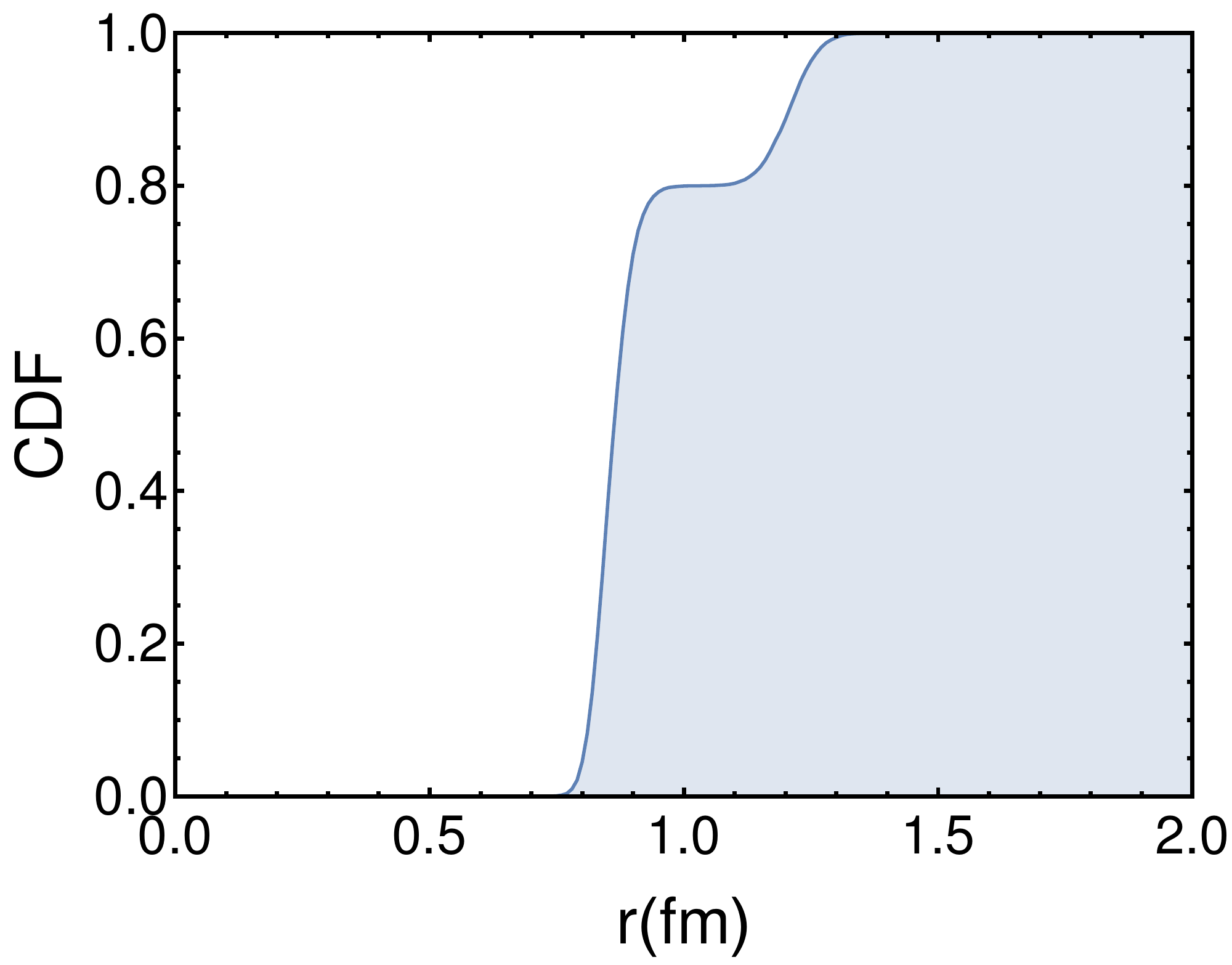}
\caption{Top panel: Histogram of the distance between nucleon pairs for $N=6$ simulation at $T=1$ MeV. Bottom panel: Cumulative distribution function.\label{fig:distN6T1}}
\end{center}
\end{figure}

In Fig.~\ref{fig:distN6T1} we can observe how already at $T=1$ MeV the two peaks are smeared out due to the thermal motion of the nucleons. Nevertheless, it is still possible to identify the octahedron
configuration.

\subsubsection{$N=8$}

  For $N=8$ we notice that the naive expectation of a cubic geometry was already ruled out in Sec.~\ref{sec_correlated} in favor of a square antiprism configuration. The later
 configuration has a lower potential energy for $N=8$ nucleons. The distribution of mutual distances is rather different from the cubic configuration case as seen in Table~\ref{tab:polyhedra}. 

\begin{figure}[ht]
\begin{center}
\includegraphics[width=6.5cm]{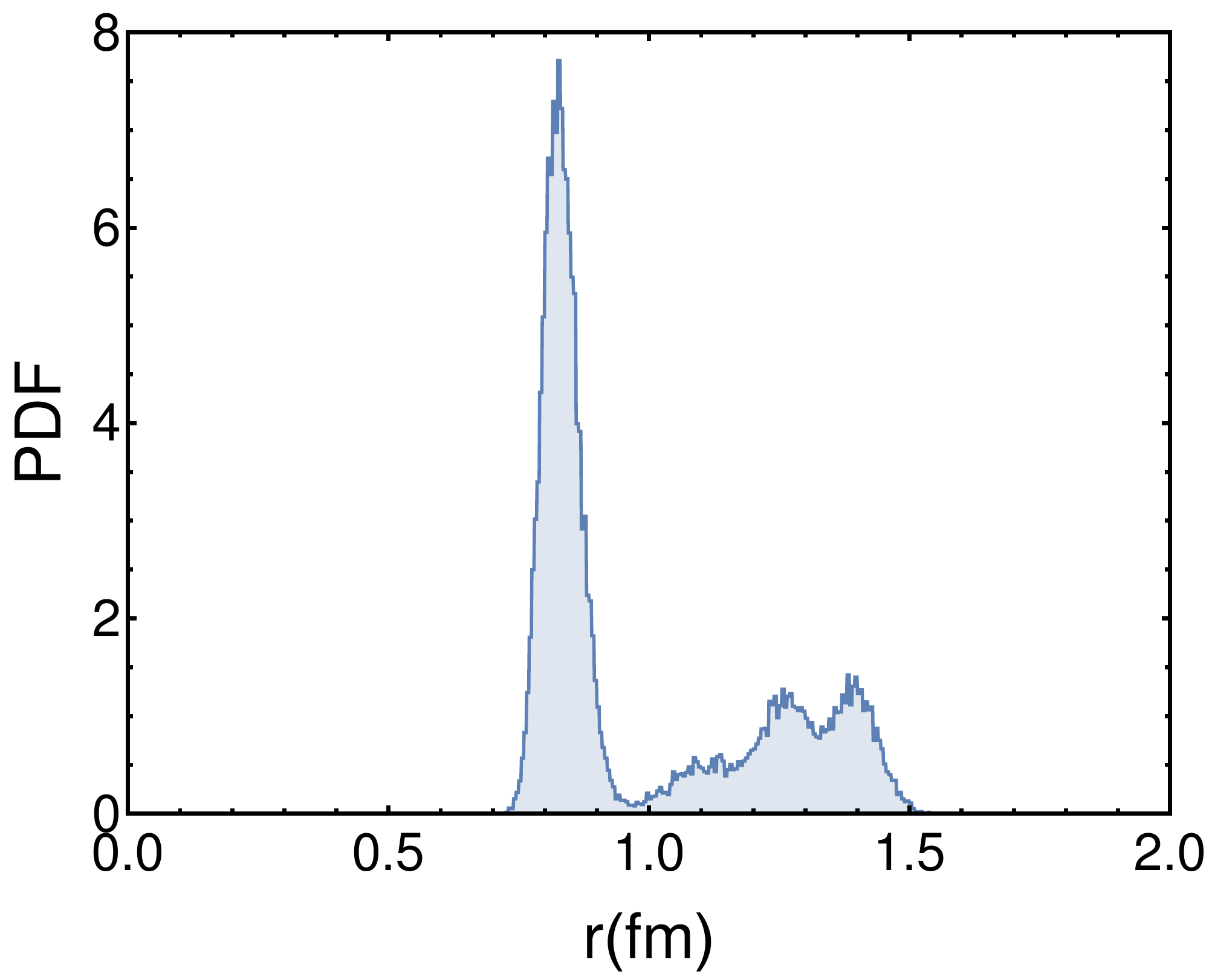}
\caption{Top panel: Histogram of the distance between nucleon pairs for $N=8$ simulation at $T=1$ MeV. This configuration is apparently signaling a square antiprism shape, when comparing 
the distribution with the ideal distribution in Table~\ref{tab:polyhedra}.\label{fig:distN8T1}}
\end{center}
\end{figure}

\begin{figure}[ht]
\begin{center}
\includegraphics[width=6.5cm]{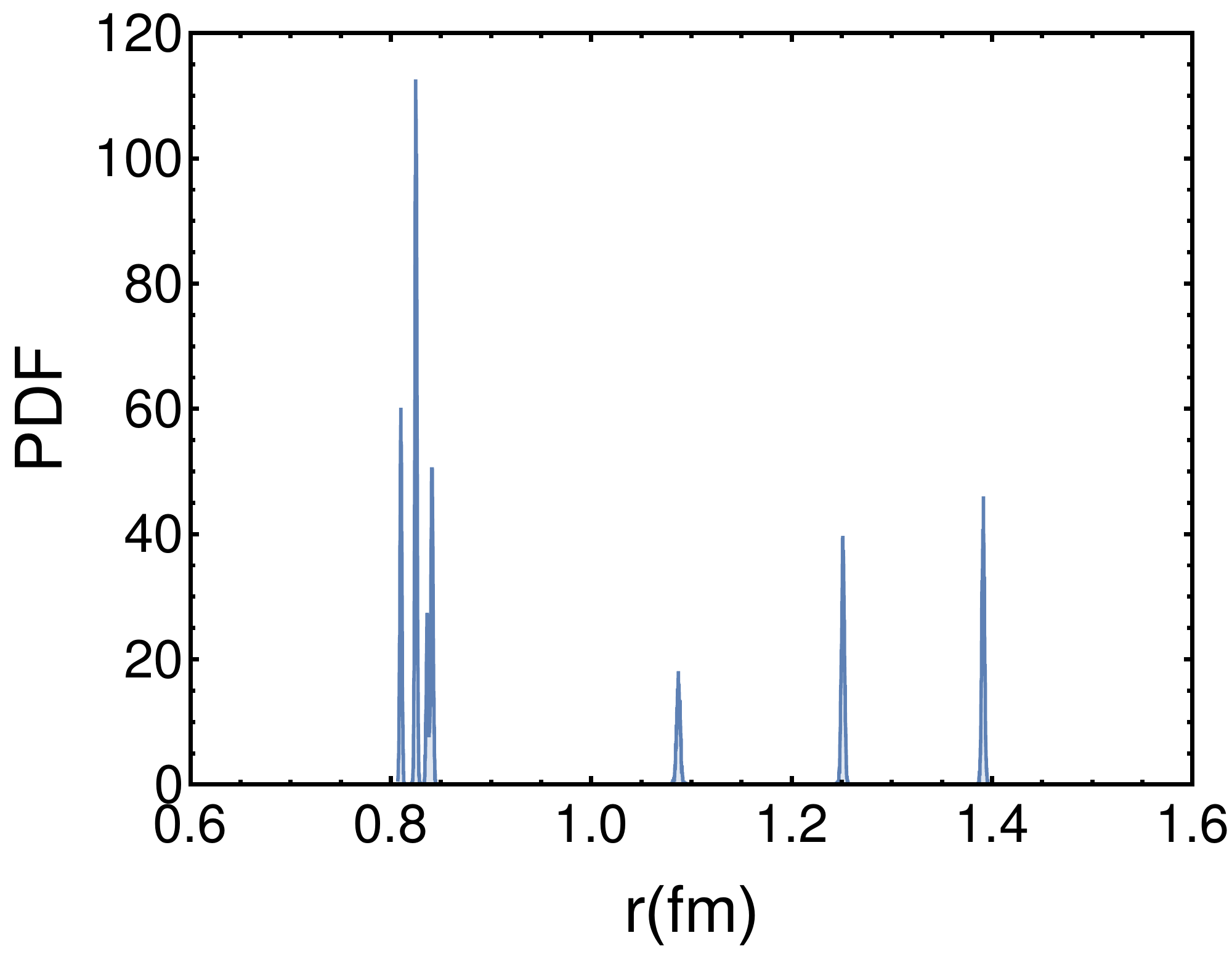}
\includegraphics[width=6.8cm]{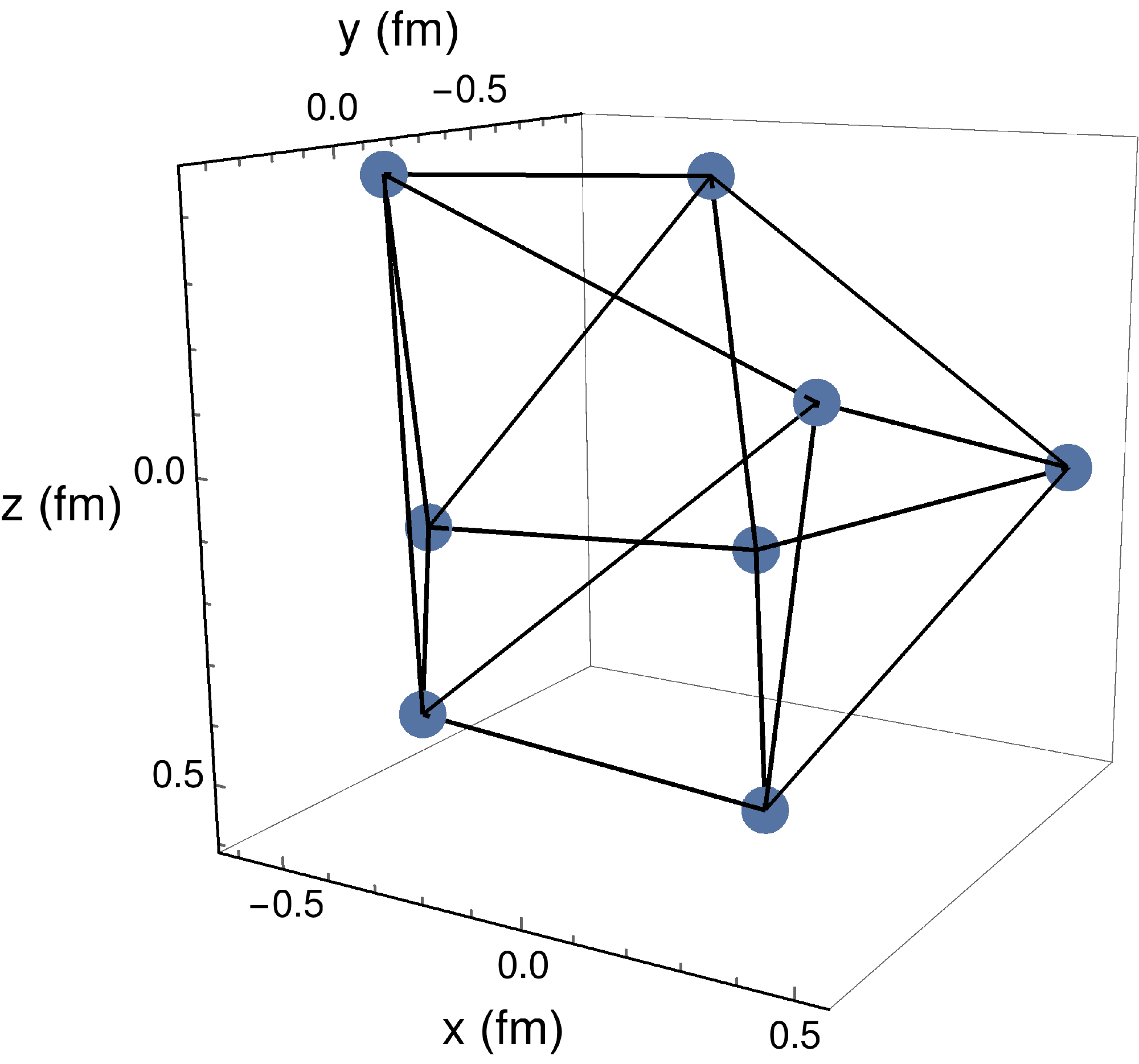}
\caption{Top panel: Histogram of the distances between nucleon pairs for $N=8$ simulation at $T=10^{-3}$ MeV. Bottom panel: Spatial configuration of the $N=8$ nucleons after equilibration.\label{fig:distN8T0001}}
\end{center}
\end{figure}

  A calculation at finite temperature $T=1$ MeV seems to be roughly consistent with this expectation. The PDF shown in Fig~\ref{fig:distN8T1} is clearly inconsistent with a cubic configuration
after comparing to the numbers in Table~\ref{tab:polyhedra}. To test the square antiprism configuration we run a calculation at $T=10^{-3}$ MeV. The resulting PDF shown in Fig.~\ref{fig:distN8T0001}
shows that this distribution is much richer and not consistent with this geometry. The potential energy per particle at $T=10^{-3}$ MeV
is $\langle V\rangle_N =-119.45$ MeV, also not consistent with neither cube of square antiprism (see Table~\ref{tab_minima}). We were not able to identify the precise geometrical shape
(shown in the bottom panel of Fig.~\ref{fig:distN8T0001}), but we have classified 7 different distances with relative weights 2:4:1:2:1:2:2.

\subsubsection{$N=12+1$: Icosahedron+1}

\begin{figure}[ht]
\begin{center}
\includegraphics[width=6.5cm]{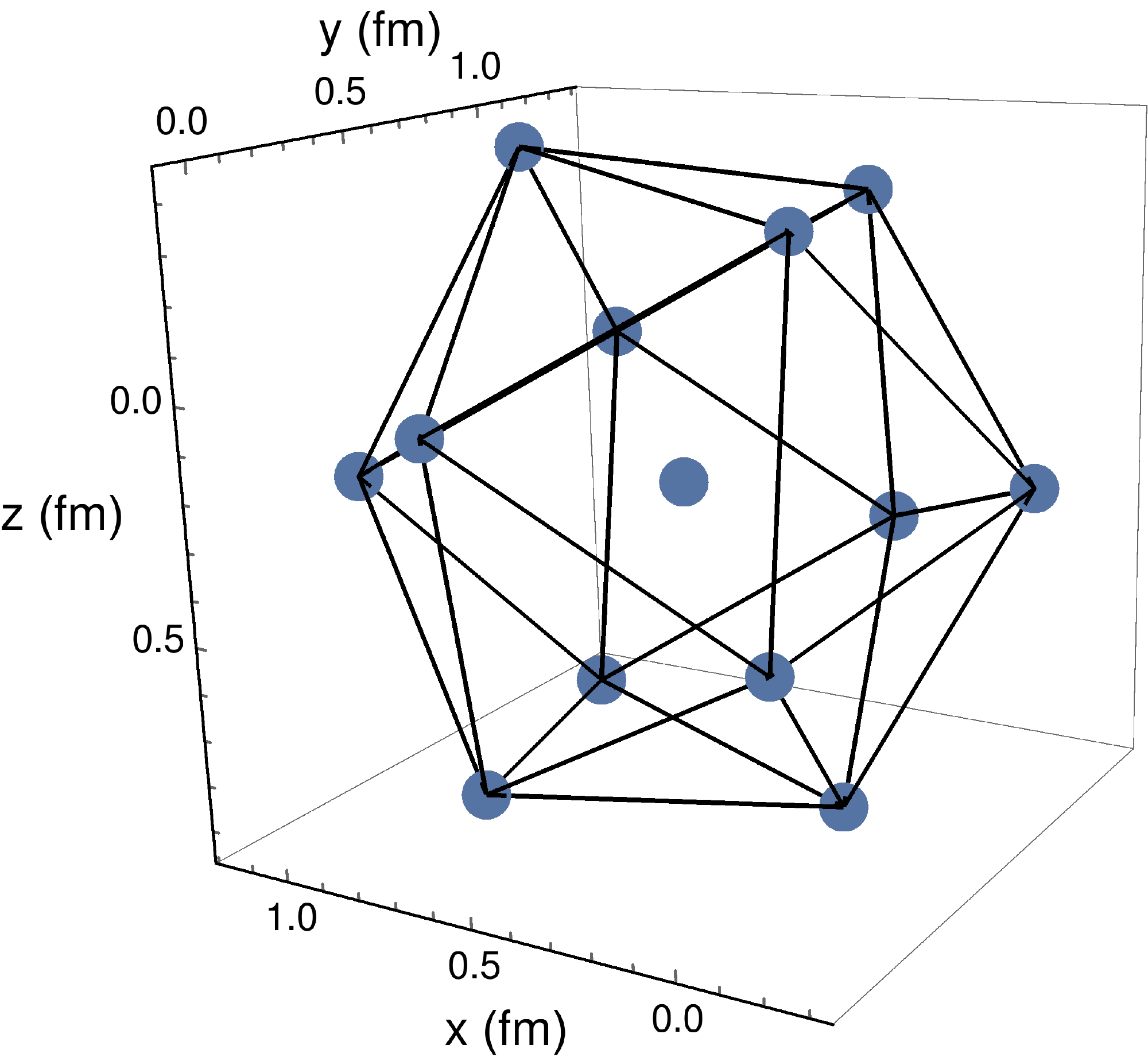}
\includegraphics[width=6.5cm]{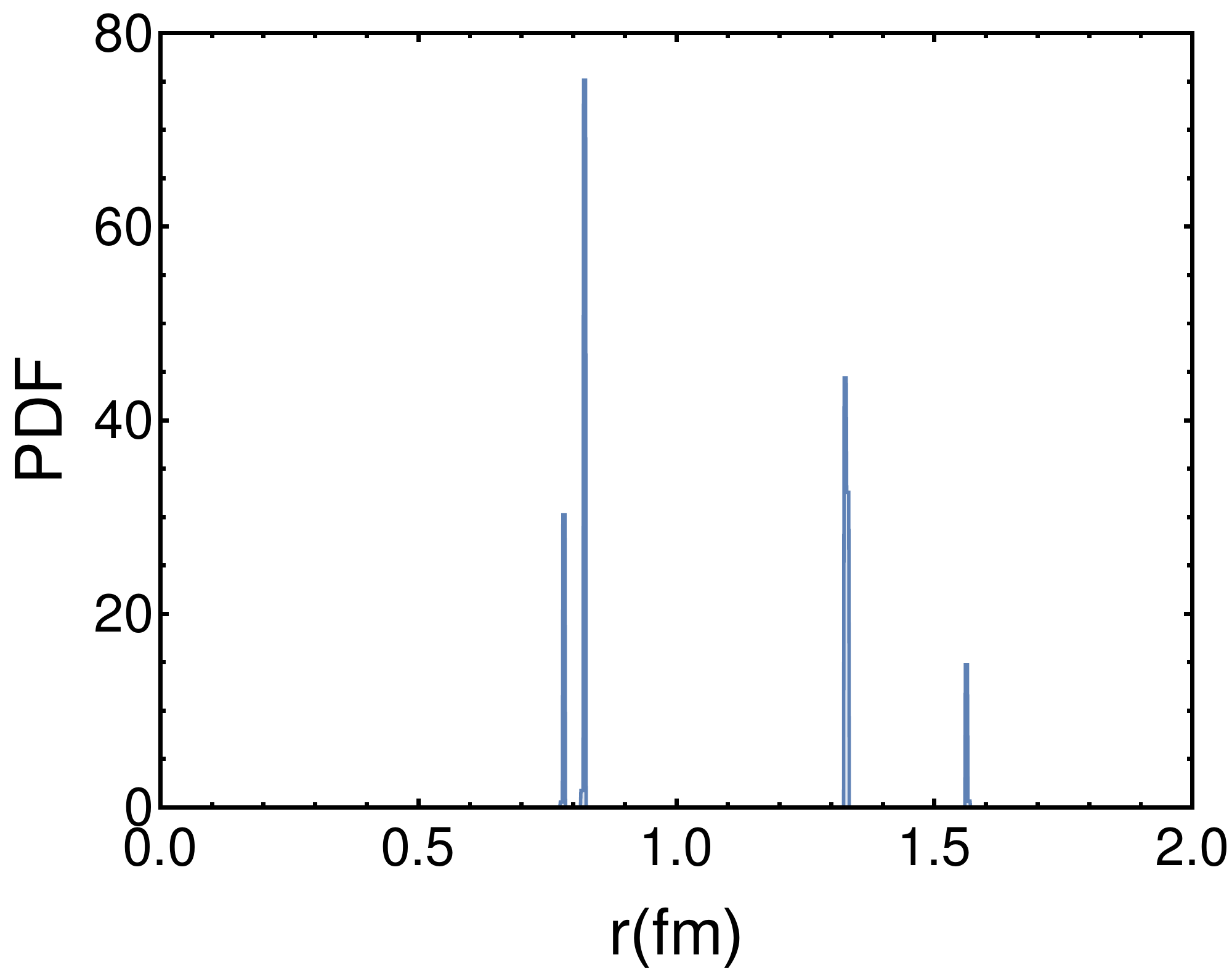}
\includegraphics[width=6.5cm]{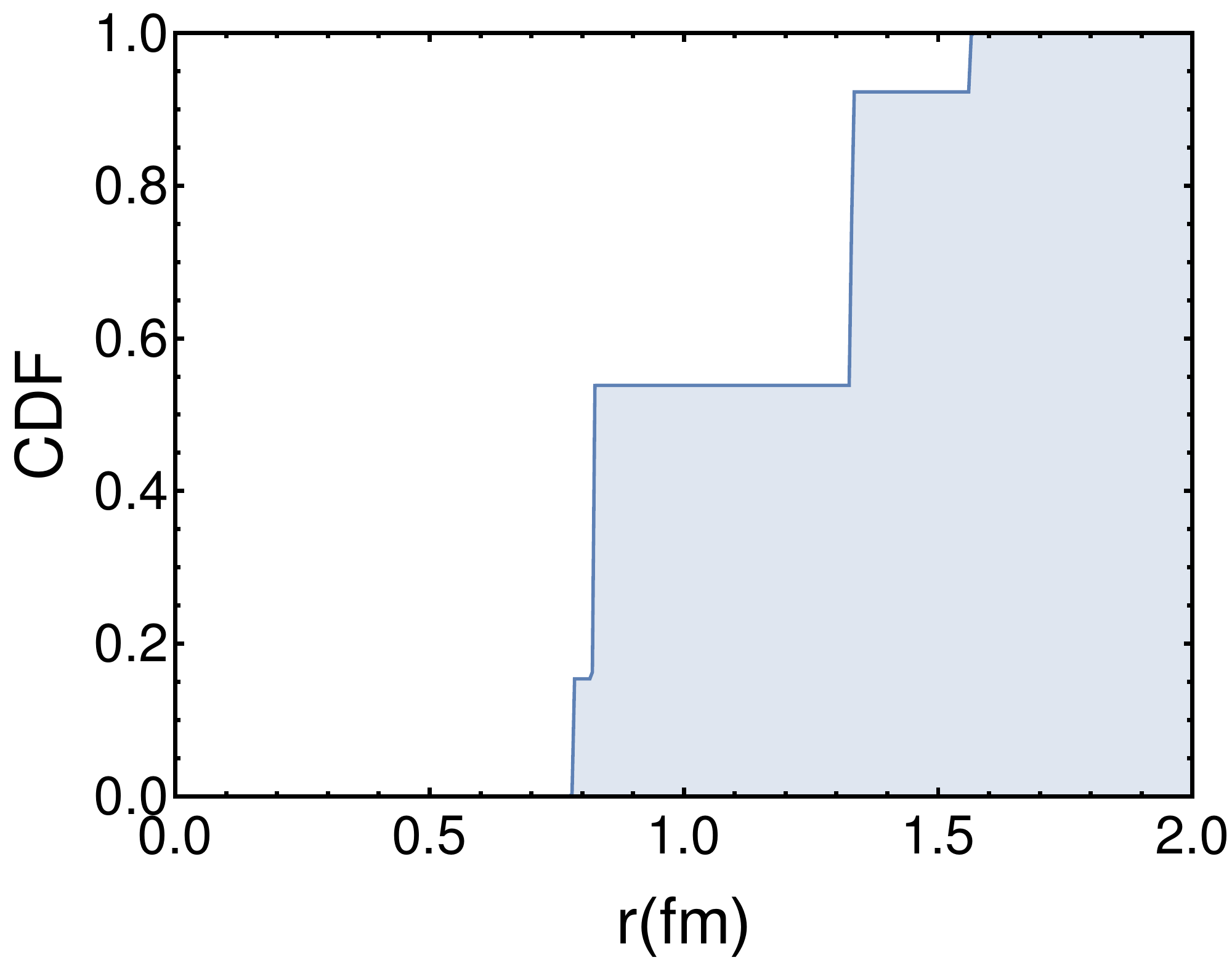}
\caption{Top panel: Spatial configuration of $N=13$ nucleons at $T=10^{-3}$ MeV at some arbitrary time after equilibration. Middle panel: Histogram of the distances between nucleon pairs
for the same simulation. Bottom panel: Cumulative distribution function.\label{fig:distN13T0001}}
\end{center}
\end{figure}

We conclude the study of small cluster by considering $N=12+1$ nucleons at $T=10^{-3}$ MeV, where the expected configuration is an icosahedron plus one nucleon at the center.
It is easy to see by naked eye that the geometrical configuration resembles this expectation. In the top panel of Fig.~\ref{fig:distN13T0001} we present a snapshot of the spatial configuration 
at some time after equilibration. In the middle panel we also present the distribution of (78) mutual distances. We observe 4 different sets of distances, and with a relative
weight (see cumulative distribution function in the middle of the same figure) in excellent agreement with the expectations of Table~\ref{tab:polyhedra}. Finally, the
minimum distance (position of the first peak of the distribution) is 0.782 fm, and the potential energy per nucleon 
obtained is $\langle V\rangle_N=-177.32$ MeV, both in agreement with the values in Table~\ref{tab_minima}.

\subsection{Clustering  at freeze-out temperatures}

   In this section we describe simulations following the scheme presented in the previous section (MD+Langevin with modified Walecka potentials). The 
number of nucleons is large $N=128$, and the temperature is fixed at the typical freeze-out temperatures $T_{kin}=120$ MeV~\cite{Adamczyk:2017iwn}.

\begin{figure}[ht]
\begin{center}
\includegraphics[width=6.8cm]{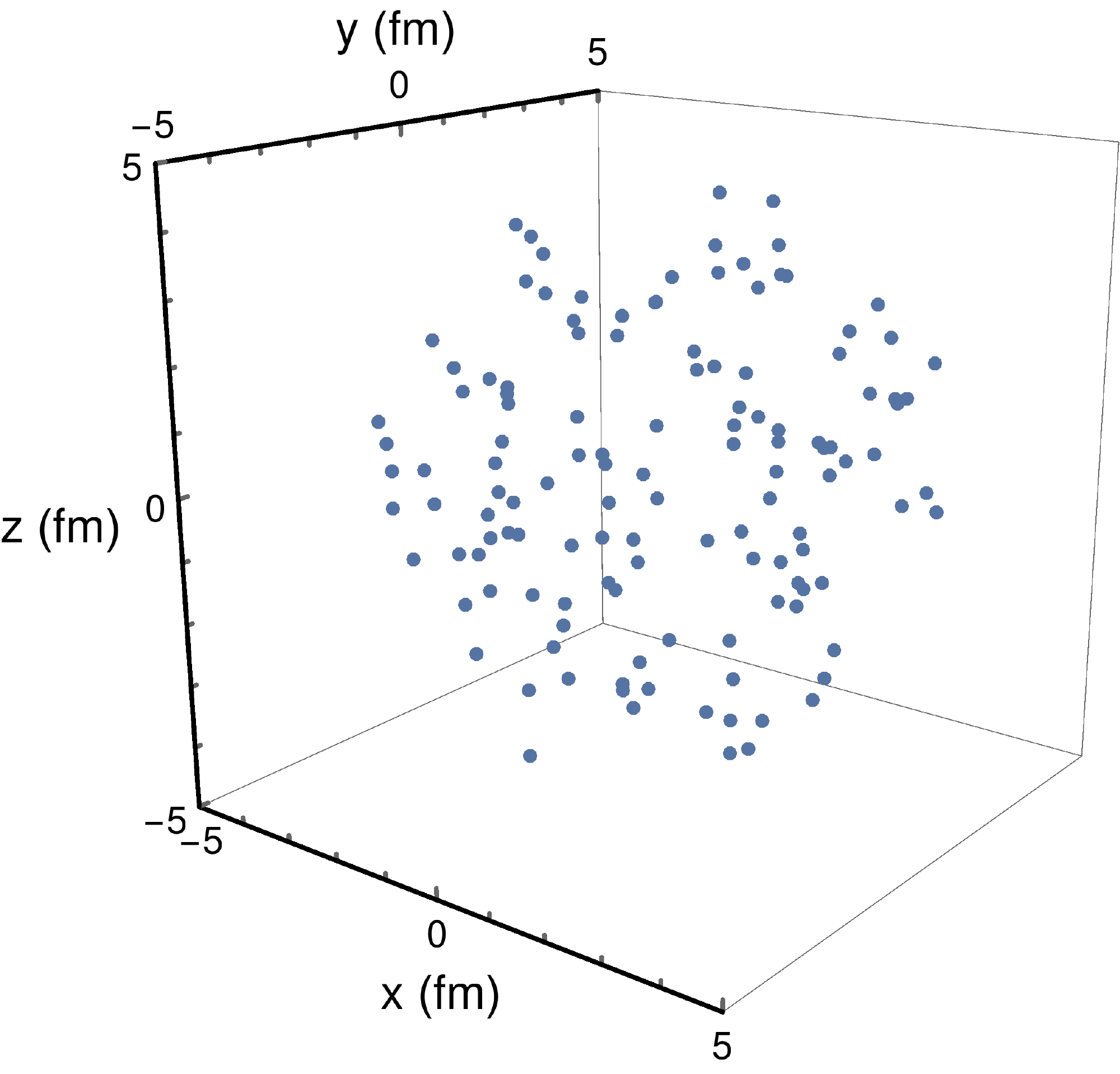}
\includegraphics[width=6.8cm]{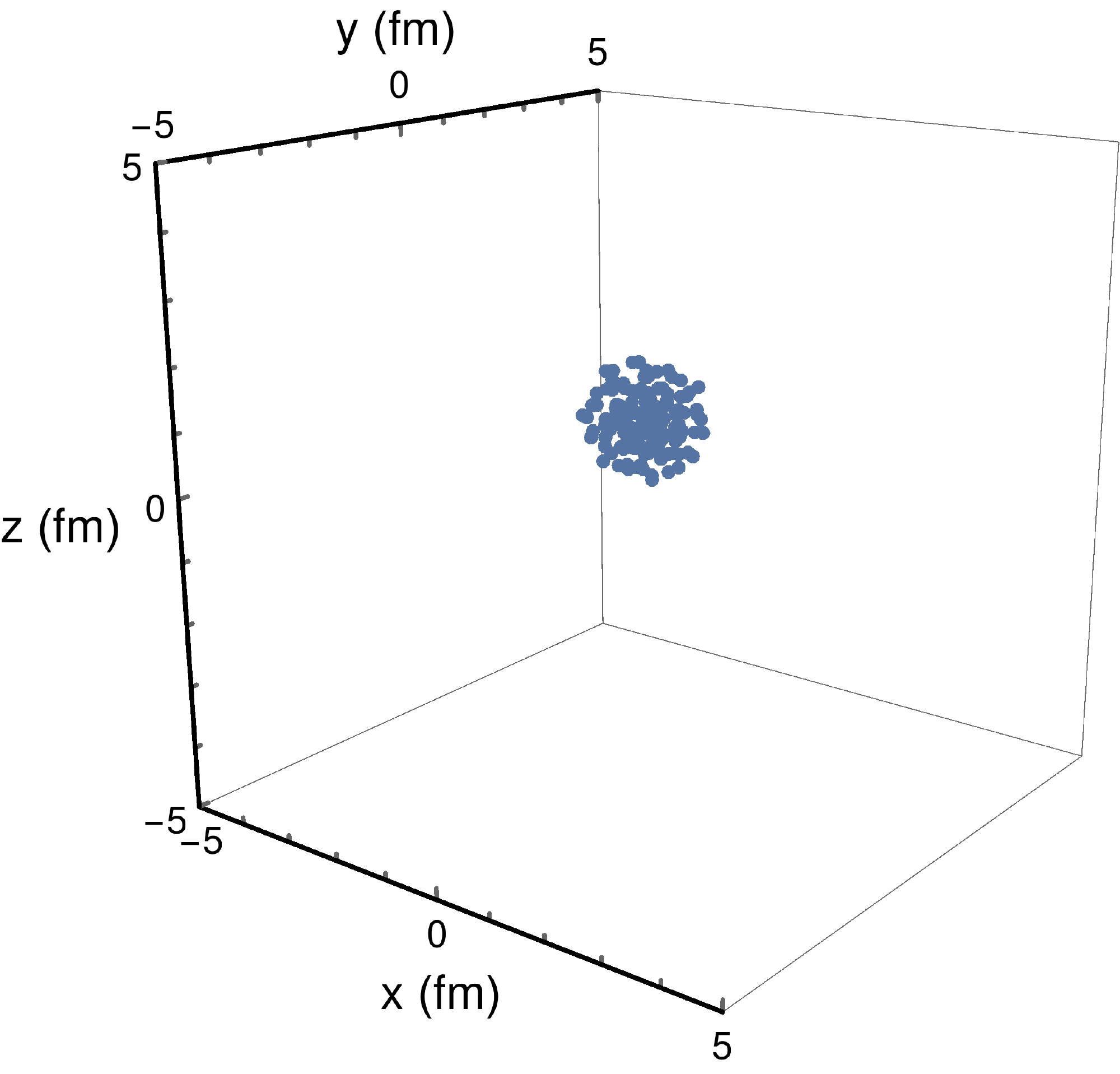}
\caption{Configuration of $N=128$ nucleons in coordinate space at initial time $t=0$ fm (top panel) and at some arbitrary time after full equilibration $t>250$ fm (bottom panel).}
\label{fig:coorsN128}
\end{center}
\end{figure}

   In this section we use the potential $V_{A}$ to see how a deep potential can bind nucleons and eventually produce a large 
cluster. In this example we look for a clear example of such a cluster, and to apply various systematic procedures to analyze its internal structure. The initial state and a
configuration after its equilibration are shown in Fig.~\ref{fig:coorsN128}. The initial geometry is spherical with a density of $n=0.16$ fm$^{-3}$.

\begin{figure}[ht]
\begin{center}
\includegraphics[width=6.5cm]{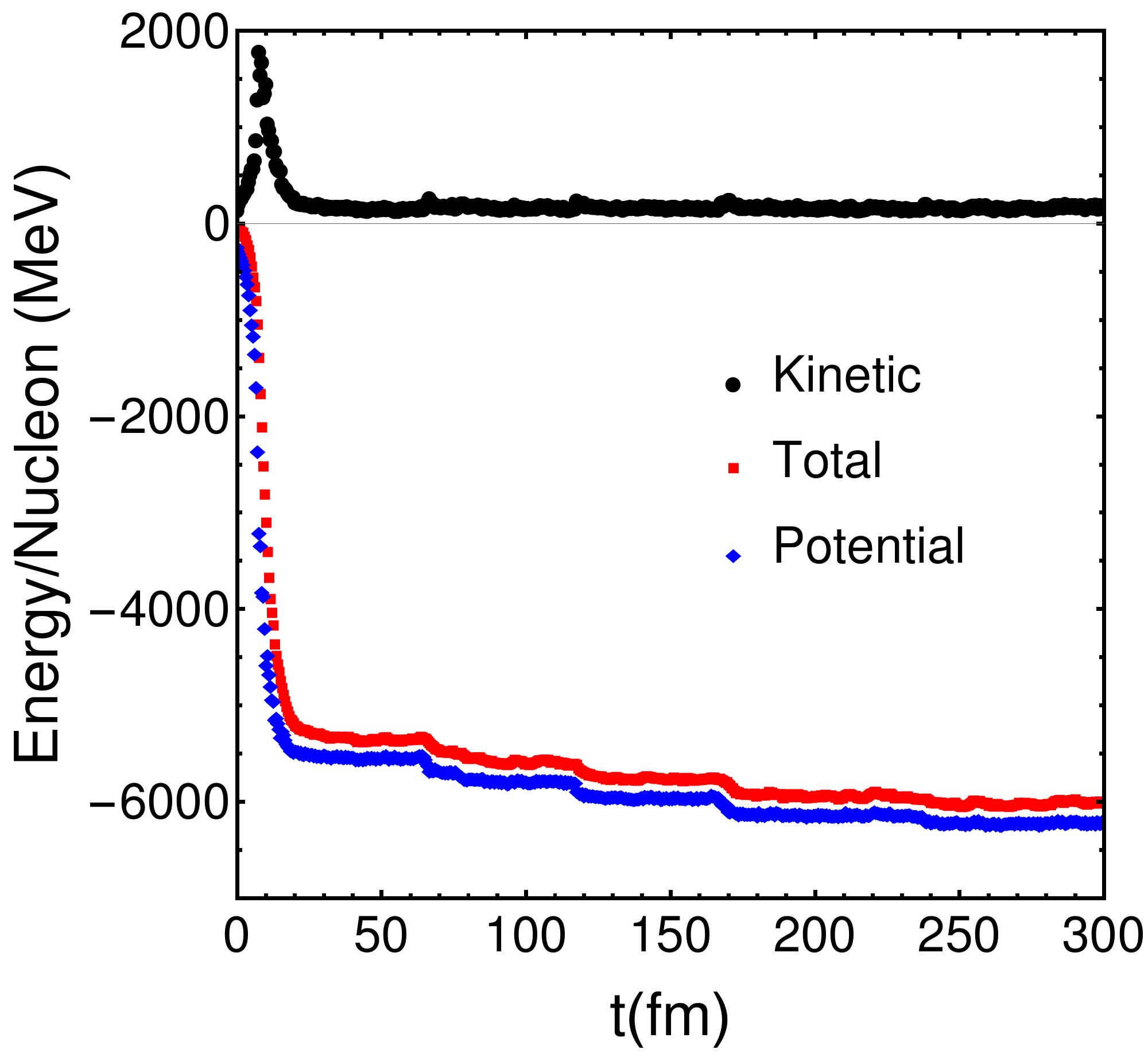}
\includegraphics[width=6.5cm]{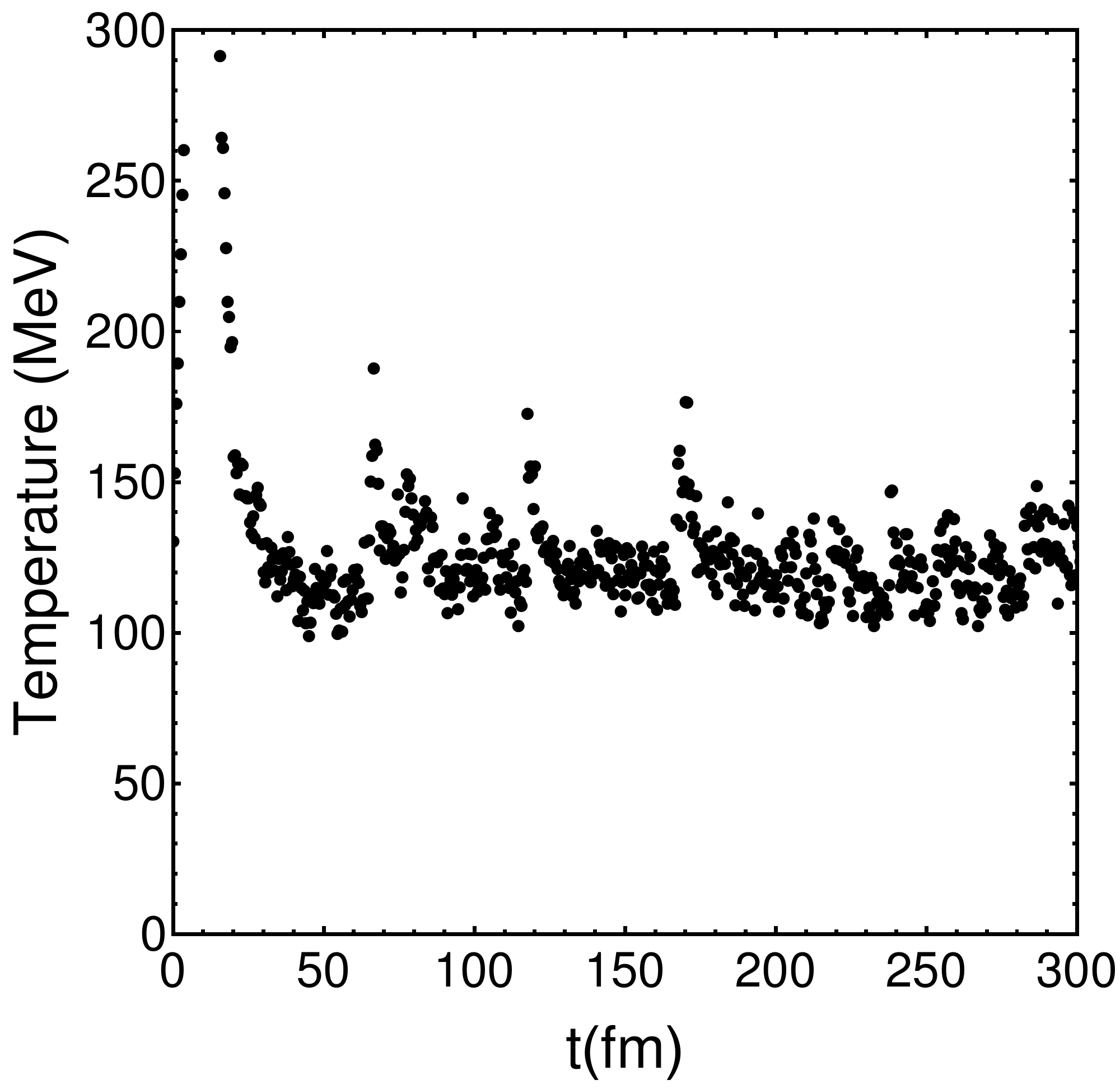}
\caption{Top panel: Kinetic (black dots), potential (blue diamonds) and total energies (red squares) as a function of time for a configuration of $N=128$ nucleons using the Walecka potential $V_A$. Bottom
panel: temperature and a function of time for the same simulation.}
\label{fig:energN128}
\end{center}
\end{figure}

  When equilibrium is reached we obtain a big cluster which includes all $N=128$ particles. In Fig.~\ref{fig:energN128} we show the time evolution of the kinetic, potential
and total energies (top panel) as well as the temperature evolution (bottom panel). We observe that the total energy of the system is
dominated by large negative potential energy, so to see one cluster structure is not surprising. The temperature at equilibrium (plateaux formed after $t\sim 50$ fm in the bottom panel) fluctuates around the 
value $T_{kin}=120$ MeV. The sudden kicks in temperature and the steps in energies occur when one more particle is captured by
the cluster, and falls to the deep potential well.

  Following the mean-field approach, and the King's solution a decreasing distribution of particles is expected as a function of the radial distance. We want to analyze the internal arrangement of nucleons in this cluster by
finding the radial distribution of nucleons starting at $r=0$ (defined as the centroid of the cluster). As the centroid evolves in time, we monitor its position
at each time step, and perform the radial distribution of the nucleons. To have independent events in the distribution, and to avoid spurious correlations we choose time steps well separated 
to perform the average. We measure the number of nucleons per unit volume/radial distance, and plot it versus the distance from the centroid.
  
\begin{figure}[ht]
\begin{center}
\includegraphics[width=6.5cm]{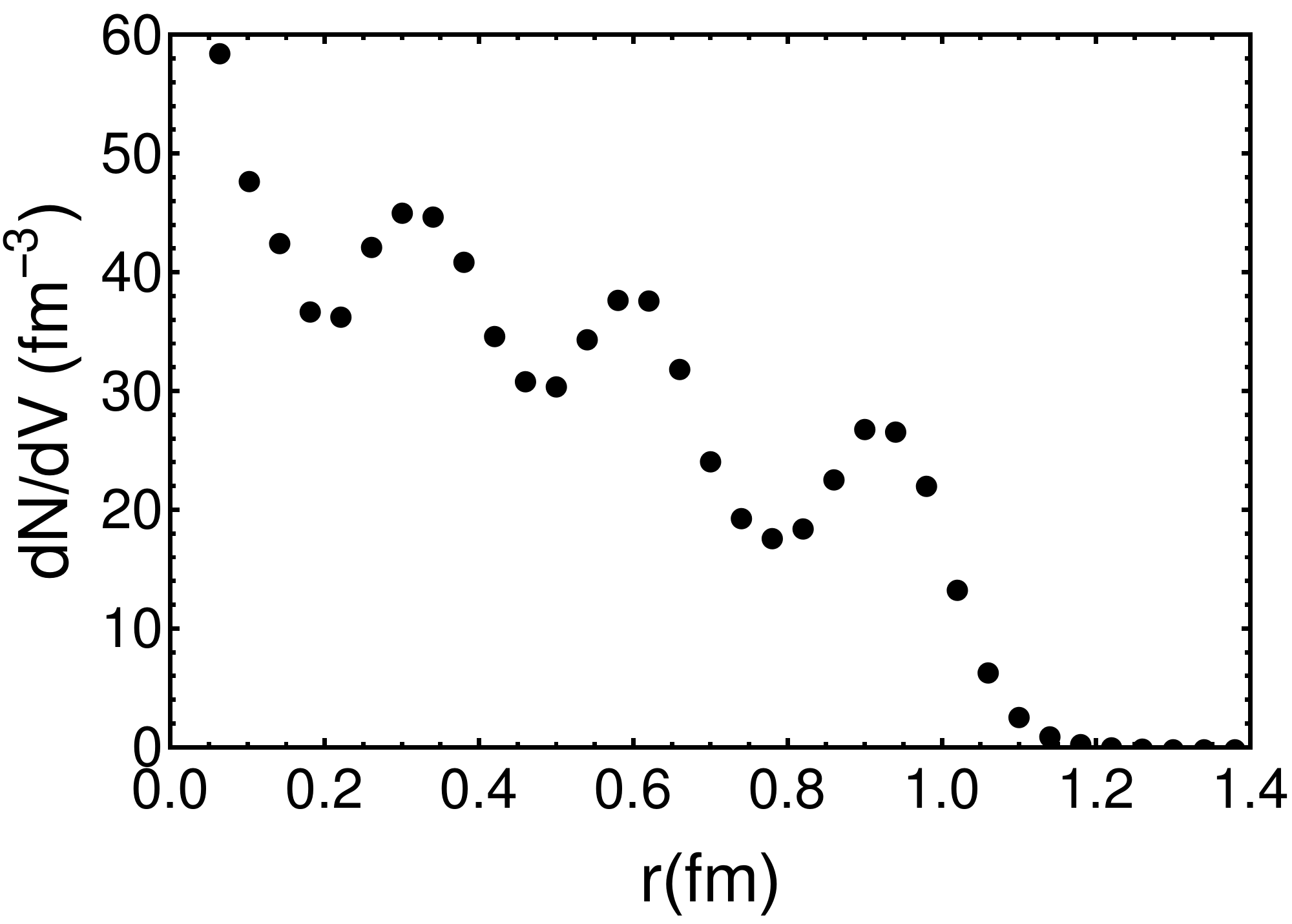}
\includegraphics[width=6.5cm]{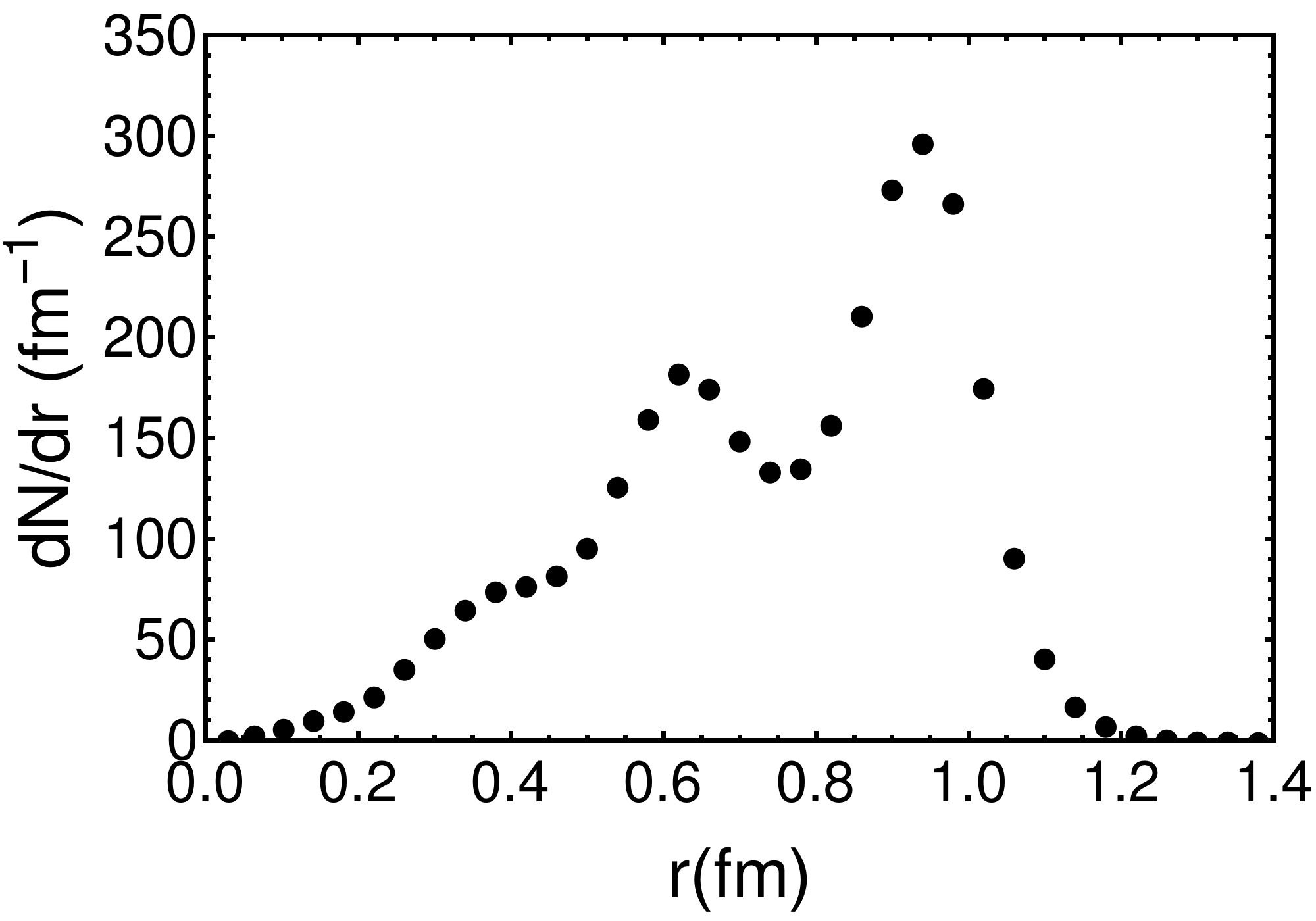}
\caption{Density of nucleons per unit volume (top panel) and per unit distance (bottom panel) inside a cluster of $N=128$ nucleons after equilibration.}
\label{fig:denN128_T120}
\end{center}
\end{figure}

  The distributions $dN/dV$ and $dN/dr$ are represented in Fig.~\ref{fig:denN128_T120} showing a non monotonous structure suggesting a shell like
organization with accumulations of nucleons every 0.3 fm in the radial direction.

  
  For clarity, let us note that this study is only done for investigative purposes. The time scales considered in the plot above, up to $t\sim 300$ fm$/c$,
are much longer than those available in heavy-ion collisions, $t \sim 10$ fm$/c$. Furthermore, this analysis was done in static, rather than exploding, heat bath.  
So, by no means we suggest that such clusters are actually produced in experiment: at best we hope to find evidences of the very beginning of the clustering process.

\section{Baryonic clusters near the chiral transition} \label{sec_clustering_at_fo}

  In this study we continue the study of big cluster formation, and their time scales as the clustering process becomes more and more important. 
This will happen when the original parameters of the Walecka potential are modified as a consequence of the changes in the properties of the $\sigma$ mode.
  
 We will compare the potentials $V_{B1}$, $V_{B2}$, $V_{C}$, each one thought to be acting closer and closer to the chiral transition.

\subsection{Formation of clusters}

  All simulations begin with randomly placed nucleons. Naturally the cluster formation starts with small clusters, which then assemble 
into larger and larger ones. We decided to follow the process by defining variables in which one can separate
clustered and non-clustered baryons in the most direct way, and then histogram the distributions at different times.

We performed a number of such studies, demonstrating here one example, for 4-particle variable. The variable $S$ (from sum) is defined as the normalized sum of all mutual distances
between particles in the system,
\be
S  =  \frac{1}{N_d} \sum_{\substack{i=1 \\ j>i}}^{N_d} |\vec{x}_{ij}| \label{eqn_S} \ ,
\ee
where $i,j=1,...,N$ run over all nucleons, $\vec{x}_{ij}$ is the vector joining pairs, and $N_d=N(N-1)/2$ is the number of mutual distances between different nucleons.

\begin{figure}[ht]
\begin{center}
\includegraphics[width=6.5cm]{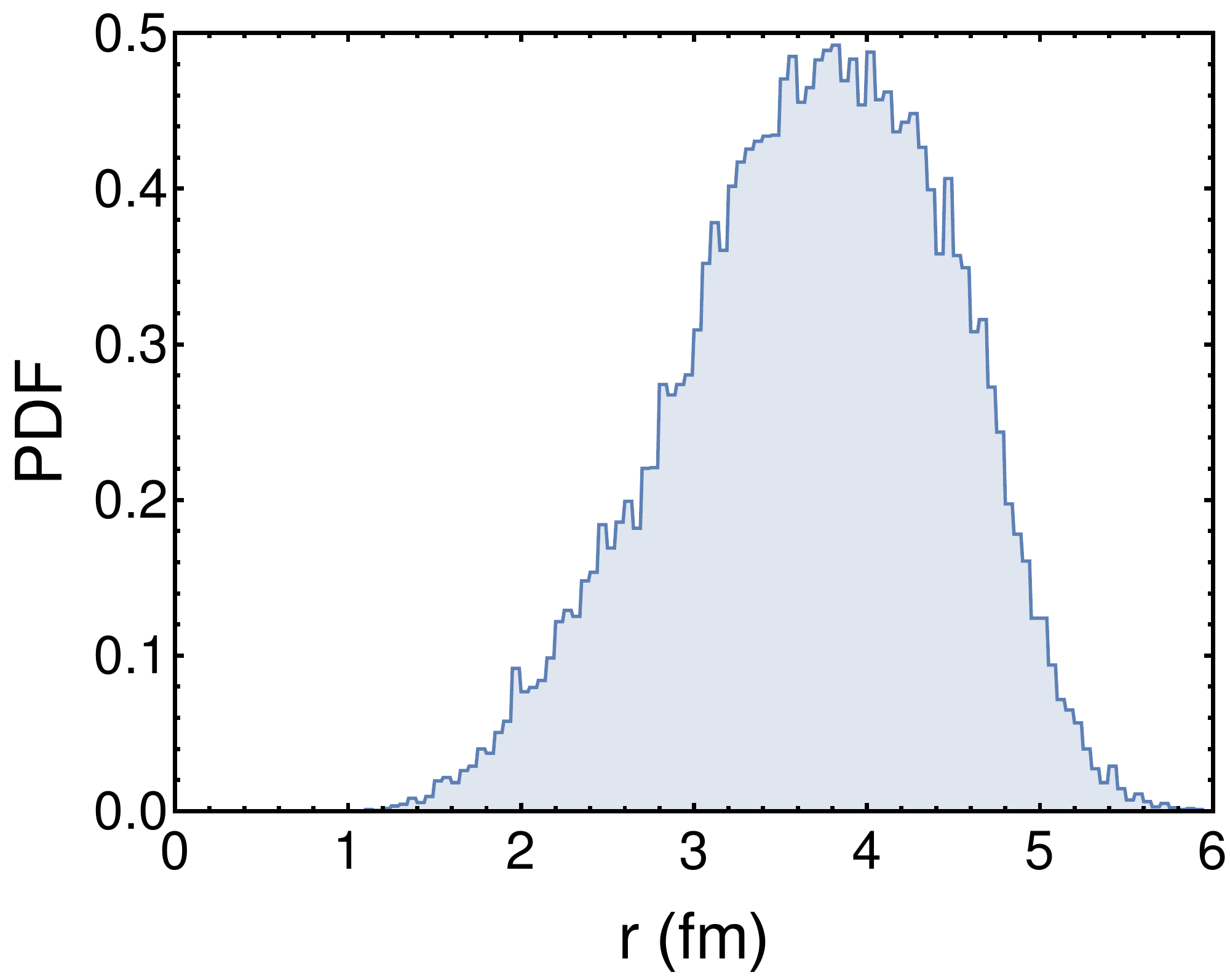}
\includegraphics[width=6.5cm]{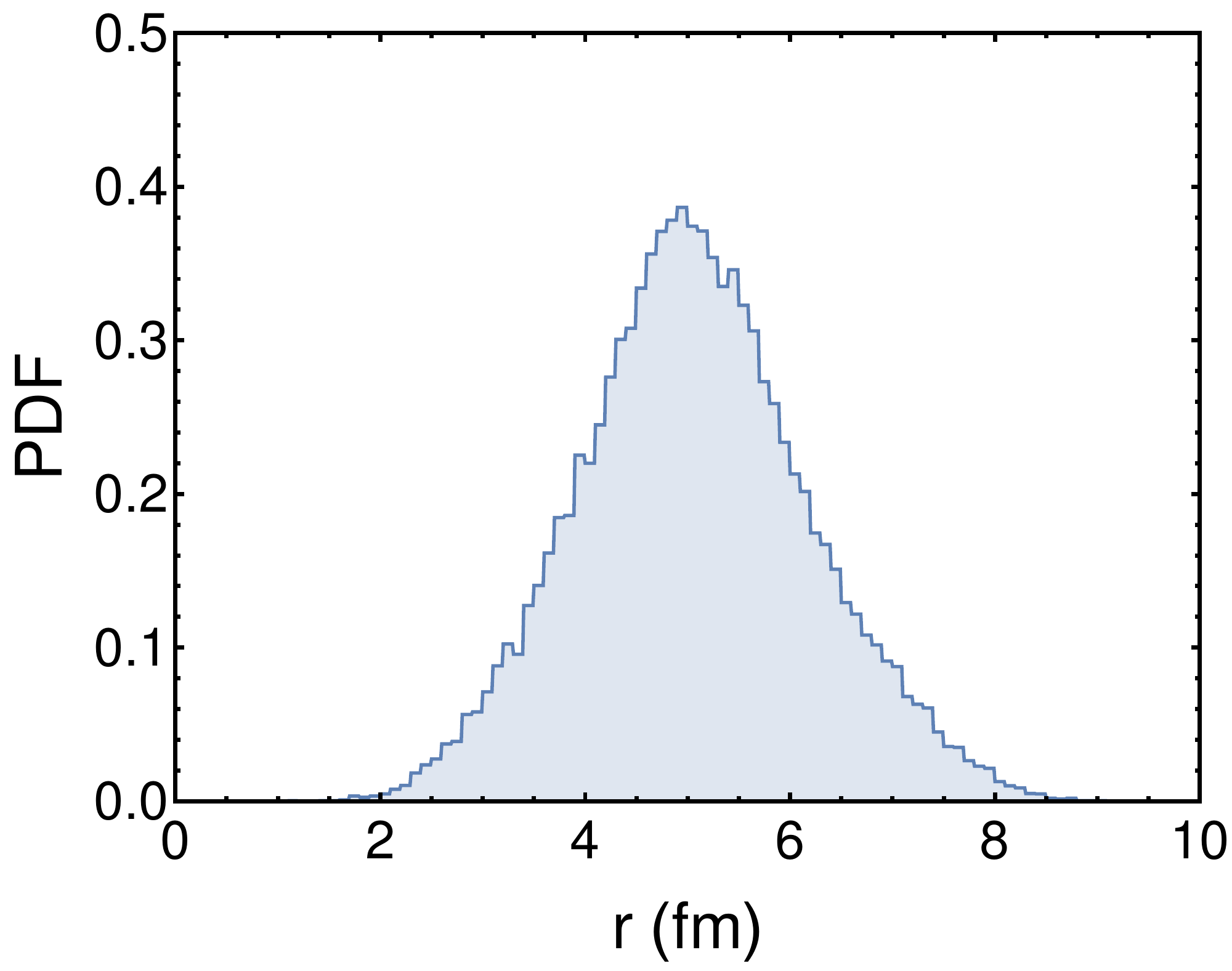}
\caption{Distribution over variable $S$ Eq.~(\ref{eqn_S}), for time equal $t=0$ and $4$ fm$/c$, respectively.
The calculation is done for 32 particles and the Walecka $V_{A'}$ potential.\label{fig_S4_A}}
\end{center}
\end{figure}

As one can see from an example shown in Fig.~\ref{fig_S4_A} (note the scales), for potentials $A,B1,B2$ we observe that the entropy wins over the energy. With time the distribution slowly become wider due to the diffusion of baryons. 

\begin{figure}[htbp]
\begin{center}
\includegraphics[width=6.5cm]{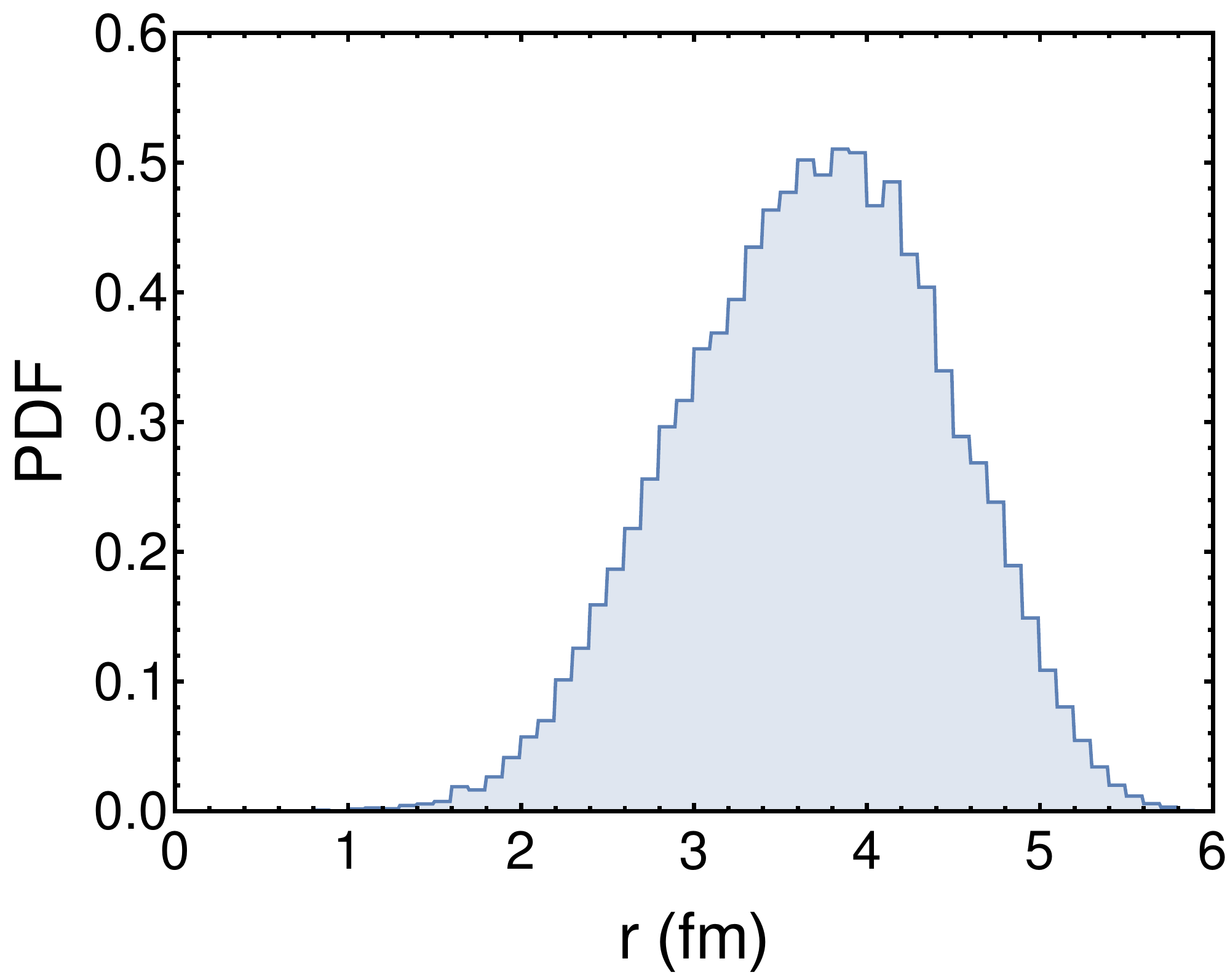}
\includegraphics[width=6.5cm]{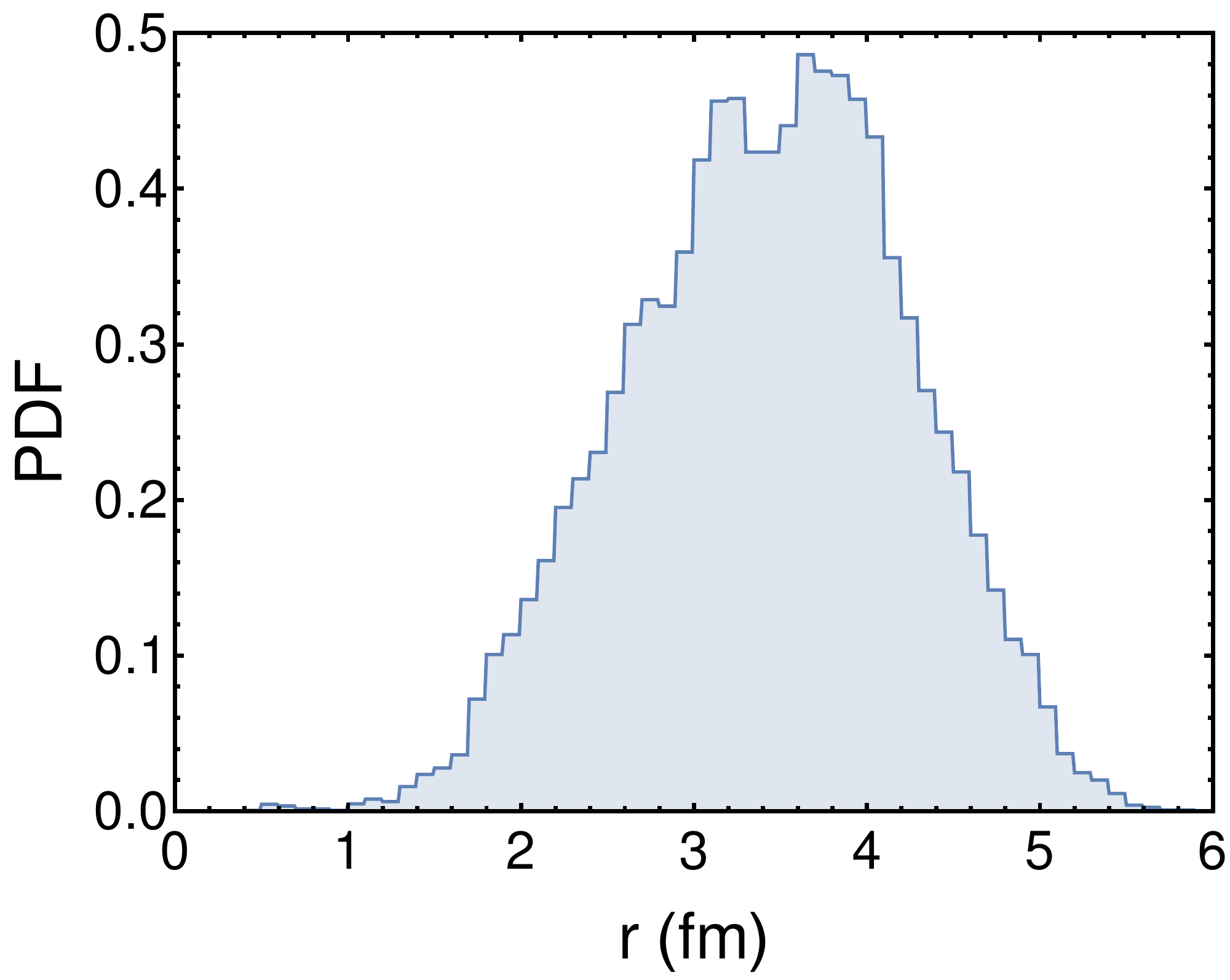}
\includegraphics[width=6.5cm]{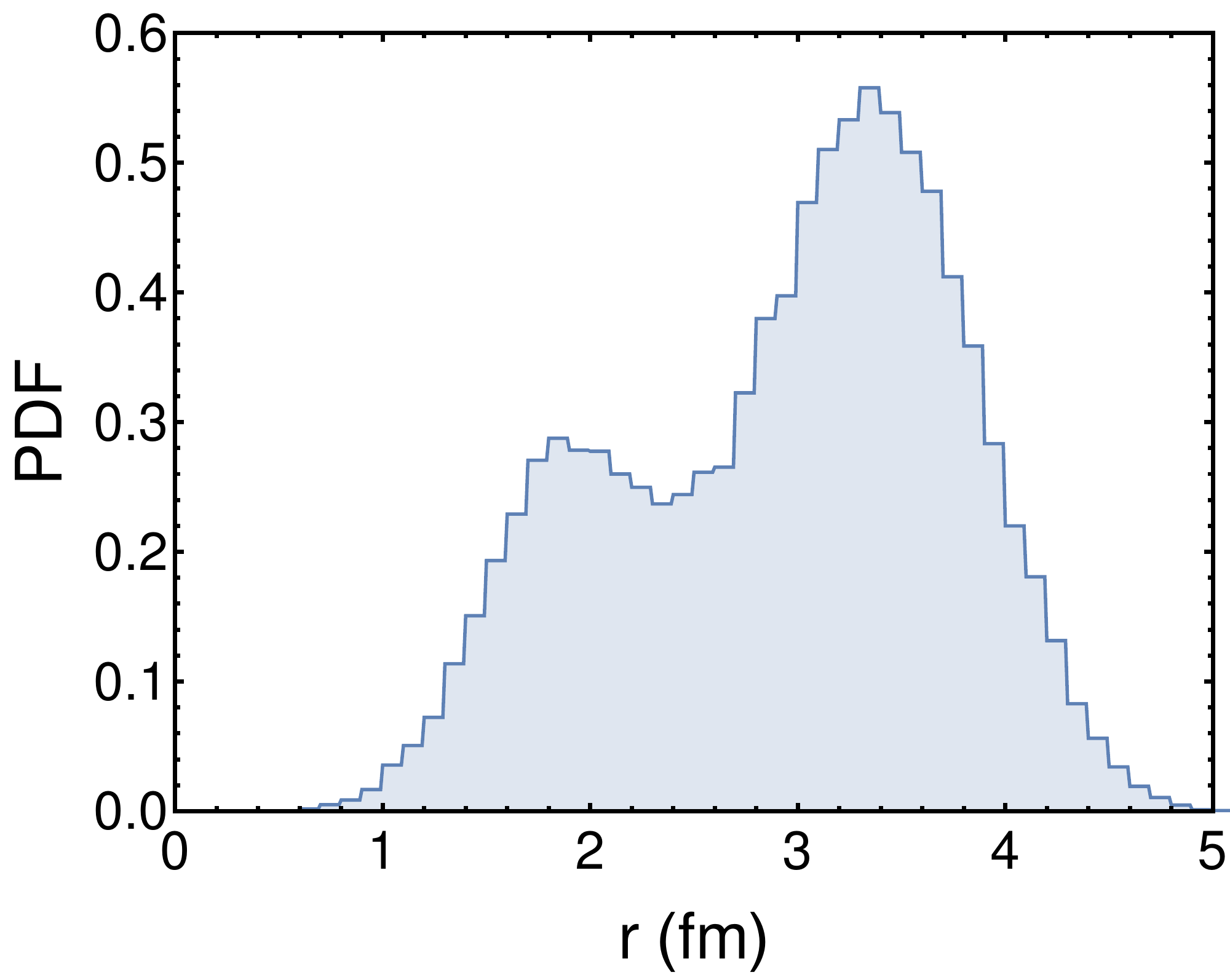}
\includegraphics[width=6.5cm]{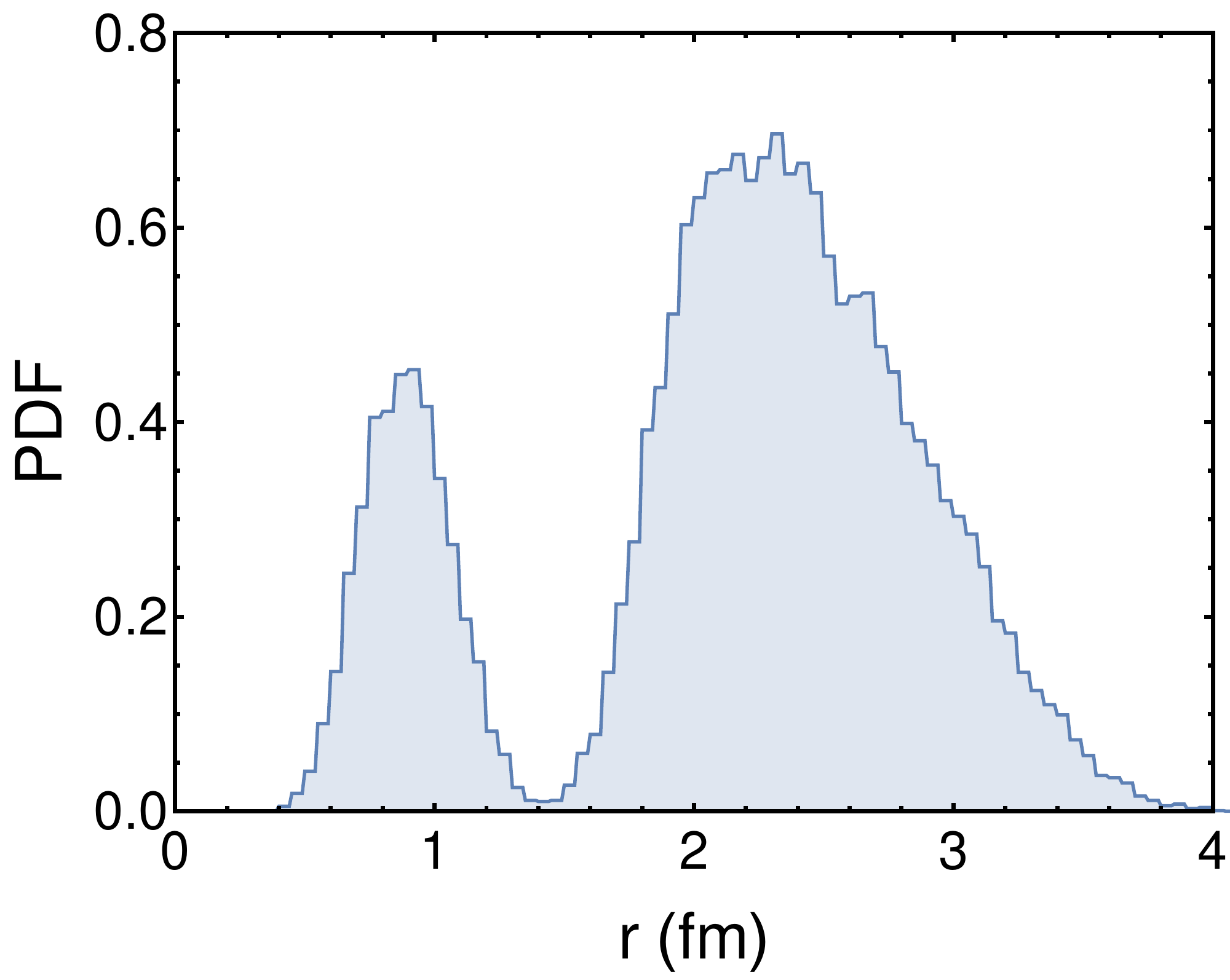}
\caption{Distribution over variable $S$ (\ref{eqn_S}), for time equal $t=0,1,2,4$ fm$/c$,respectively.
The calculation is done for $N=32$ particles and the $V_C$ potential with $x=1$.\label{fig_S4}}
\end{center}
\end{figure}

In contrast to that, the potential $C$ with longer-range attraction shows the opposite trend,
the potential wins over the entropy, leading to
a rather robust clustering. An example of the time evolution for the $C(x=1)$ potential is shown in Fig.~\ref{fig_S4}.
The clear separation of the distribution into two peak structure, in this one particular event,
corresponds to a formation of two clusters (in this event, those have sizes of 9 and 22,
with only one particle evaporating out). The first peak corresponds to intra-cluster distances in both clusters, whereas the second peaks reflect inter-cluster distances.   

\subsection{Time scales}

\begin{figure}[ht]
\begin{center}
\includegraphics[scale=0.38]{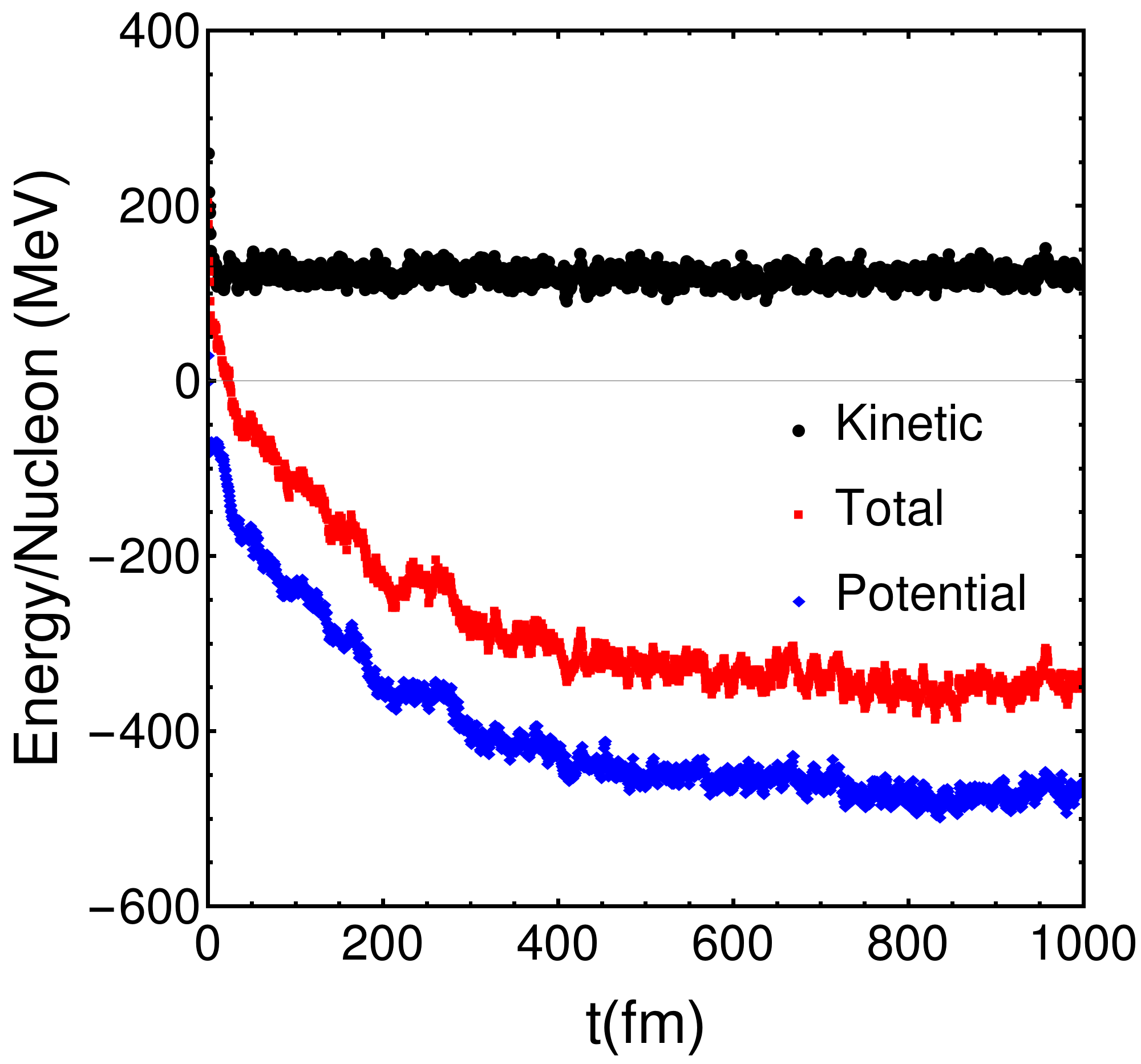}
\includegraphics[scale=0.38]{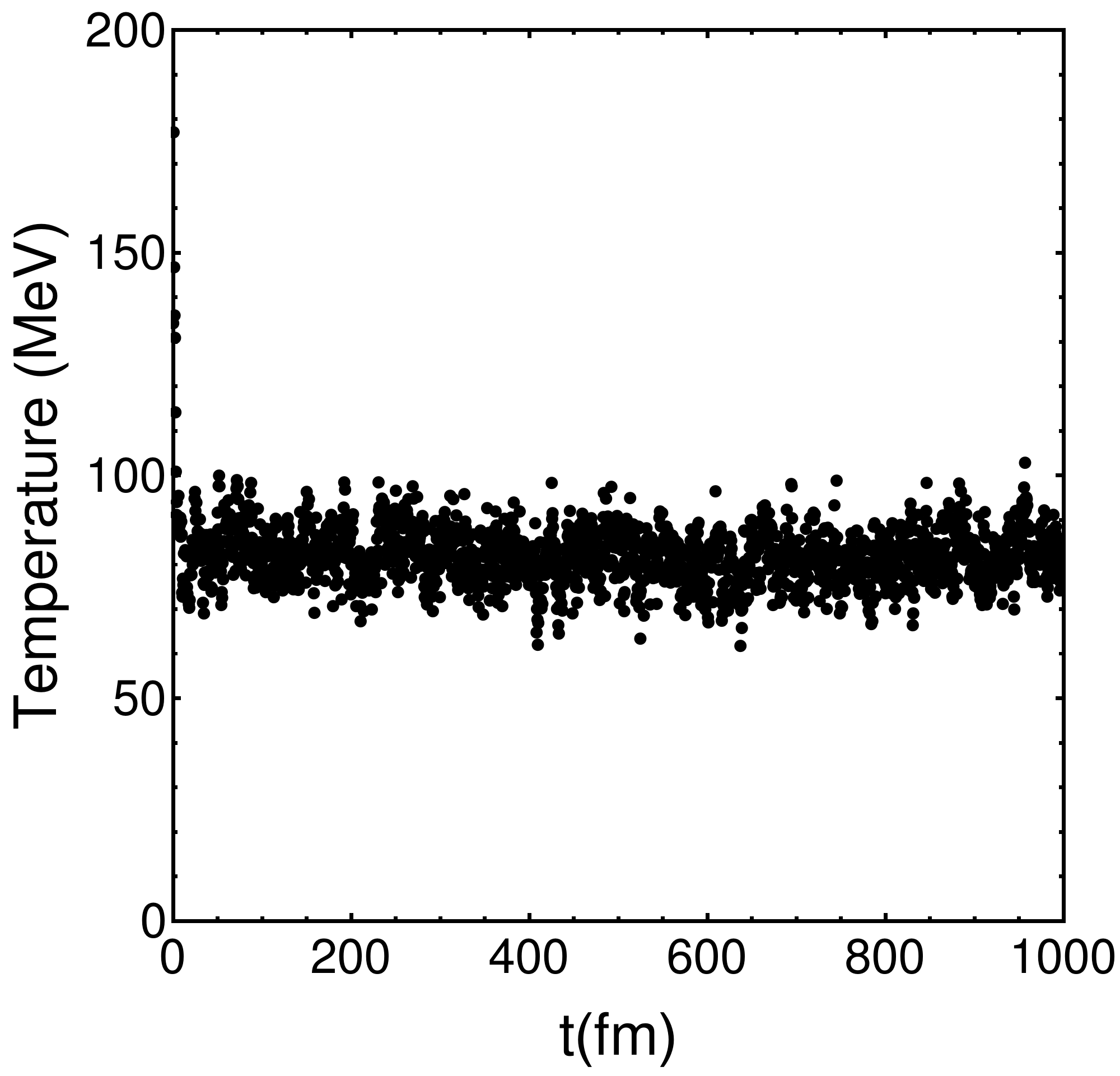}
\caption{Top panel: Kinetic (black dots), potential (blue diamonds) and total energies (red squares) as a function of time for a configuration of $N=128$ nucleons using the potential $V_{B1}$. Bottom
panel: temperature and a function of time for the same simulation.\label{fig:denN128_T80_B1}}
\end{center}
\end{figure}

  We consider a system of $N=128$ nucleons at temperature $T=80$ MeV, with an initial nuclear density $n=0.13$ fm$^{-3}$ and finite size. In Fig.~\ref{fig:denN128_T80_B1}
we show the time dependence of the energies per particle (top panel) and the temperature (bottom panel). After a fast thermal equilibration the temperature is approximately
constant, while the total energy is not conserved in the evolution as dissipation occurs due to the drag force.

The potential $V_{B1}$ is able to produce full clustering after long times. From the example in Fig.~\ref{fig:denN128_T80_B1}, the full equilibration time is
of the order of $\sim 800$ fm/$c$, and clustering has taken place. Individual particles can escape the cluster
thanks to their kinetic energy, however, we avoid the lose of particles with the external trapped potential in Eq.~(\ref{eq:WS}).

We can define an equilibration time by noticing that the total energy has an approximate exponential decay $\exp \left(- \frac{t}{t_{eq}} \right)$. We obtain $t_{eq} = 187$ fm in this particular example.

\begin{figure}[ht]
\begin{center}
\includegraphics[scale=0.38]{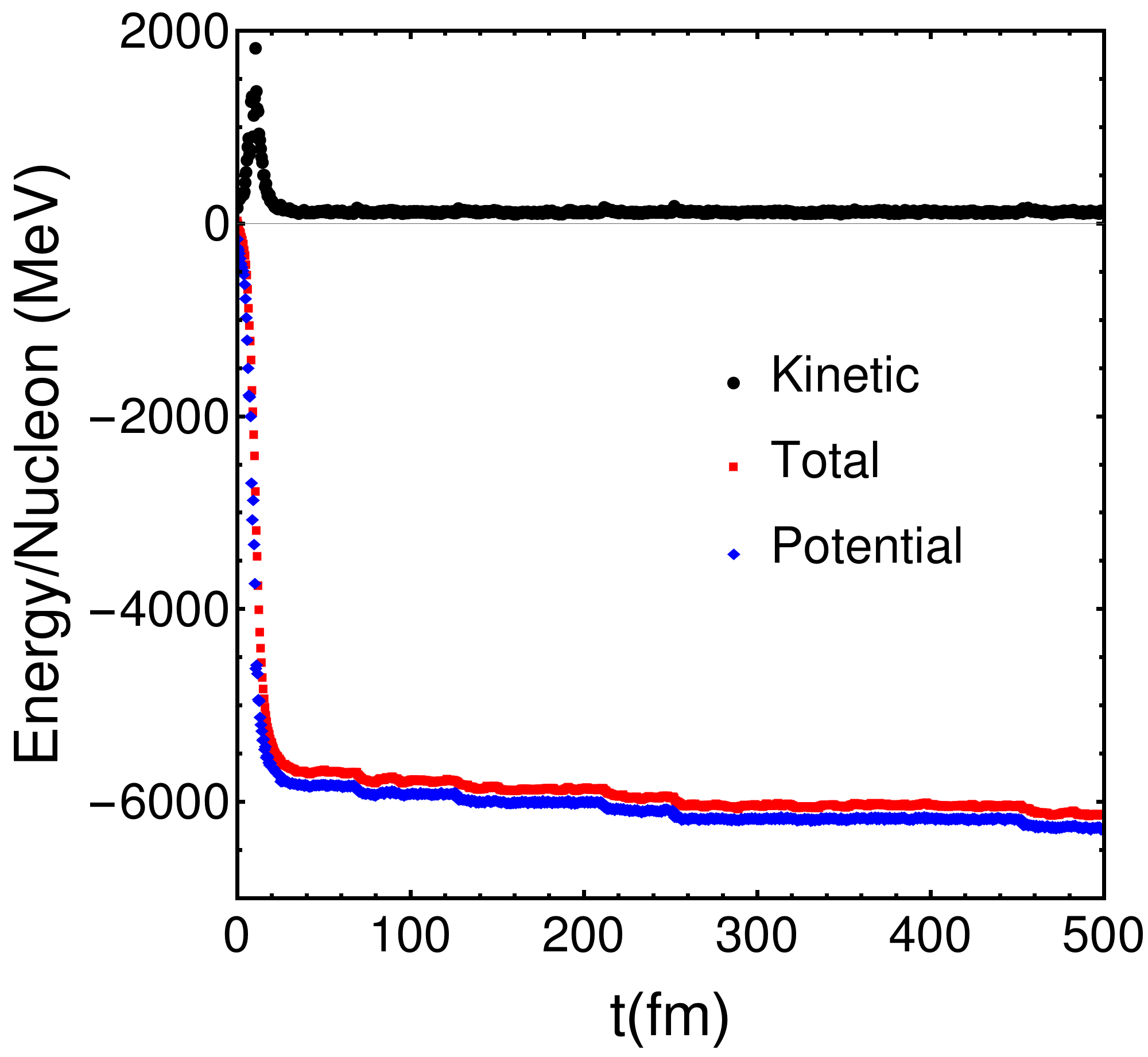}
\includegraphics[scale=0.38]{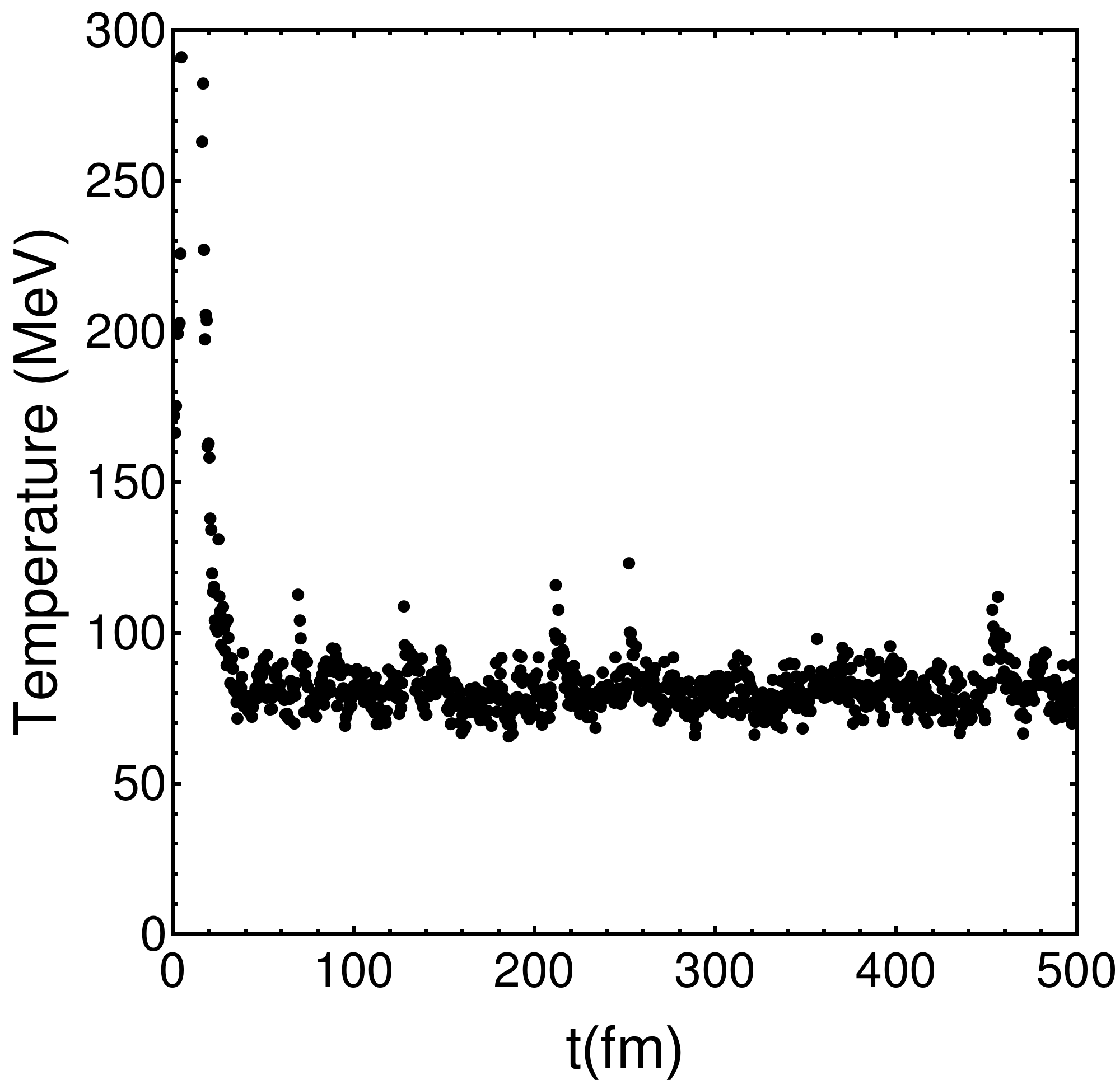}
\caption{Top panel: Kinetic (black dots), potential (blue diamonds) and total energies (red squares) as a function of time for a configuration of $N=128$ nucleons using the potential $V_{B1}$. Bottom
panel: temperature and a function of time for the same simulation.\label{fig:denN128_T80_B2}}
\end{center}
\end{figure}

  Although the $V_{B1}$ is enough to form a big cluster in several hundreds fm/$c$, these scales are totally irrelevant for HICs. A slightly critical potential $V_{B2}$ produces the clustering in a 
much faster way $\sim 40$ fm/$c$. We present the time dependence of the energies and temperature in Fig.~\ref{fig:denN128_T80_B2}.

In this case, the exponential decay is much less evident. We find an initial regime of $\sim 10$ fm/$c$ where the energy is approximately constant. Between 10 fm and 17 fm we find a good exponential decay with
an equilibration time of $t_{eq}=3$ fm/$c$. After this transient exponential decay the relaxation is much softer, reaching equilibration in around 40 fm/$c$. 
The time scales for the clustering with this potential are much closer to those in heavy-ion collision, so it seems reasonable to consider this mechanism as potentially important close
to the critical point (where the equilibration time is even reduced using a deeper potential like $V_C$).
  

  Nevertheless, it seems clear that the time for full clustering is still large to take place completely in heavy-ion collisions. We only hope to have a potential effect close to the critical point where the signatures
of initial clustering might certainly occur (perhaps clusters of few nucleons as $^4$He). Starting from a system away from the critical point, we will calibrate our model with noncritical potential $V_A$
to experimental data at energies where Poissonian fluctuations are observed. Then, we will modify our potential to increase criticality and compute observables like higher-order moments of the (net-)proton distribution.

\section{Baryonic clusters in BES conditions} \label{sec_exp}

  In this section we apply our model to heavy-ion collisions in the condition of BES. Rather than providing a quantitative result, for what 
one would need a more sophisticated evolution code, we contempt ourselves to show that the effect is consistent with what is observed experimentally
using the closest experimental conditions we are able to implement.

  At high collision energies above  $\sqrt{s_{NN}}=19.7$ GeV STAR data shows approximate Skellam distribution for net protons~\cite{Adamczyk:2013dal,Luo:2015ewa} consistent 
with thermal equilibrium fluctuations. This will be our baseline energy.

For collisions at this energy and below we can safely neglect antiprotons. In Table~\ref{tab:protontoanti} we show the ratio of proton/antiproton in the kinematic cut $|y|<0.5, 0.4
<p_\perp<0.8$ (GeV/$c$) for the most central collisions at different energies considered in this work.

\begin{table}[ht]
\caption{Proton-to-antiproton ratio for $|y|<0.5$ and $0.4<p_\perp<0.8$ (GeV$/c$) at centrality bin 0-5\% for collision energies $\sqrt{s_{NN}} \le 19.6$ GeV.
From Ref.~\cite{Adamczyk:2013dal}.}
\begin{center}
\begin{tabular}{|c|c|c| c|}
\hline
$\sqrt{s_{NN}}$ (GeV) & 7.7 & 11.5 & 19.6 \\
\hline
\hline
proton/antiproton & $114.4\pm 0.6$ & $30.64 \pm 0.07 $  & $9.89 \pm 0.01$ \\
\hline
\end{tabular}
\end{center}
\label{tab:protontoanti}
\end{table}%

 In our simulations, Poisson statistics is achieved when measuring the distribution of protons in a sub-volume or the order $\%$ of the initial volume.
This is consistent with the fact that experimental net-proton distribution in the narrower $p_T$ cut is 5\% of the total net-proton multiplicity~\cite{talkllope}, 
and matches very well Poisson expectations.

  Our first task is to achieve similar multiplicities in a noncritical scenario, where Poissonian fluctuations dominate, which we identify with experimental 
data at $\sqrt{s_{NN}}=19.6$ GeV. For that energy we will be running the potential $V_{A'}$, i.e. Walecka potential with an additional repulsion.
The kinetic freeze-out temperature for low energies is roughly $T_{kin}=120$ MeV. In our code, the MD will simulate a few Fermi/$c$ between hadronization and freeze-out, so
we set the temperature to $T=150$ MeV (although the calculation is not very sensitive to this parameter).

  The baryon density is close to $n_{kin} \sim 0.12$ fm$^{-3}$ at freeze-out, but at earlier stages it can take a few times this value~\cite{Ivanov:2018vpw}. We use a value of
$n=0.3$ fm$^{-3}$. For numerical convenience we use a reduced number of protons $N=32$ and then scale up the different cumulants as suggested in Ref.~\cite{Karsch:2010ck}, to 
be able to compare with the experimental cumulants. In particular, we note that scaled skewness $S\sigma$ and kurtosis $\kappa \sigma^2$ do not depend on volume, so
we can compare them without the need of scaling. The number of events for each potential is $N_{ev}=10^5$.
 
   We summarize the parameters used in our MD+L simulations for this section in Table~\ref{tab:params}. These parameters will be common for all potentials, as we would like 
to isolate the only effect of the potentials. In addition, we checked that the parameters from thermal fits do not change too much within these energies (the most sensible 
parameter would be the baryochemical potential).

\begin{table}[ht]
\caption{Parameters used in the simulations of protons for kinetic freeze-out conditions of STAR collisions at $\sqrt{s_{NN}} \le 19$ GeV.
Respectively: temperature, nucleon density, number of nucleons, number of events, drag coefficient, and time duration.}
\begin{center}
\begin{tabular}{|c|c|c|c|c|c|c|c|}
\hline
$T$ & $n$ & $N$ & $N_{ev}$& $\lambda$ & $\Delta t$ \\
\hline
\hline
150 MeV & 0.3 fm$^{-3}$ & 32 & $10^5$ & 0.256 fm$^{-1}$ & 5 fm/$c$ \\
\hline
\end{tabular}
\end{center}
\label{tab:params}
\end{table}

  We run our MD+L a total time of $\Delta t=5$ fm$/c$, corresponding to an approximate time between hadronization and kinetic freeze-out. While this time can be extended, perhaps up to a factor 2, 
we prefer to be conservative not to overestimate the effect of clustering (as we have seen larger times help to create more bound clusters).

  The calculation is performed in a non expanding medium, i.e. without radial flow implemented. This is convenient for the use of nonrelativistic dynamics at all times. A final boost in rapidity and 
transverse momentum will take care of the mapping of the particles into the appropriate kinematic domain, consistent with experiment.

  A first simulation using $V_{A'}$ generates the expected Gaussian distributions of rapidity and transverse momentum for nonrelativistic dynamics. The fitted $p_\perp$ distribution is perfectly consistent
with the temperature used (providing an intermediate check of the code). Then, we perform a mapping of these distributions to mimic the experimental findings and take into account radial flow. Otherwise, the distribution of particles cannot match 
experiment. First, the experimental $p_\perp$ distribution of net-protons is well fitted to a double exponential~\cite{Adamczyk:2017iwn}. However, we have found that within the same kinematic cut a Gaussian form is 
already very approximate. Therefore we opt by simply rescaling the transverse momenta of our particles by a single factor of 1.7 to match experimental distribution. 

  The rapidity distribution is a relative narrow Gaussian one centered at zero. Therefore, most of the particles live at midrapidity, which is not consistent with experiment, where one expects $\sim 10\%$ of particles
at midrapidity. This is normal, as in our simulation we do not account for any longitudinal boost. Then, we also transform our Gaussian distribution into a uniform distribution in rapidity between the
kinematic limits for that energy. The uniform rapidity distribution is quite a crude approximation, but it does not involve any additional parameters. 

   Once fixed our final distributions, we proceed to test the proton distribution by performing the kinematical cuts considered in the literature. In what follows we will denote {\it Cut 1} as the one with 
rapidity $|y|<0.5$ and transverse momentum $0.5 \textrm{GeV}/c < p_\perp < 0.8 \textrm{GeV}/c$; whereas {\it Cut 2} extends the $p_\perp$ coverage up to 2 GeV$/c$, thanks to the time of flight detector for the 
particle identification~\cite{Luo:2015ewa}.

\begin{figure}[htbp]
\begin{center}
\includegraphics[width=7cm]{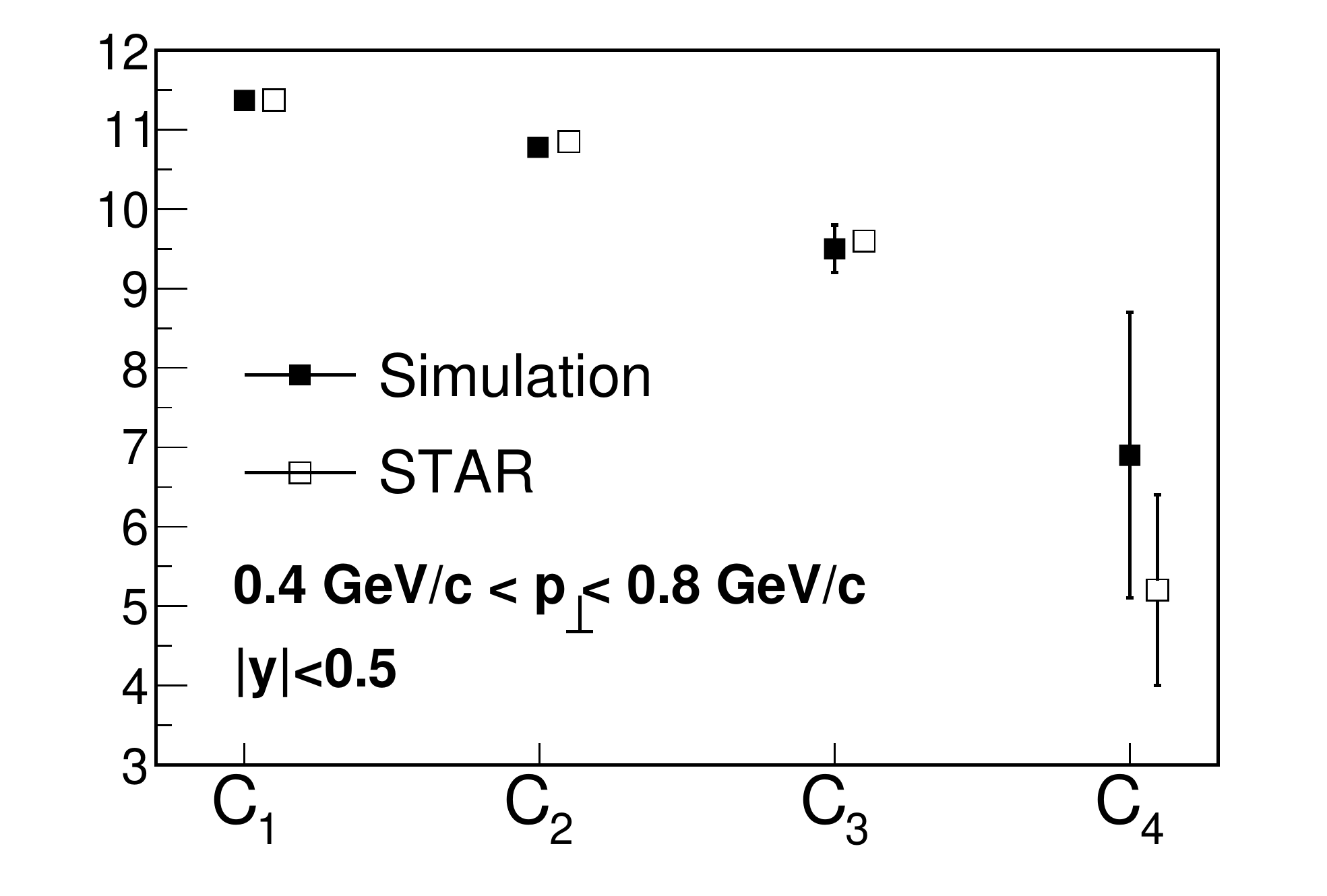}
\includegraphics[width=7cm]{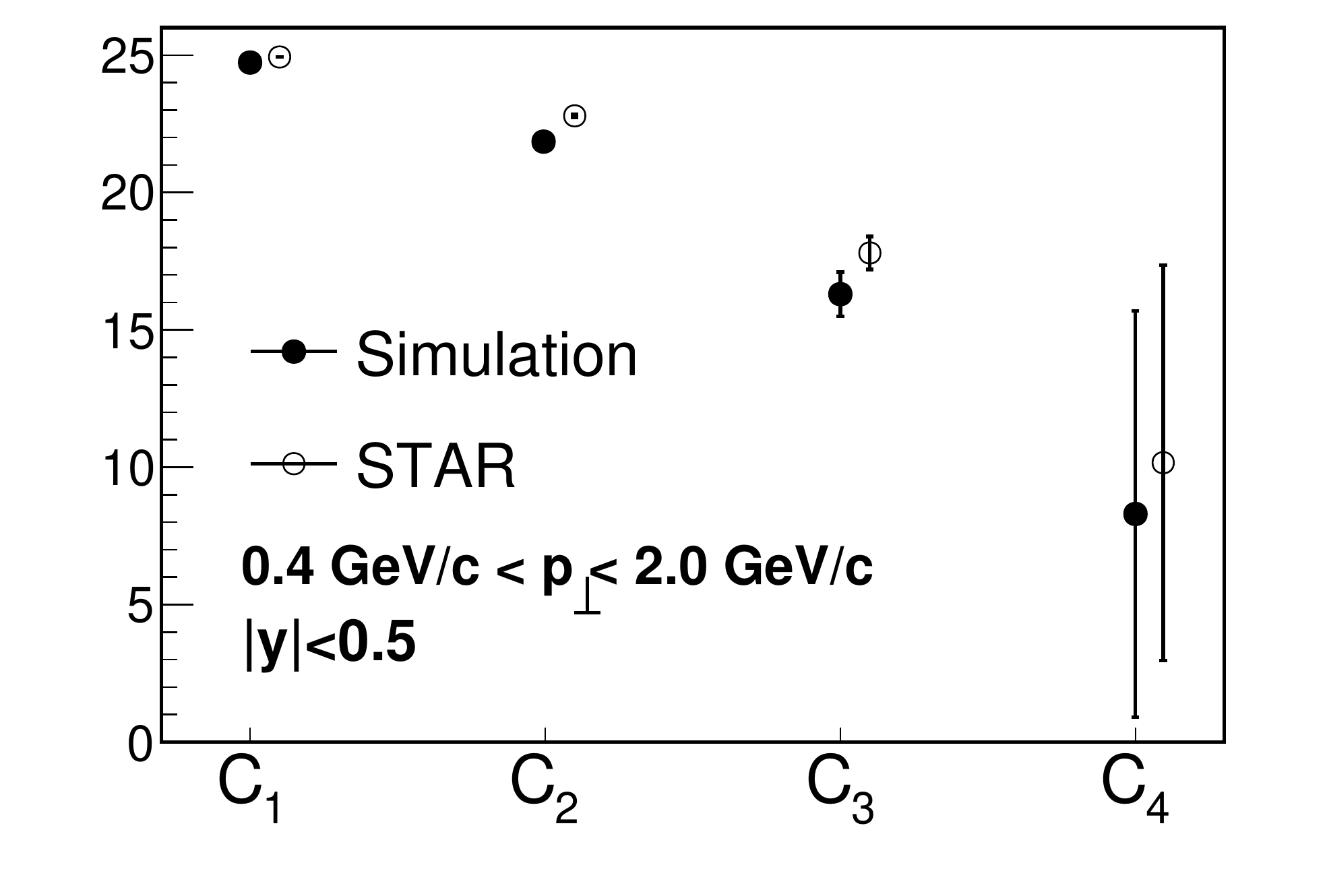}
\caption{Mean and central moments for proton distribution ($C_2$ is the variance) for 2 different kinematic cuts at kinetic freeze-out conditions of STAR collisions 
at $\sqrt{s}_{NN} \le 19.6$ GeV. Our simulation results have been scaled by a constant factor, which is fixed by matching the experimental value of $C_1$ in {\it Cut 1}
(the upper left point in the left figure). Experimental data for {\it Cut 1} is taken from~\cite{Adamczyk:2013dal} and for {\it Cut 2} from~\cite{Luo:2015ewa}.} 
\label{fig:cumulants}
\end{center}
\end{figure}

   We summarized our results for the proton cumulants in Fig.~\ref{fig:cumulants} compared to experimental data for the two mentioned cuts. In our simulation we simply perform the analysis within the kinematic cuts 
including the $N_{ev}=10^5$ events. The statistical uncertainty is coming from the Delta theorem, as explained in Ref.~\cite{Luo:2011tp}. We need to make a final scaling from our $N=32$ to match the absolute number
of protons observed in experiment. For this we choose the average number of proton $C_1$ in {\it Cut 1} and scale up our value to the experimental results. As it is well known, all the cumulants scale with volume, so
all the other cumulants (up to order fourth) are scaled up by the same amount.


We observe that despite our crude model, we can match in a good degree of accuracy the cumulants for proton number, in {\it both} kinematical cuts for $\sqrt{s}_{NN} \le 19.6$ GeV. Therefore, the noncritical scenario
is reasonably under control. As a last check, after multiplication of $dN/dy$ by the same scaling factor we obtain $dN/dy=30.5$ at midrapidity. The experimental value given for the most central collisions
is $dN/dy=34.2 \pm 4.5$.



   It is also possible to compute ratios of these cumulant to obtain the skewness $S\sigma=C_3/C2$ and kurtosis $\kappa \sigma^2=C_4/C_2$ and compare with experiment. 
As being directly related to cumulants these moments for proton distribution will be in accordance within the same levels at those in Fig.~\ref{fig:cumulants}. 

A step forward would correspond to compare our skewness and kurtosis with the same quantities for the net-proton distribution. However, at this energy there is an additional 10 \% systematic 
error because antiprotons (not accounted in our simulation) are still important (see Table~\ref{tab:protontoanti}).


  Once the obtained multiplicities and cumulants for the proton distribution are calibrated to those as in experiment, we simply repeat the calculation with different potentials 
at our disposal. Each of them are supposed to encode the modification of the NN potential closer and closer to the QCD critical point. Notice that a precise matching between the
experimental collision energy and the particular NN potential is far from clear, so we cannot compare directly to our results with real data. We can nevertheless observe the qualitative 
effect on the skewness and kurtosis after increasing the criticality of our model. We use Models $A, B1,B2$ and Model $C$ with $x=0.1,0.5,1$.

\begin{figure}[ht!]
\begin{center}
\includegraphics[width=6.5cm]{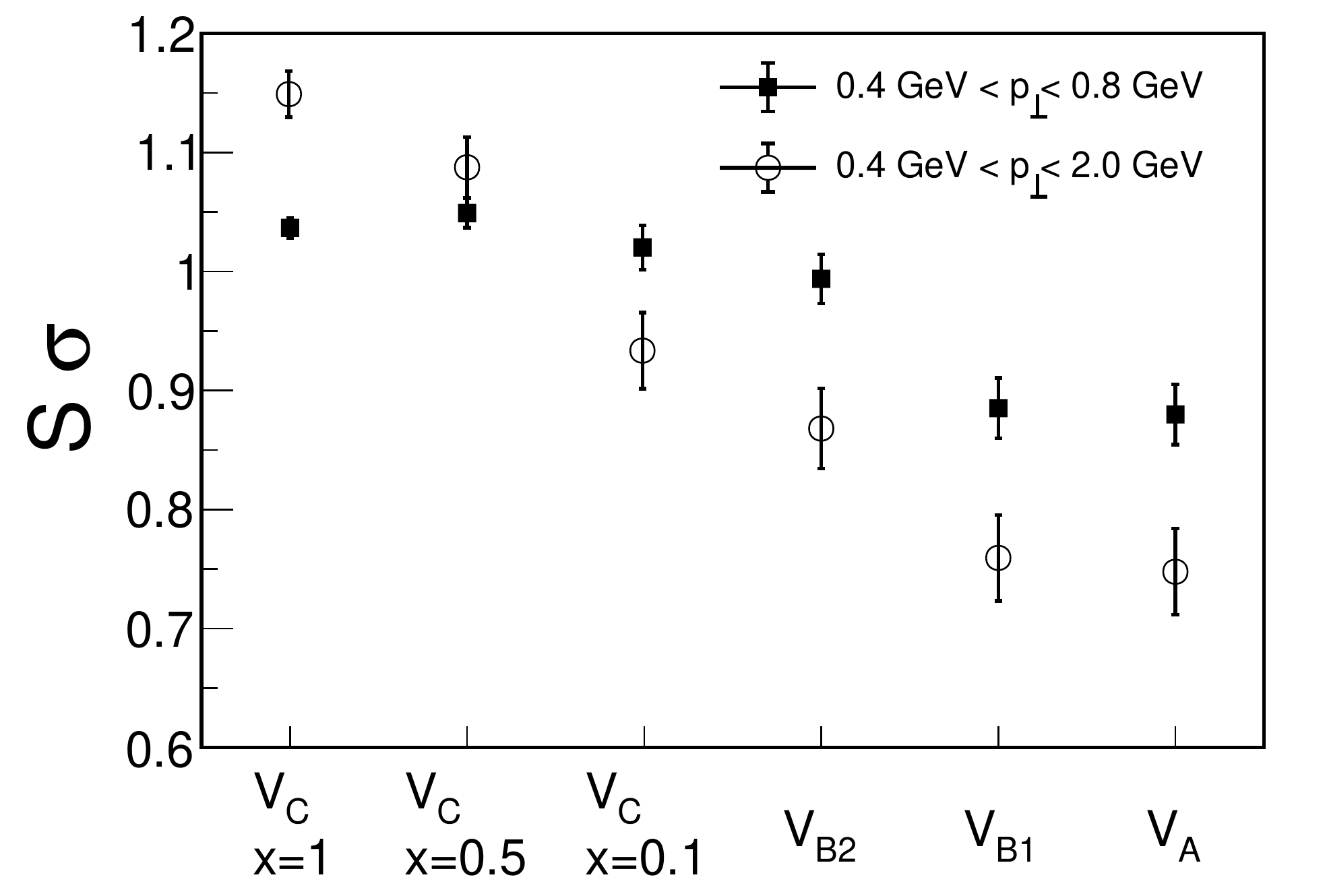} \\
\includegraphics[width=6.5cm]{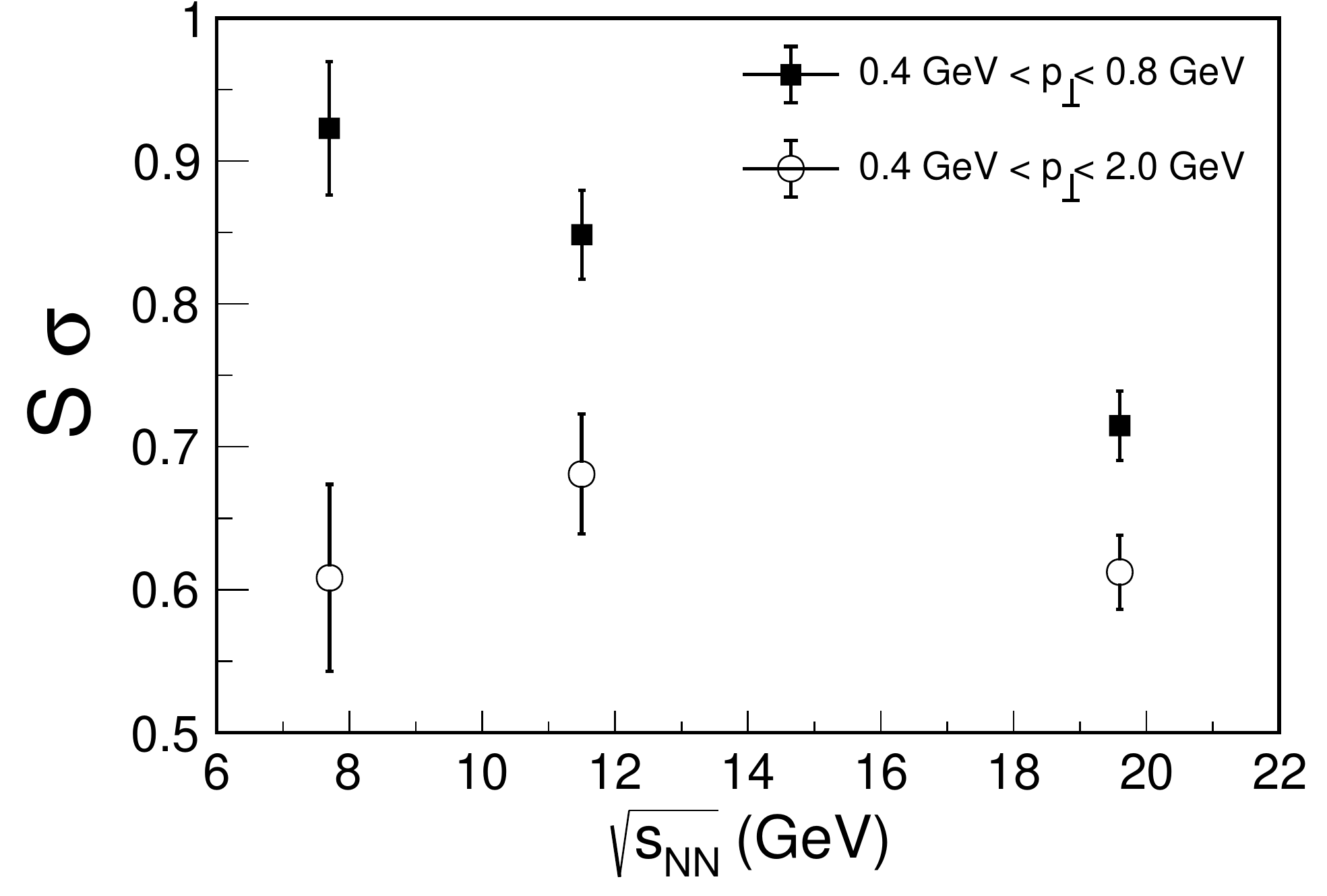} \\
\includegraphics[width=6.5cm]{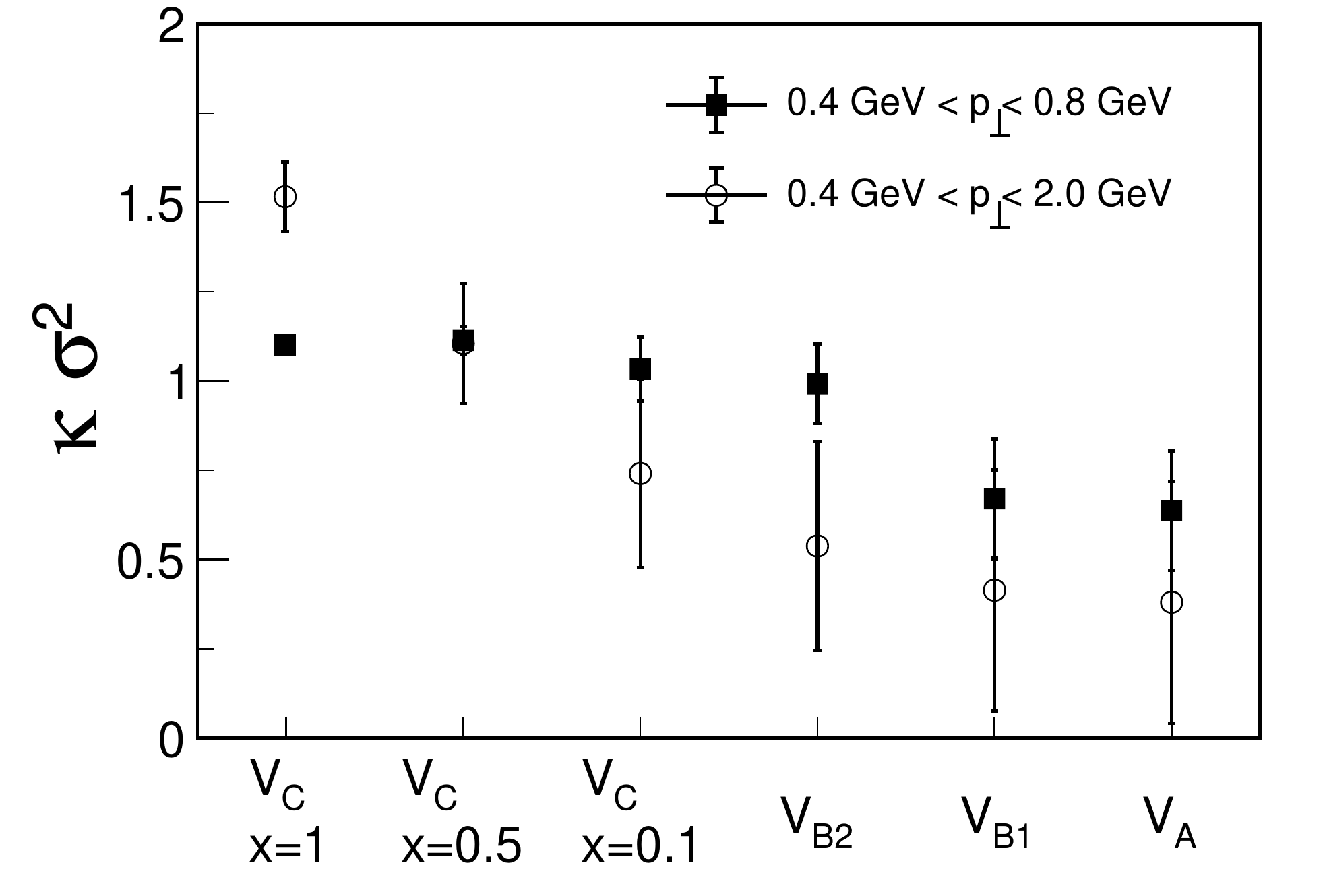} \\
\includegraphics[width=6.5cm]{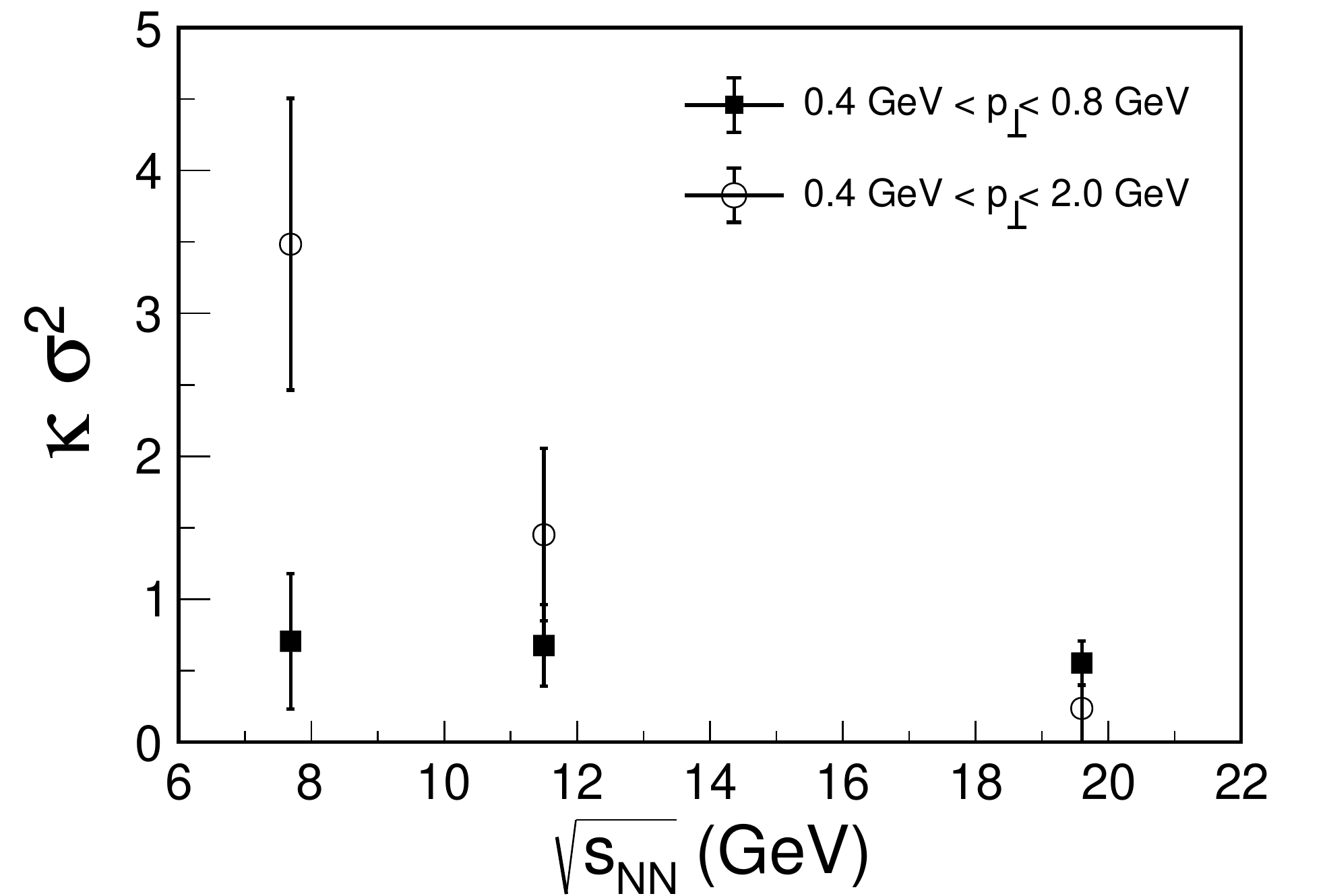}
\caption{Scaled skewness (two top panels) and kurtosis (two lower panels) as a function of the potential used and from experimental data~\cite{Adamczyk:2013dal,Luo:2015ewa} from STAR collaboration.}
\label{fig:skewandkur}
\end{center}
\end{figure}

  In Fig.~\ref{fig:skewandkur} we our present our results for  $S\sigma$ and $\kappa\sigma^2$ in our simulations. From top to bottom we show the theoretical skewness as a function of the 
  NN potential, the experimental skewness as a function of the collision energy, and the same dependences for the kurtosis, respectively. In all cases we consider both {\it Cut 1} and {\it Cut 2}.
As mentioned, a direct comparison is not possible due to the difficulty of matching a given potential to a precise collision energy. However, we base our study in the idea that lowering the collision energy
from high energies, should necessarily approach the expanding system to the critical region, until some particular value of $\sqrt{s_{NN}}$. In our setup this is achieved by increasing the attraction of
the NN potential towards a more critical one. 

  One important result is that the increase of the kurtosis is consistent with the presence of a critical point. Therefore, clustering of nucleons close to $T_c$, and the increase of NN correlations,
translated into an increase of higher moments.

  In this document we have proven that the nuclear clustering is a solid phenomenon with relatively conservative assumptions if the system is left for large amount of time. 
In this respect, we do not find realistic to experimentally find big clusters of nucleons due to this phenomenon, as the required time for this is much longer than the hadronic phase. 
However, early signatures of clustering can be reflected in higher-order cumulants of (net-)protons, and we showed that these signatures are compatible with what has been preliminarily observed in experimental data.

\subsection{Light-nuclei clusters at freeze-out}

  The results of the previous sections show that not only a strong correlations among nucleons, but also a clustering of few of them might be possible during
a time interval of several fm/$c$. A mechanism producing such effects between nucleons is required, according to Ref.~\cite{Bzdak:2016jxo}, to explain STAR experimental data. In that
reference the third and fourth order cumulants cannot be explained by a model with nucleon stopping and baryon global conservation alone. The conclusion of~\cite{Bzdak:2016jxo} is that some 
sort of clustering is needed to describe the data. In this work we provide such a natural mechanism for clustering, if the $NN$ interaction is attractive enough close to the critical point. 

While for the calculation of the higher-order moments of the proton distribution we have included the contribution of all protons---within the corresponding kinematic cuts---we will now extract the nuclear clusters 
which may give rise to light nuclei at the end of the evolution. We will denote these clusters as ``pre-nuclei'' as they are products of the nucleon coalescence at freeze-out. If such ``pre-nuclei'' are able to survive 
as bound objects until the final
stage of the fireball at very low temperatures, then it will form states like $^3$H, $^3$He,$^4$He...resulting in an excess of light nuclei over the expected thermal production.

  Let us focus on $^4$He, which has a binding energy of 28.3 MeV~\cite{wanghe4}. Freeze-out temperatures are larger than this energy, but the modification of the nuclear potential 
can provide extra binding to it. We will look for candidates of $^4$He nuclei at the final time of our simulation. We just need to find 4 nucleons close in the phase space at the moment 
of the freeze-out. If a pair of nucleons are separated by a large distance in the phase space, then they are assumed not to belong to the same cluster. We run our code using several versions of the $NN$ 
potentials and identify configurations of 4 nucleons, or ``pre- $^4$He''. We apply the following criteria:
  
\begin{enumerate}
 \item We only search clusters of 4 nucleons. If any nucleon also belongs to a different cluster, the whole set is ruled out (as the nucleons belong to a bigger nuclei). 

\item  The relative position between pairs of nucleons should be small. The rms of $^4$He is 1.67 fm, and the rms for proton is 0.87 fm. Assuming a tetrahedron configuration (see Sec.~\ref{sec_simulations}), one obtains that the 
 distance between the center of 2 nucleons should be 1.69 fm. Giving some freedom to this value due to the thermal motion (deformation of the tetrahedron), we assign a maximal distance of $\Delta r=2$ fm.
 
 \item The momenta should also be similar. Taking a typical thermal momentum for $^4$He of $\sqrt{mT}=0.77$ GeV, this gives a momentum 
 of 0.11 GeV to each of the Cartesian components of each individual nucleon. We impose the condition that any component of the relative momentum cannot be larger than $\Delta p =0.22$ GeV.
 \end{enumerate}
 These two numbers satisfy $\Delta p \Delta r ={\cal O}(1)$, so this choice seems reasonable.

\begin{figure}[ht!]
\begin{center}
\includegraphics[width=7cm]{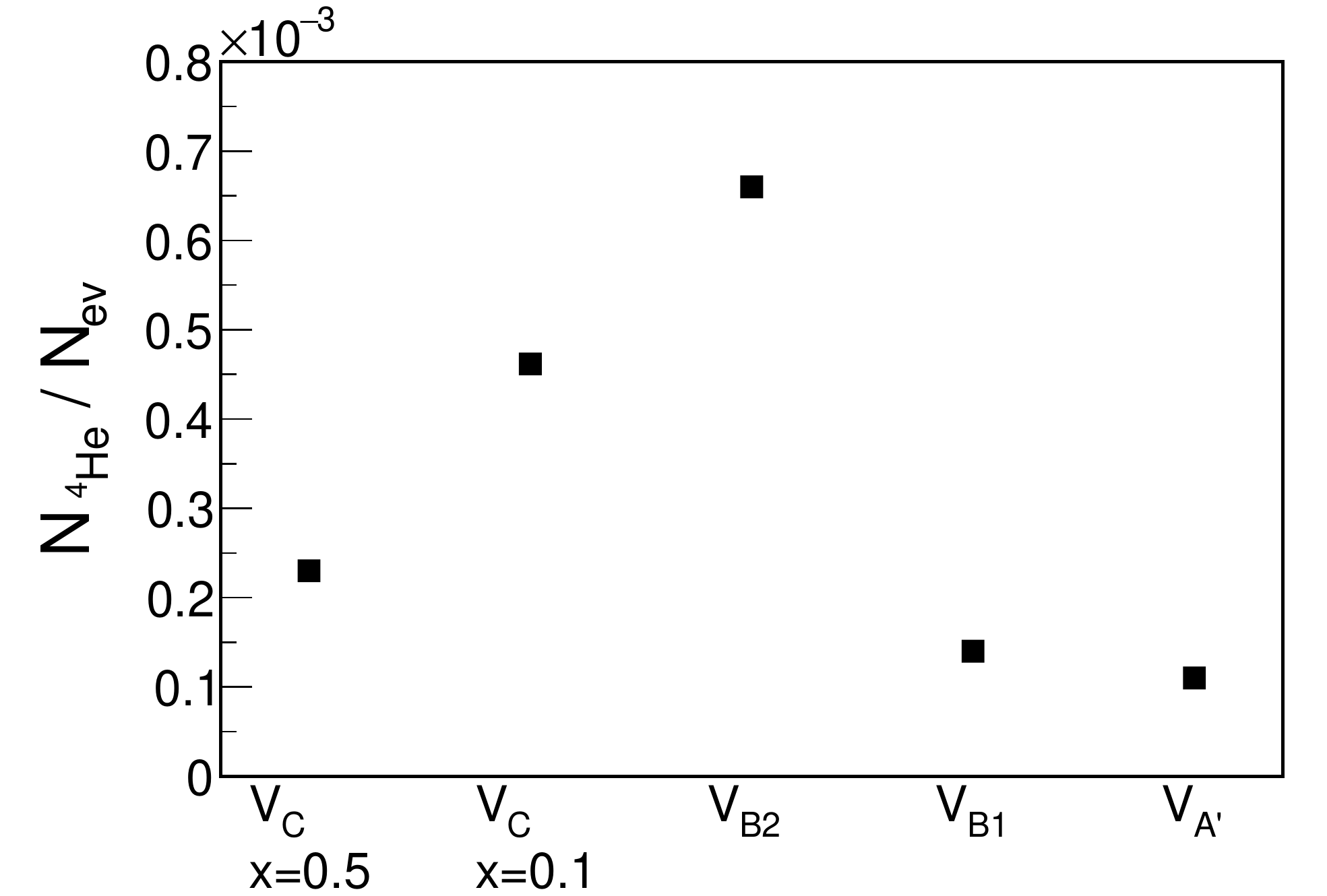} 
\caption{Multiplicity of ``pre- $^4$He'' (isolated clusters of 4 nucleons at freeze-out) per event as a function of the $NN$ potential (atraction among $NN$ increases from right to left).}
\label{fig:prehe4}
\end{center}
\end{figure}

  In Fig.~\ref{fig:prehe4}  we present the number of clusters of 4 nucleons (``pre- $^4$He'') per event (we use $N_{ev}=10^5$ events for each $NN$ potential). Similarly to the calculation of the moments of the 
proton distribution, we associate the noncritical potential $V_{A'}$ to the experimental collision energy of $\sqrt{s_{NN}}=19.7$ GeV. Then, using this potential in our MD+Langevin code we count the number of such clusters
defined by the previous criteria. 

  The results illustrate the effect of larger clustering formation with the $NN$ potential used. Going from right to left in Fig.~\ref{fig:prehe4} we find that the number of ``pre- $^4$He''  
increases with the attraction of the nuclear potential. Surprisingly, the most attractive potential $V_{C}$ presents a decrease of the number of clusters. The explanation is that because the huge attraction the 
nucleons belong to  	bigger clusters, i.e. it is more difficult to find 4 nucleons isolated from the rest.
  
  In spite of the qualitative motivation we can check that the numbers are not unrealistic for the $V_{A'}$ potential (identified with the collision energy of $\sqrt{s_{NN}}=19.7$ GeV). The multiplicity of $^4$He 
is not measured at this energy by STAR. However, other light nuclei have been measured by the NA49 experiment~\cite{Anticic:2016ckv}. For a close energy of $\sqrt{s_{NN}}=17.3$ GeV ($E_{beam}=158 A$GeV) we
know that after increasing the mass number in one unit (from $A=1$ to $A=2$, and from $A=2$ to to $A=3$), the $dN/dy$ at midrapidity decreases a factor of 100~\cite{Anticic:2016ckv}. Assuming that this scaling 
holds also up to $A=4$ we then expect one nucleus of $^4$He for each $10^6$ protons. Using that in our simulation we have $N_{ev}=10^5$ events with $N=32$ nucleons we expect a total of 3 nuclei of $^4$He for
the $V_{A'}$ case. We have numerically obtained 11 ``pre- $^4$He''. Given the number of simplifications made in our study, this number seems reasonable due to the following argument. Notice that our 4-nucleon clusters might not necessarily become $^4$He at the end of the evolution. Assuming a rather sharp freeze-out process, one would need to project the Wigner function of these 
``pre-nuclei'' configurations to the actual wave function of $^4$He. This study---which would give a more precise prediction for the produced light nuclei---is left for a future work, and here we restrict ourselves to
show the qualititative increase of clusters with the reduction of the $\sigma$ mass close to $T_c$.

\section{Summary and Outlook}

In this work we have studied baryonic clustering at the freeze-out conditions corresponding
to baryon-rich heavy-ion collisions. More specifically, we
have observed that both the clustering rate and the properties of the resulting clusters 
are very sensitive to the magnitude of the effective inter-nucleon potential, and suggest
that detailed studies of the baryon distributions will be able to fix such
potentials, and ultimately tell us whether the QCD critical point exists or not.

In Sec.~\ref{sec_forces} we have defined a set of inter-nucleon
effective potentials, which are modifications of the Walecka-Serot model, some with the addition of a 
long-range component related to massless critical mode at the (hypothetical) critical point.
Then in Sec.~\ref{sec_binding} we performed some initial studies of baryonic clusters which such potentials can support.
The main tool we used is classical molecular dynamics,
complemented by Langevin stochastic terms accounting for the mesonic heat bath, and also
by additional repulsive potential modeling quantum Fermi effects for the case of cold infinite nuclear matter in App.~\ref{sec_quantum}. 

If the matter is not exploding and the system evolves long enough, we do observe
that the initial stage, with random baryon positions, is always clustering,
in one or few large clusters. If the time is not so long, corresponding to $\Delta t \sim 5 \textrm{ fm}/c$
available for the hadronic phase of heavy-ion collisions, the degree of clustering is very strongly
dependent on the version of the potential used. Our main result is thus the {\em high sensitivity} of this phenomenon to the inter-nucleon potential.

We also tried to imitate an experimental fireball, mapping it to an expanding system. We also
impose similar cuts to the experimental acceptance of STAR papers, and calculated
the baryon number distribution. We do observe an increase of kurtosis, by about
a factor of 3, from the original Walecka potential to our most attractive version. 

Our main qualitative conclusion is that while the evolution time available is insufficient to
produced fully developed ``nucleosynthesis" with large clusters, one definitely
should find the baryon distribution in the final state far from thermal equilibrium. 
Indeed, the confidence in this statement is also provided by similar studies in atomic systems and globular clustering in galaxies
(briefly outlined in the corresponding appendices). We therefore suggest 
to look at possible deviations from thermal equilibrium in the yields of light nuclei, such as d,t,$^3$He,$^4$He.

While in this paper we cannot directly compare our results to the STAR BES data,
we do focus on one important finding: a growth of the {\it kurtosis} of the 
proton distribution near mid-rapidity, at the lowest collision energies~\cite{Luo:2015ewa}.

Although the specific critical enhancement of the multi-particle fluctuations remain the major goal of this program, 
one needs to also study other phenomena which can lead to those. In this paper we focused
on the {\it clustering of baryons} due to their attractive interaction.  
As we detailed above, significant clustering should in fact occur due to
the usual nuclear forces.

\appendix

\section{Mean-field approach to the Serot-Walecka model} \label{app:meanfield}

In this appendix we remind the reader a simplified form of nuclear forces, following a model by
Serot and Walecka~\cite{Serot:1984ey}. One important simplifying characteristic is that it only includes the isoscalar mesons:
scalar $\sigma$ and vector $\omega$, so there is no difference between coupling to protons and neutrons.  

 Its Lagrangian density is shown in Eq.~(\ref{eq:lagrangian}), and the inter-nucleon potential is written in 
Eq.~(\ref{eq:WalPot}), which we reproduce here again for convenience,
  
\be \tilde V_A(r)=- \frac{g_\sigma^2}{4\pi r}e^{-m_\sigma r}+ \frac{g_\omega^2}{4\pi r}e^{-m_\omega r} \ , \ee
with parameters in (\ref{eq:WalPotPar}) chosen by mean-field calculations.

\begin{figure}[htbp]
\begin{center}
\includegraphics[width=6.5cm]{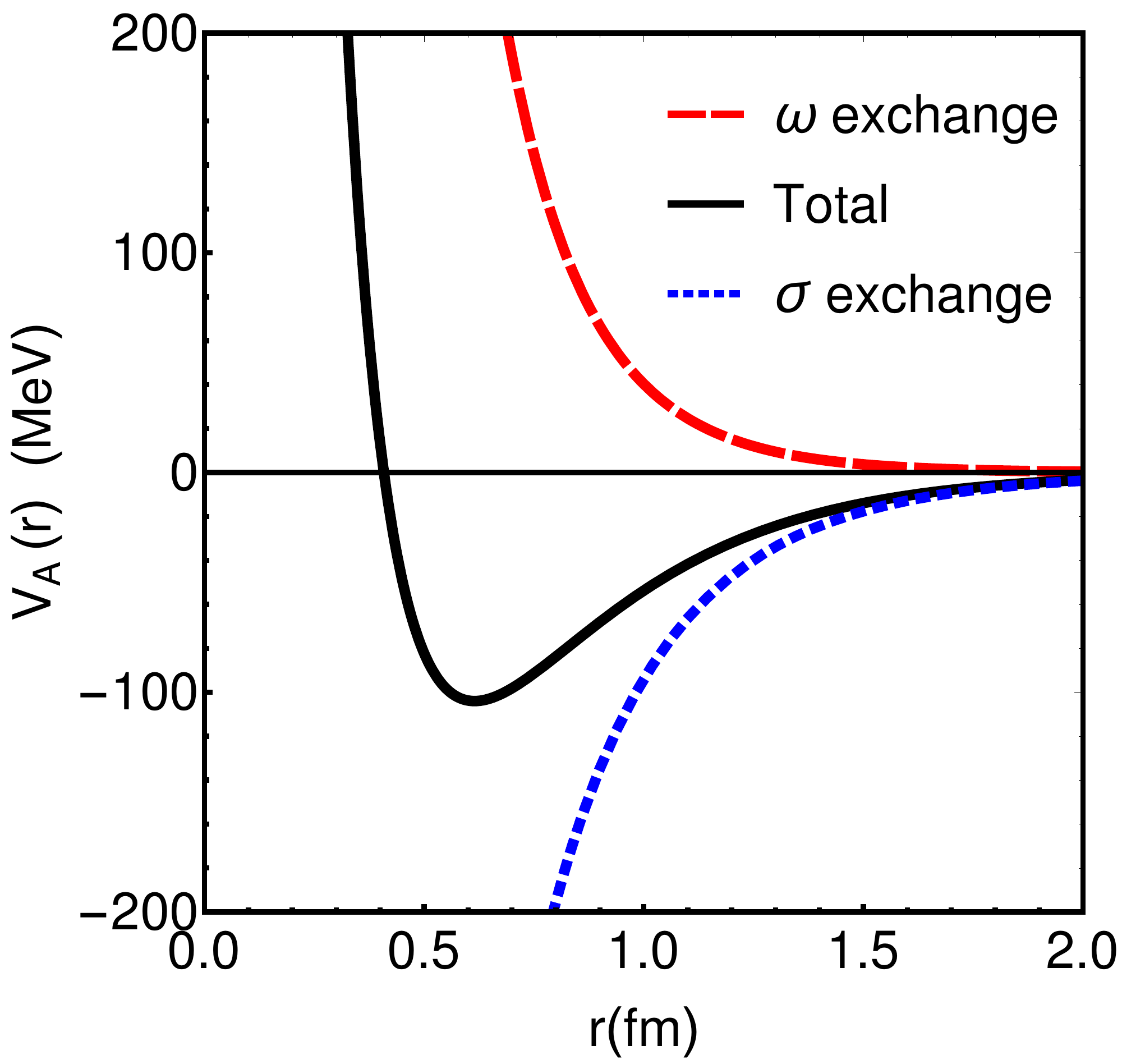}
\caption{Solid line: Serot-Walecka potential as given in Eq.~(\ref{eq:WalPot}). Dot-dashed line: Attractive part of the potential given by the first term in Eq.~(\ref{eq:WalPot}).
Dashed line: Repulsive contribution of the potential described by the second term of Eq.~(\ref{eq:WalPot}).}
\label{fig_Walecka_potAbis}
\end{center}
\end{figure}

In Fig.~\ref{fig_Walecka_potAbis} we illustrate the partial cancellation of the attractive and repulsive terms of this potential.

Considering the case of infinite homogeneous  matter of density $n$ and ignoring correlations
between the nucleons, one gets the mean potential energy 
\be \langle V \rangle=\frac{n}{2}  \ \left( - \frac{g_\sigma^2}{m_\sigma^2}+\frac{g_\omega^2}{m_\omega^2} \right) \ . \ee
If matter is cold, $T=0$, the baryons are in a form of degenerate Fermi gas
of quasiparticles with dispersion relation 
\be E_k= k+g_\omega V_0+ \sqrt{k^2 +M^2_*} \ , \ M_*=m_N-g_\sigma \phi_0 \ . \ee
 
Note that if one expands the square root, the leading term of the mean potential
$g_\omega V_0-g_\sigma \phi_0 $ will be the same as the one from the usual non-relativistic 
theory, but the kinetic energy term would be $k^2/2M_*$ rather than   $k^2/2m_N$.
The total energy of the gas is
\begin{eqnarray} 
  E_{mfa} &=& \frac{g_\omega^2}{2 m_\omega^2} n_B^2 + \frac{m_\sigma^2}{2 g_s} (m_N-M_*)^2 \nn \\
&+& \frac{\gamma}{(2\pi)^2} \int^{k_F} d^3k  \sqrt{k^2 +M^2_*} \ , 
\end{eqnarray}

where the statistical weight $\gamma=4$ for symmetric nuclear matter, and $2$ for neutron matter in neutron stars.   
 Two densities, the vector and scalar, can now be written as integrals over the Fermi sphere
\begin{eqnarray}
 n_B & = & \frac{\gamma}{(2\pi)^2} \int^{k_F} d^3k \ , \\
 n_s &= &  \frac{\gamma}{(2\pi)^2} \int^{k_F} d^3k \frac{M_*}{\sqrt{k^2 +M^2_*}} \ . 
 \end{eqnarray}
Note that the latter has scalar mass $M_*$ in the numerator and the energy 
in the denominator, which is needed because Lorentz invariant integration measure is $d^3k/E_k$.  
 
At this stage all is fixed except the scalar mean field (or alternatively $M_*$): this is a parameter of the homogeneous-field
trial function, which as any variational parameter, it should be found from 
{\em minimization} of the ground state energy. This leads to the following equation
\be M_*=M- \frac{g_\sigma^2}{m_\sigma^2} \frac{\gamma}{(2\pi)^2} \int^{k_F} d^3k \frac{M_*}{\sqrt{k^2 +M^2_*}} \ , \ee
for $M_*$ to be solved numerically. As shown in the original work~\cite{Serot:1984ey}, such mean-field result can be fitted to reproduce the nuclear matter density and nuclear binding. 

For finite spherical nuclei the procedure includes the solution of the
mesonic equations of motion 
\be \left\{ 
 \begin{array}{ccc} 
  \left(\partial_\mu \partial^\mu +m_\sigma^2 \right)\phi & = & g_\sigma \bar \psi \psi \ ,
   \\
   \partial_\mu F^{\mu\nu} + m_\omega^2 V^\nu &=& g_\omega \bar \psi \gamma^\nu \psi \ ,
 \end{array} \right.
\ee
supplemented by Thomas--Fermi-like treatment of baryons. For heavy nuclei the
results are rather good.
   
While this model is only a stripped-down version of nuclear forces and 
the mean field is only the first of various approximations used for nuclear matter description,
we will use it below due to its simplicity. In particular, this model only includes isoscalar 
exchanges, which means that $pp$ and $pn$ forces are the same. As a result, the
only place where isospin matters is in the quantum kinetic energy, since it depends on
the number of species.
We are however fully aware of the fact that Walecka model parameters are only good for
mean field treatment, and the resulting forces do not describe elastic $NN$ scatterings or
the deuteron binding. To improve on this it is possible to increasing the repulsion, via higher $\omega$ coupling $g_\omega^2 \rightarrow 1.4 g^2_\omega$, so that the
resulting potential gets very similar to the Bonn potential~\cite{Machleidt:2000ge} (see right panel of Fig.~\ref{fig_Walecka_potA}). 
This is what we will call ``modified Walecka potential'', which will be used in this work and denoted by $V_{A'}$.

\section{$\sigma$-meson dependence of the $NN$ potential from the functional renormalization group.}

  In Sec.~\ref{sec_forces} we have analyzed the modification of the attractive part of the $NN$ potential due to the $\sigma$ mass modification.
In this work we have not dealt with the precise dependence of this mass with the temperature/density. This would imply an additional uncertainty dependent on the model used e.g. linear sigma model, quark-meson model,
Nambu-Jona--Lasinio model... On the other hand a more rigorous treatment would involve the modification of the whole spectral function of this state.

  In this appendix we will illustrate how these two issues can be addressed using results of the $\sigma$-meson properties in the $N_f=2$ quark-meson model, approached by the application 
of the functional renormalization group (FRG)~\cite{Tripolt:2013jra,Tripolt:2015mtd}.

  The version of the quark-meson model presented in~\cite{Tripolt:2013jra,Tripolt:2015mtd} contains quarks, antiquarks, pion and $\sigma$ degrees of freedom. After the evolution of the FRG equations one is able 
to obtain the medium-modified properties of these states in the infrared limit. In particular, the spectral function of the $\sigma$ meson can be obtained at different temperatures and chemical potentials.

  In Ref.~\cite{Tripolt:2013jra} the critical point of the quark-meson model is located arount $T\simeq 10$ MeV, which seems to be quite low from the phenomenological point of view (this critical temperature is supposed to increase
when extending the calculation to $N_f=3$ flavors and after introducing the effects of the Polyakov loop potential). Therefore, we will consider two cases: at $T=\mu=0$ where the $\sigma$ screening mass is very
close to our vacuum mass $m_\sigma \simeq 500$ MeV for the potential $V_A$; and $T=150$ MeV where the $\sigma$ screening mass drops to values around $m_\sigma \simeq 280$ MeV.

  The attractive part of the static $NN$ potential is computed as a Fourier transform of the $\sigma$-meson exchange diagram, 
  \be V_\sigma ({\bf r}) = g_\sigma^2 \int_{-\infty}^\infty dt \int_{-\infty}^\infty \frac{d^4p}{(2\pi)^4} e^{i p\cdot x} \ D^R_\sigma(p_0,{\bf p}) \ , \ee
  where $p\cdot x \equiv p_0 t- {\bf p} \cdot {\bf r}$, and the $\sigma$ retarded propagator is used in the Lehmann representation to account for the complete spectral function $\rho_\sigma(\omega,{\bf p})$,
  \be D^R_\sigma(p_0,{\bf p}) = - \int_{-\infty}^\infty d\omega \frac{\rho_\sigma (\omega,{\bf p})}{\omega-p_0-i\epsilon} \ . \ee

  The data from Ref.~\cite{Tripolt:2015mtd} is given between $\omega \in (-1,1)$ GeV and $p \in (-1,1)$ GeV. For the case at $T=\mu=0$ MeV the $\sigma$-mass pole lies away from the real energy axis (i.e. its
real part is above the $\pi-\pi$ unitary threshold). Therefore, the mass appears in the spectral function as a broad pole, which can be numerically integrated in energy and momentum. However, at $T=150$ MeV, $\mu=0$ MeV
the $\sigma$ mass goes down below the unitary threshold and the pole is located on the real axis. In such a case the mass appears in the spectral function as a Dirac delta, which we need to add by hand to
the spectral function as the discretized data cannot capture it. Following the conventions in~\cite{Tripolt:2015mtd} we add to the $\rho_{\sigma} (\omega,{\bf p})$ a term like:
\be Z^{-1} \textrm{sgn} (\omega) \delta (\omega^2-{\bf p}^2 -m_\sigma^2) \ , \ee
where $m_\sigma$ is the pole mass of the $\sigma$ and $Z^{-1}$ is the pole weighting factor (we refer to \cite{Tripolt:2015mtd} to see how to compute these from the spectral function).

\begin{figure}[htbp]
\begin{center}
\includegraphics[width=7.5cm]{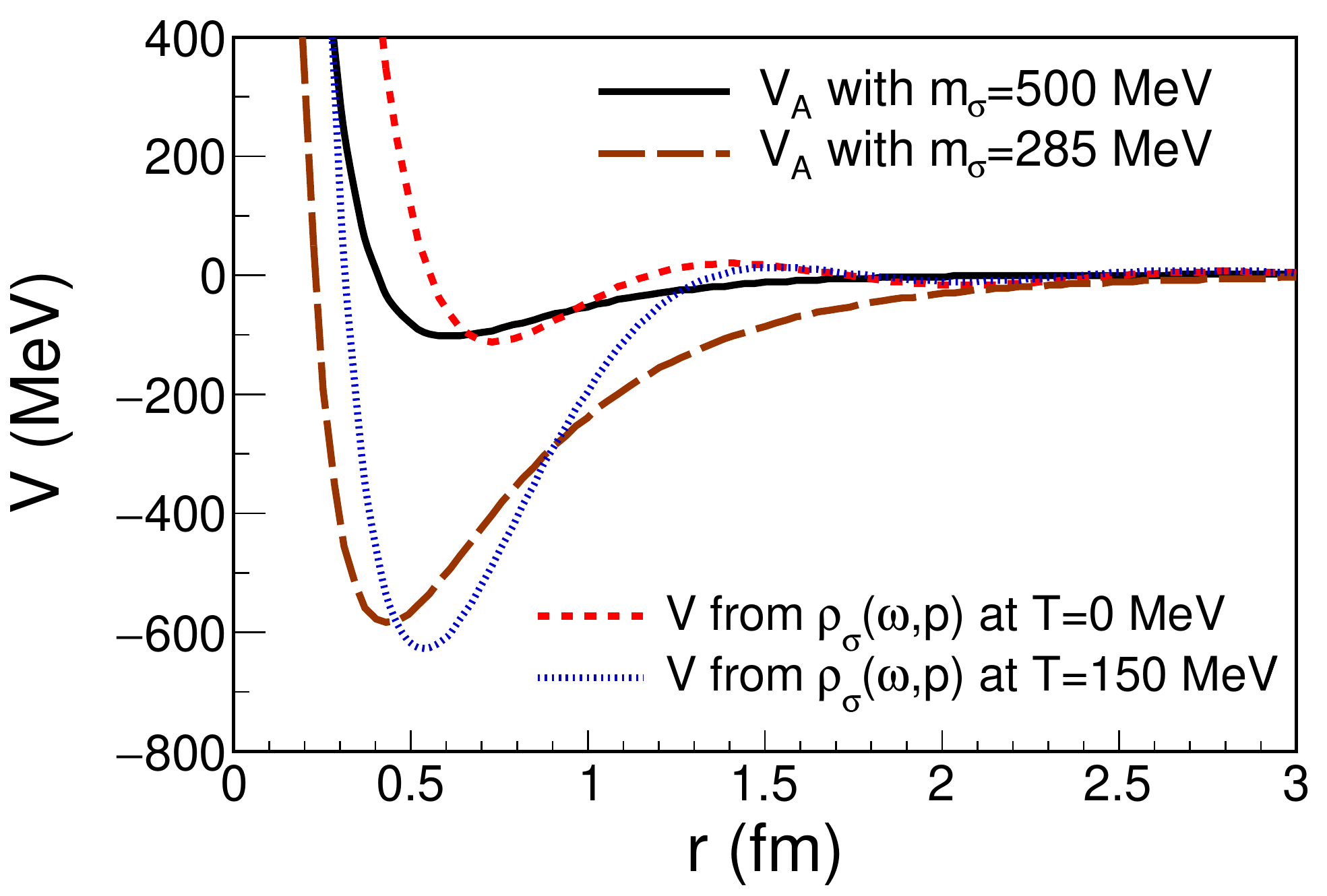}
\caption{Solid line: Serot-Walecka potential Eq.~(\ref{eq:WalPot}) with our vacuum parameters. The $\sigma$ mass is $m_{\sigma}=500$ MeV.
Short-dashed line: Result coming from the $\rho$ spectral function in vacuum taken from \cite{Tripolt:2015mtd}. The $\sigma$ screening mass in vacuum is $m_\sigma = 500$ MeV.  
Long-dashed line: Potential as in Eq.~(\ref{eq:WalPot}) with a $\sigma$ mass of 285 MeV, keeping the rest of the parameters as in vacuum.
Dotted line: Potential after using the spectral function at $T=150$ MeV, $\mu=0$ from \cite{Tripolt:2015mtd}. The $\sigma$ screening mass for this temperature is $m_\sigma = 285$ MeV.}
\label{fig_pot_from_specfunc}
\end{center}
\end{figure}

  In Fig.~\ref{fig_pot_from_specfunc} we compare the results coming from the Fourier transform of the spectral function of the $\sigma$ meson, and the simple potential as given in our Eq.~({\ref{eq:WalPot}})
  (the repulsive part 
  due to the $\omega$ meson is kept the same). Even at the quantitative level the two sets of potentials look similar. The main difference of the $\sigma$ potential occurs at small distances, but
in this limit the full potential is dominated by the $\omega$ repulsion. In the results using the $\sigma$ spectral function we observe some spurious oscillations around zero, which are nothing but the Gibbs effect
coming from the inverse Fourier transform of the spectral function performed within a compact support of energy and momentum.

\section{Kinetics and clustering in atomic systems} \label{app:atomic}

The simplest atomic systems are those of the noble gases, with spherical atoms and
forces depending solely on distances. For a large enough atomic weight, one can neglect
quantum effects. For all these reasons, the object of choice is argon, with its $A=40$ (for the most abundant argon isotope)
being ten times heavier than $^4$He. By tradition, theoretical studies of it use the simple
potential
\be V(r)=4\epsilon \left[ \left( \frac{\sigma}{r} \right)^{12}  - \left( \frac{\sigma}{r} \right)^{6} \right] \ .
\ee
Its minimum is at $V(2^{1/6} \sigma)=-\epsilon $. Following one of the classic MD simulations
from 1960's~\cite{Rahman}, one can use parameter values $\sigma=3.4$ \AA, $\epsilon=120$ K.

\begin{figure}[ht!]
\begin{center}
\includegraphics[width=6.5cm]{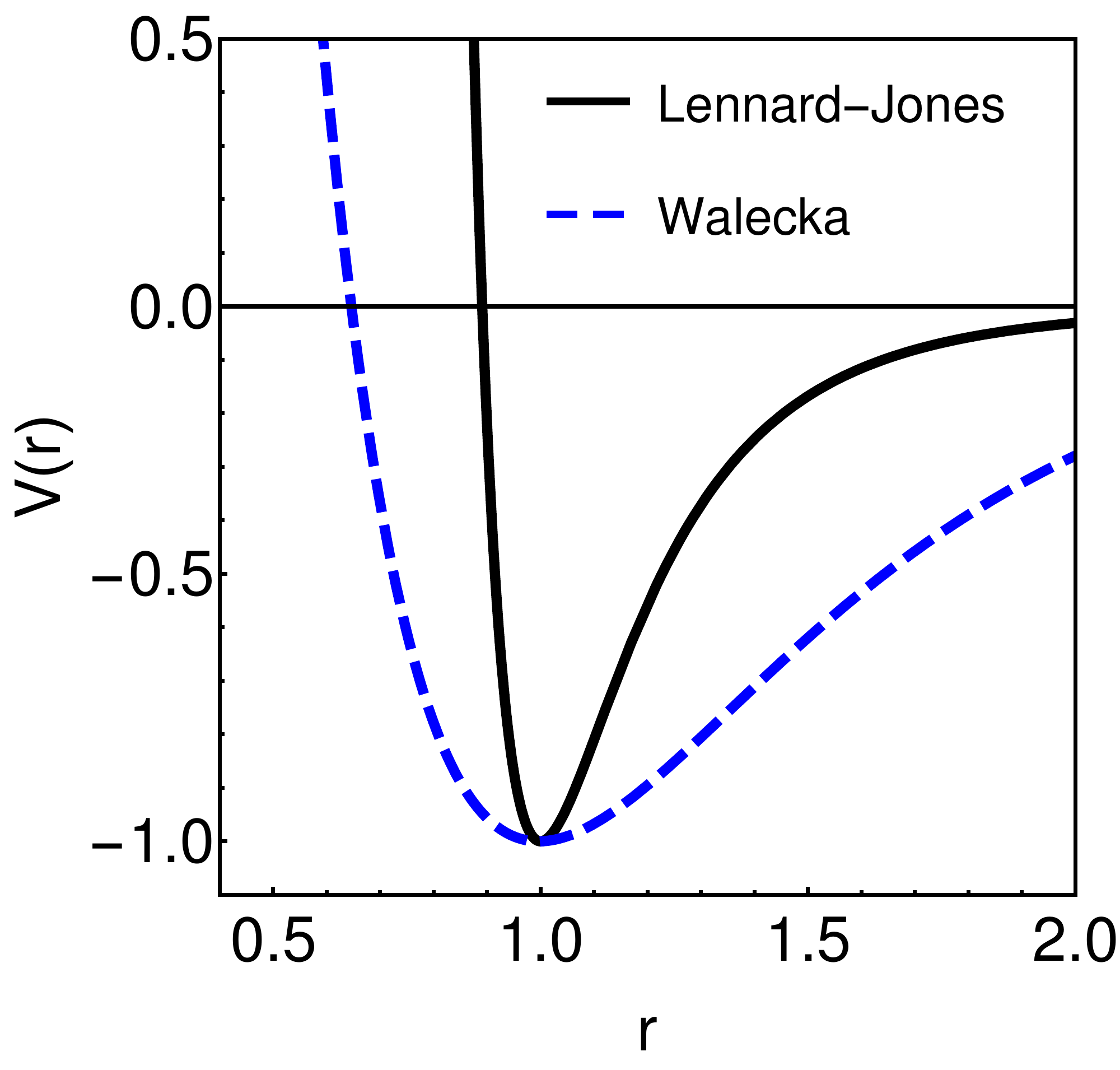}
\caption{Similarly normalized Lennard-Jones (black solid) and Walecka (blue dashed) potentials.}
\label{fig:lj_w}
\end{center}
\end{figure}

The shapes of this potential and that of the nuclear forces (Walecka model) are compared  
in Fig.~\ref{fig:lj_w}. It shows that Lennard-Jones potential is much more narrow. 
The ratio of the potential to the temperature are similar to the problem we study,
provided the temperature of argon is $T\sim 100$ K. 

\begin{figure}[ht!]
\begin{center}
\includegraphics[width=6.5cm]{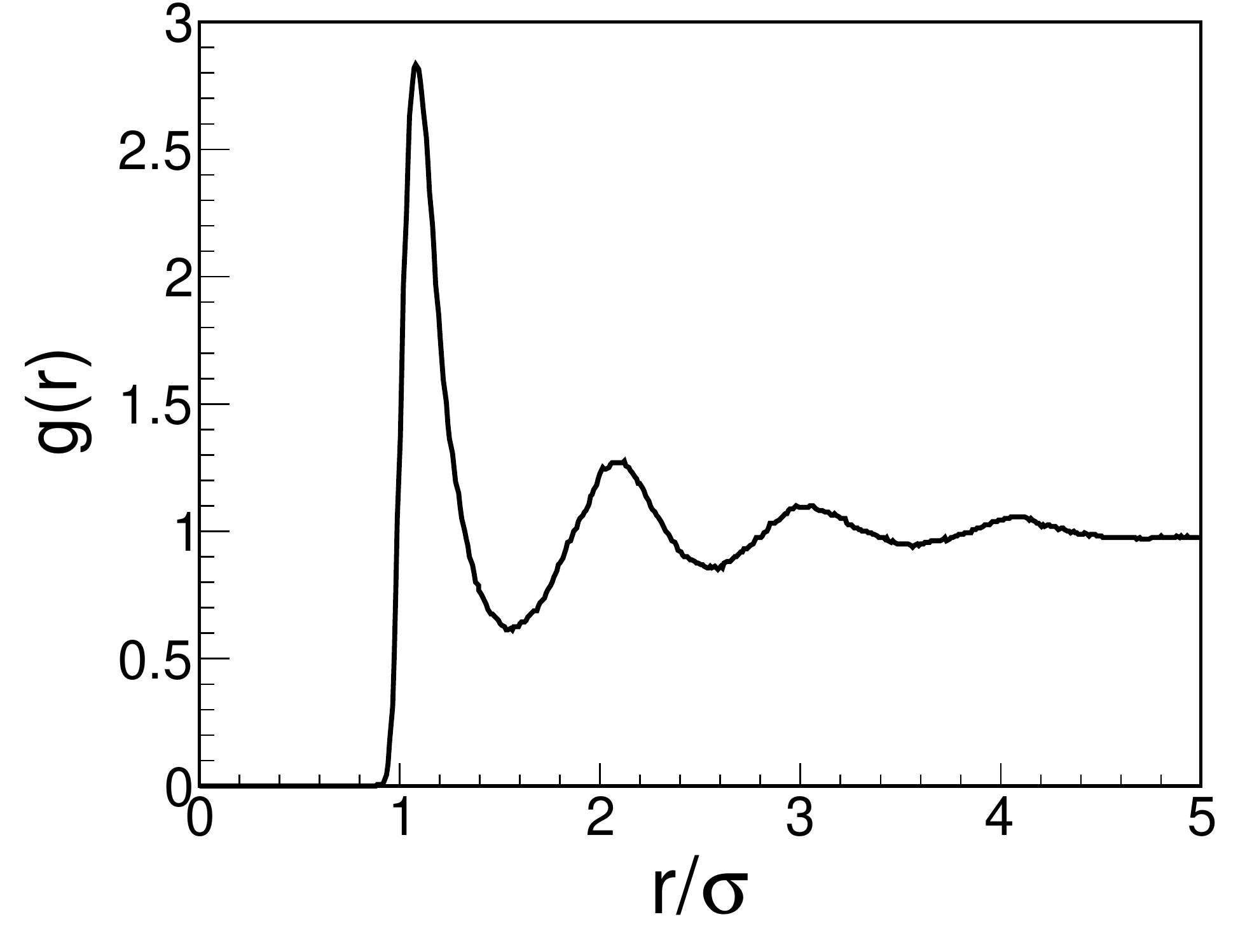}
\includegraphics[width=6.5cm]{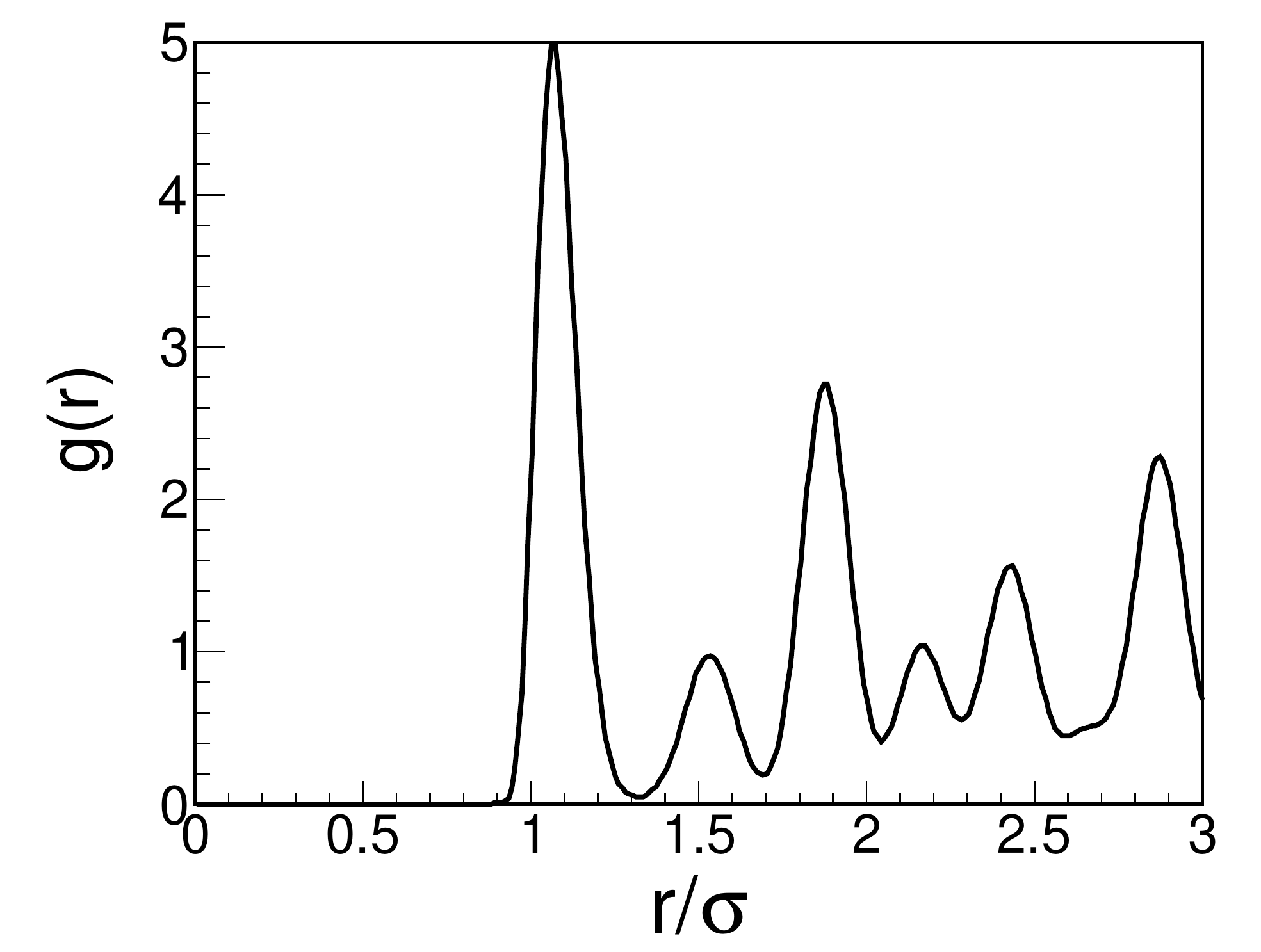}
\caption{Two-body correlation function for argon from MD simulations with $N=108$ particles. Left panel: $T^*=T/\epsilon=0.783, n^*=N/V\sigma^3=0.814$ (liquid phase).
Right panel: $T^*=0.783, n^*=1.1$  (solid phase).}
\label{fig:lj_gr}
\end{center}
\end{figure}

The work~\cite{Rahman} focused on one temperature $T=94$ K and one density $\rho=1.37$ g/cm$^3$, which is well
in the liquid phase. We minimally modify our MD (without Langevin dynamics) to run an isocanonical simulation (by rescaling of instantaneous temperature).
We use $N=108$ and similar conditions with a reduced temperature of $T^*=T/\epsilon=0.783$ and
a reduced density of $n^*=N/V\sigma^3=0.814$. The radial two-body correlation function $g(r)$ is shown in the upper panel of Fig.~\ref{fig:lj_gr}, presenting several peaks,
indicating strong correlations between the atoms at particular distances.

  Increasing the density one crosses the phase transition to a solid like phase. A new simulation with $n^*=1.1$ gives the radial distribution function in the lower panel of
Fig.~\ref{fig:lj_gr}. The amount of very pronounced peaks is a signature of the solid (crystalline) structure of the system. In this case the distribution of peaks can be
identified with a face-centered cubic distribution (which is the configuration used to initialize our simulation).

The standard MD simulation, unlike Monte Carlo ones, have not just static
(fixed time) but also the time-dependent information,
such as velocity-velocity and other correlation functions, Using standard Green-Kubo
formulas one can calculate diffusion constant, viscosity, etc.
 We however would not go into vast literature on the kinetic properties,
 except to note that liquid argon, like other liquids, has a second order critical point,
 and studies of the singularities of kinetic coefficients there remain to be
 better understood.
 
Finally, we would like to mention instead a particular large-scale MD simulation~\cite{Diemand},
 using as many as a billion atoms, and focusing on transition from homogeneous particle
 distributions to liquid phase, at supersaturated conditions. As it is well known,
 the process can be divided into two stages: (i) creation of {\em  critical clusters}, with $i^*$ 
 particles in them; and (ii) their subsequent linear growth as a function of time with certain rate.
 Large scale of the simulation had allowed to cover a range of temperatures and densities,
 in which the clustering rates change over many decades, and cluster sizes
 grow to well over 100 particles. However, what is most important, is that 
 in all cases the critical clusters
 are relatively small, ranging from $i^*\sim 12$ to about $100$ atoms. Therefore, the classical
 theory of nucleation---treating these clusters as macroscopic drops with a surface and
 volume free energies---needs to be corrected. After the actual energies of these
 clusters are used, the corrected theory was shown to work well. 
 Equilibrium configuration of small and medium size cluster in Lennard-Jones interaction has been studied e.g. in~\cite{hoare1971statics}.

\section{Globular clusters in galaxies} \label{sec_globular}

Gravity is the simplest attractive interaction, and the stars in the galaxy---which can be
well approximated by structureless point masses---are the simplest classical objects
one can think of. Galaxies themselves, and the globular clusters are products of instabilities induced by long-range attractive interaction,
and all of them appear from the homogeneous cosmological plasma at a certain temperature.

We will not be discussing here those instabilities and complicated paths which
lead to globular cluster formation, focusing at the classical theory of quasi-stationary
clusters. Since this field belongs to astronomy and is rather far from nuclear physics, we include in this summary 
its main elements.

Globular clusters are approximately spherically symmetric bound states of many stars.
Their typical number $N$ varies from $10^3$ to $10^6$, which is much smaller than
that in the whole galaxies $\sim 10^{11}$. For definiteness, we will mention 
numbers for $N=10^5$. The clusters are believed to possess black holes at their centers,
intermediate in mass between those due to a star collapse and those at the centers of the galaxies. In any case, their masses are way too small to play any role in what follows.

The main parameters of the clusters can be inferred from their
size $\sim 10$ pc and the typical velocity  $v\approx 10$ km/s, resulting in the smallest of relevant time scales, the crossing time
\be t_{crossing} =\frac{r}{v} \sim 10^6 \, \textrm{ yr} \ .
\ee
Scattering leads to equilibration of the system, relaxing it to certain
virial equilibrium in which we see the observed clusters. The relaxation time of a cluster is
\be t_{relaxation}  \sim 10^9 \, \textrm{ yr} \ .
\ee
This equilibrium is however a quasi-equilibrium, since collisions make a small
fraction of the stars venture above the escape velocity and leave the cluster.
The largest time scale is called the ``evaporation time" (assuming cluster is
not surrounded by any matter) which is
\be t_{evaporation}  \sim 10^{10} \, \textrm{ yr} \ .
\ee
It qualitatively coincides with the age of observed clusters and the lifetime of the Universe.

Considering an object with a unit mass, we define its energy by
 \be \epsilon=- \frac{v^2}{2} -\Phi(r) +\Phi_0 \ . \ee
Note the minus signs compared to
the usual definition: so positive $\epsilon$ corresponds to binding. The gravitational potential at distance $r$ from the center $\Phi(r)$ is, as usual, defined up to a constant, which we will select later. Note that $\epsilon=0$ defines the (coordinate dependent) escape velocity
$v_e=\sqrt{-2\Phi(r)+2\Phi_0}$.

The first step is to satisfy the stationary Boltzmann equation for the
star distribution function $f(\vec x,\vec v)$.
Setting $\partial f/ \partial t =0$ and neglecting the collision term, one has
\be (\vec v \cdot \vec \nabla_x) f- (\vec \nabla_x \Phi)\cdot \frac{\partial f}{\partial \vec v}=0 \ .
\ee
This however is achieved rather easily, for any distribution of the form
$ f\left( \epsilon(\vec x,\vec v)\right)$. 

Step two is the selection of a particular distribution of such kind.
We will discuss the so-called King distribution, in which $f=0$ for negative $\epsilon$
values (that is, the cluster has no unbound stars), and for positive $\epsilon$ it is
\be f_K(\epsilon)=const \ (2\pi \sigma^2 )^{-3/2}\left[ e^{ \frac{\epsilon}{\sigma^2}} -1\right] \ ,
\ee
which is a $shifted$ Maxwell-Boltzmann distribution with temperature $T=\sigma^2$.

\begin{figure}[h]
\begin{center}
\includegraphics[width=6.5cm]{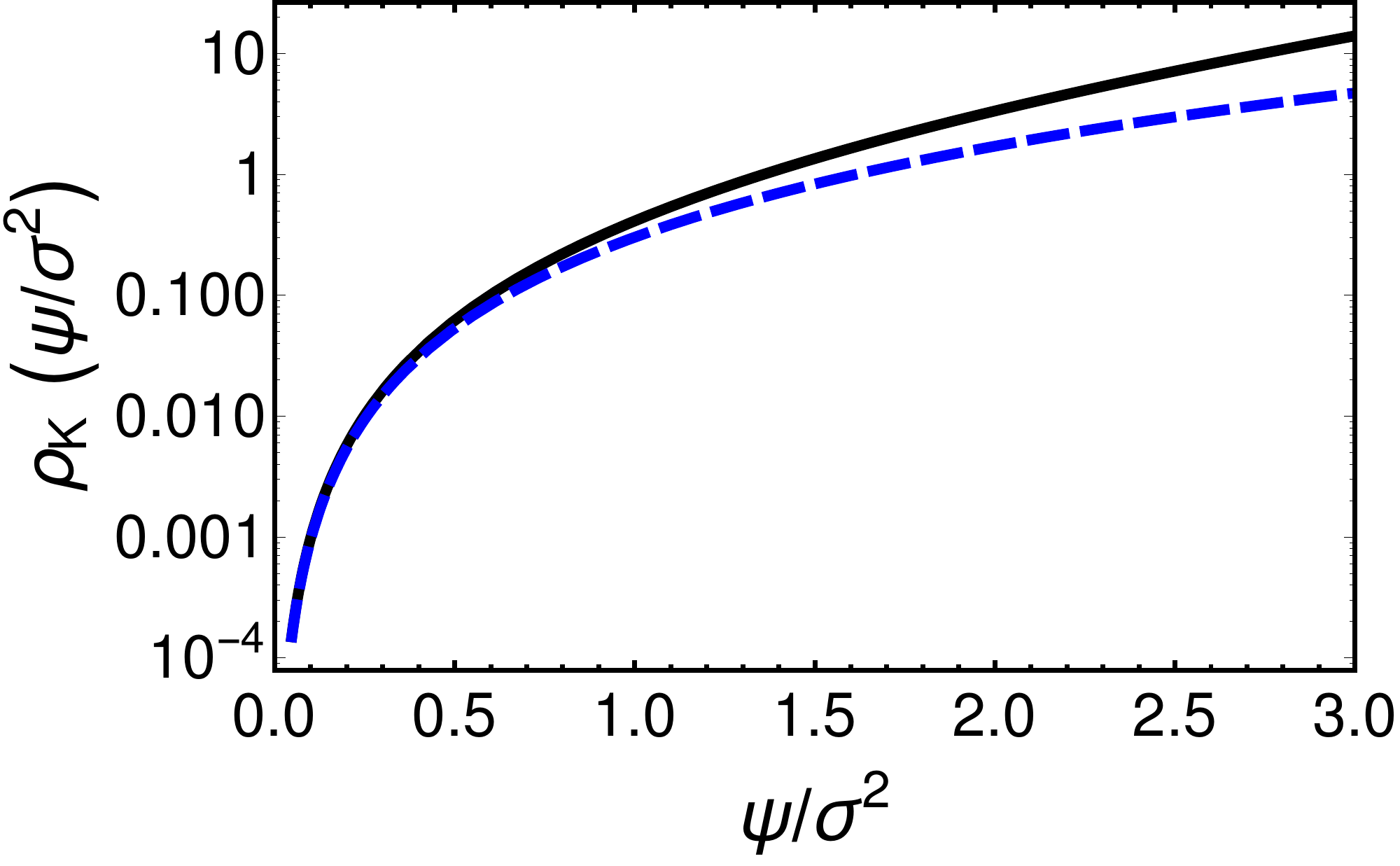}
\caption{The function $\rho_K(\psi/\sigma^2) $ defined in Eq.~(\ref{dens_function_of_pot}) is shown
by the black solid line, together with its asymptotic form at small values
of the argument  $0.30 (\psi/\sigma^2)^{5/2}$, shown by the blue dashed line.\label{fig_rho_of_psi}}
\end{center}
\end{figure}

Step three is a calculation of the corresponding density of stars, which includes the
integration over the velocity. Note that it is limited by the escape velocity defined via the potential, so the density obtained is {\em the function of the potential} $\psi=\Phi-\Phi_0$,

\begin{eqnarray} 
 \rho_K(\psi) &= & \frac{const}{(2\pi \sigma^2 )^{3/2}} \int_0^{\sqrt{2\psi}} \left(e^{(\psi-v^2/2)/\sigma^2}-1 \right) d^3 v  \nn \\
& = & - \frac{4}{3 \sqrt{\pi}}  \frac{\sqrt{\psi}}{\sigma} \left( \frac{\psi}{\sigma^2}+\frac{3}{2} \right) + e^{\frac{\psi}{\sigma^2}} \textrm{Erf} \left(\frac{\sqrt{\psi}}{\sigma} \right) \ . \nn \\
& & \label{dens_function_of_pot} 
\end{eqnarray}
This complicated function is plotted in Fig.~\ref{fig_rho_of_psi} , and one can see that it is a monotonously rising one.

The density is the source of the potential itself, so now we come across the main
dynamical equation to be solved, the Poisson equationn for the potential. In case of spherical symmetry it is
\be \frac{1}{r^2} \frac{d}{dr} \left(r^2  \frac{d \psi}{dr} \right)+ 4 \pi G_N \rho_K(\psi(r) ) =0 \ ,
\label{eqn_Poisson}
\ee 
which can be solved numerically starting from the center. The value $\psi(0)$ is the
single input parameter, the derivative needs to be  vanishing at the center $\psi'(0)=0$.
Solution can be followed until the point where $\psi=0$: and as it is clear
from the expression above for the density, at that point the density vanished as well
since the integration region till the escape velocity shrinks to zero. 
Substituting the resulting $\psi(r)$ into the universal $\rho(\psi)$ one
finally  obtains the spatial distribution of the stars in the cluster.

\begin{figure}[htbp]
\begin{center}
\includegraphics[width=6.5cm]{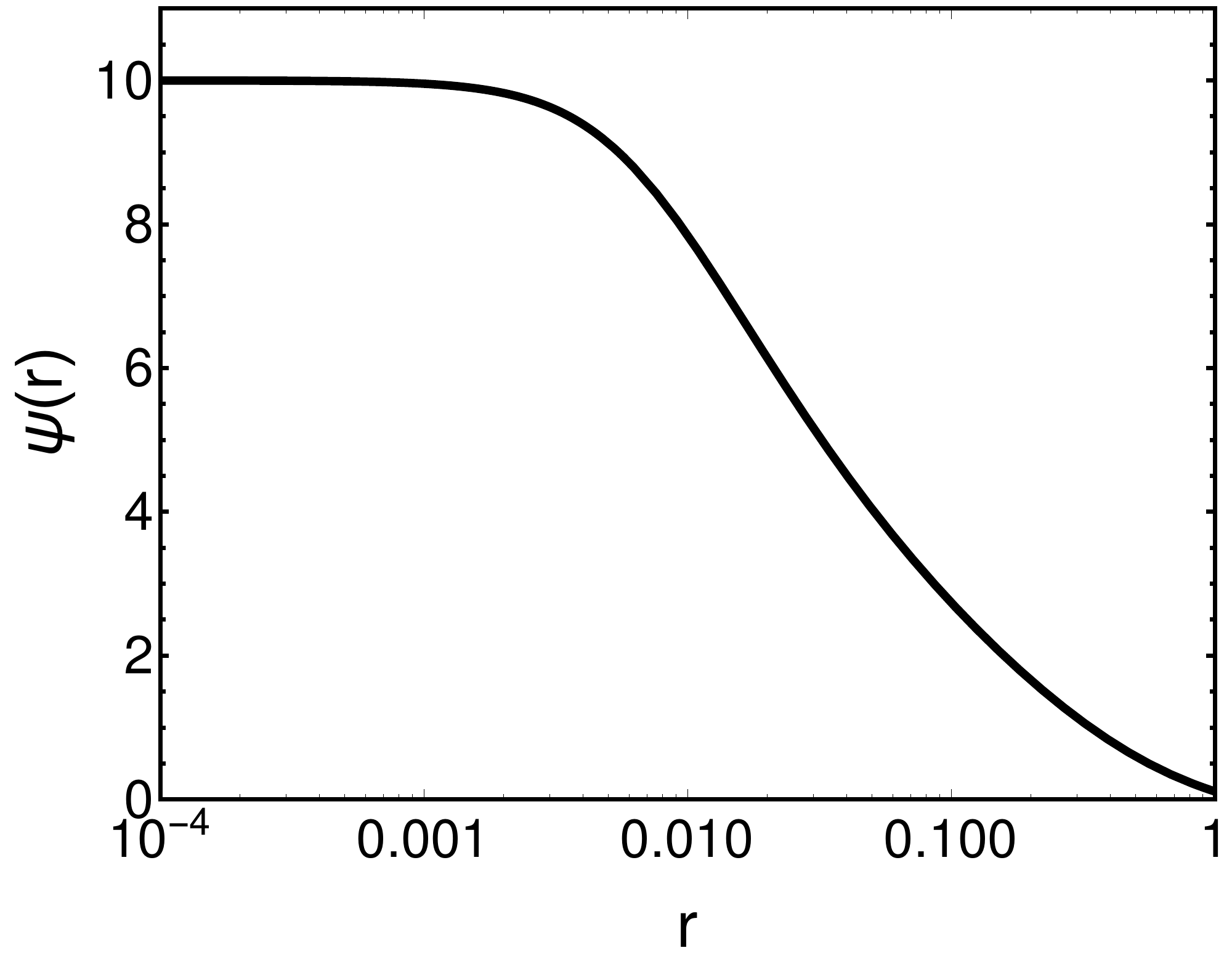}
\caption{The solution of the Poisson equation with the source $\rho_K(\psi)$, $\psi(r)$ versus $r$
at $G_N \times const=1,\sigma=1$ and $\psi(0)=10$.}
\label{fig_psi_of_r}
\end{center}
\end{figure}

\section{Cold nuclear matter and quantum Fermi repulsion} \label{sec_quantum}

To account for quantum repulsion in the simulations, the simplest thing one can do is adding the Fermi energy, evaluated in spirit of the Thomas-Fermi approach from the density profile, to
the classical kinetic and potential energy.

Full account for both quantum and thermal fluctuations can be done in
approaches called ``quantum open systems", see e.g. Ref.~\cite{Young:2010jq} for its application to
motion and heavy quarks and quarkonia, as well as general references.

Strictly speaking, a complete account for quantum effects would requires going from
classical molecular dynamics to full path integrals. As it is well known, while for distinguishable particles and bosons 
it can be considered to be just a technical complication, for fermions the amplitude 
needs to be anti-symmetrized, with brings in the notorious sign problem.
 The effective Fermi repulsion, acting as a kind of repulsive potential,  generates
correlations between particles which depend strongly on their mutual distance.

In 1980's O. Zhirov and one of us studied paths of fermions moving in one dimension. This case
 is special because one can always enumerate fermions along the line, and thus pretend that
 the ``exchange" never happens. For a small time step $t_a$ the one-particle amplitude is
 \be  
  U(x_f, x_i, t_a) \sim \exp \left[ -\frac{m (x_f-x_i)^2}{2 t_a} - t_a V\left( \frac{x_f+x_i}{2}\right) \right] \ , \nn \\ \ee
 and for two particles it can be written as  
 \be   U(x_1^f, x_1^i, t_a) U(x_2^f, x_2^i, t_a) -  U(x_1^f, x_2^i, t_a) U(x_2^f, x_1^i, t_a) \approx \nn \\
 U(x_1^f, x_1^i, t_a) U(x_2^f, x_2^i, t_a) \ \exp[-t_a V_{Pauli} ] \ , \ \ \ \ee
 with the  ``Pauli potential"  defined as
 \be  V_{Pauli} =- \frac{1}{t_a} \log\left[1- \exp \left( \frac{-m (x_1^f-x_2^f)(x_1^i-x_2^i)}{t_a} \right)\right] \ . \nn \\ \ee
Note that when two particles get close, the exponent becomes close to 1, the argument of the log near zero and
the potential gets very high. So, a node of the amplitude can be viewed as  a repulsive potential. This is going in the right direction: indeed, 
 fermions must have a larger energy than distinguishable particles in the same setting. 
If particles never jump over the note-generated barrier, their order along the line remains  preserved,  and if the Pauli potential
is included, the simulation can be done by traditional Monte Carlo.
 We checked it for several ($n=3-5$) particles put in a harmonic potential:
for distinguishable particles the ground state energy is $\hbar \omega(1/2+1/2+...)$ but for fermions
it should be $\hbar \omega (1/2+3/2+5/2+...)$ since each must be put into the next available level. 
So, our algorithm with this ``Pauli potential" worked correctly. The work was concluded in Ref.~\cite{Tursunov:1988ke}. Description of the method
and its usage is also described in Ref.~\cite{Bakker:1995pg}, in which many tests have been successfully performed.

The next step forward, allowing to use this idea in any dimension, was made by Ceperley~\cite{Ceperley:1992zz}.
It has been applied to fermionic problems, including liquid $^3$He. The main idea can be explained if one considers various paths of one fermion, keeping 
all other fermion paths frozen. The 1-dimensional node of the amplitude gets promoted to
a ``nodal surface'', which surrounds each fermion, keeping it inside a ``nodal cell''. Paths which are not allowed
to leave the nodal cell are called ``restricted": sum over the restricted paths obviously has no sign change.

The nodal surface model corresponds to a certain constant potential well with
the location of of the wall depending on those of other particles. 
The radius of this surface can be tuned to reproduce the Fermi energy of an ideal gas.

Instead of a sharp wall we decided to include a more smooth localization potential, of the form
\be \label{eq:locpot}
 V_{loc} (x_{ij})= a\frac{\hbar^2}{m_N x_{ij}^5} \ ,
\ee
where the exponent is chosen rather arbitrary as long as Pauli repulsion as short distances is achieved. 

To normalize this effective potential we attempted to simulate properties of cold
homogeneous nuclear matter by our molecular dynamics scheme, 

 \be \left\{
\begin{array}{rcl}
 \dfrac{ d \vec x_i}{dt}   & = & \dfrac{\vec p_i}{m_N} \\
 \\
 \dfrac{ d \vec p_i}{dt} & = &  - \sum\limits_{j\neq i} \dfrac{ \partial V (|\vec x_i- \vec x_j|)}{\partial \vec x_i} \ , \label{eq:infMD}
\end{array}
\right.
\ee
where $V$ represents the pair-wise potential, sum of the localization potential plus one among the different possibilities described in the main text.
We use the Walecka potential $V_{A'}(r)$ with increased repulsion, whic is closer to the NN phenomenological potentials for nuclear matter.
To simulate an infinite system be work on a cubic box with periodic boundary conditions. In such a box the particle density is fixed to the nuclear density at saturation $n_0=0.16$ fm$^{-3}$ with $N=64$ the total 
number of nucleons. To account for the interactions of the particles in the box and those outside, we use the method of images, where in the sum of Eq.~(\ref{eq:infMD}) we consider the contributions 
from all $j$-particles within a number of copies of the box in each spatial direction (positive and negative). The number of images (or copies of the elementary box) per each direction is set to 2.

  After a transient regime, the MD simulation reaches an equilibrium state with constant potential and kinetic energies (with statistical fluctuations of ${\cal O} (1/\sqrt{N})$). For infinite nuclear matter at saturation 
an average Fermi momentum of $p_F \sim 260$ MeV translates into a kinetic energy per nucleon of $K/N \approx 25$ MeV. Lacking of quantum dynamics in the classical MD we achieve this value of $K/N$
by forcing a isokinetic simulation by rescaling the velocity of each particle by $\sqrt{K/K_{inst}}$, where $K_{inst}$ is the instantaneous value of the kinetic energy at a given time step.

\begin{figure}[ht]
\begin{center}
\includegraphics[width=6.5cm]{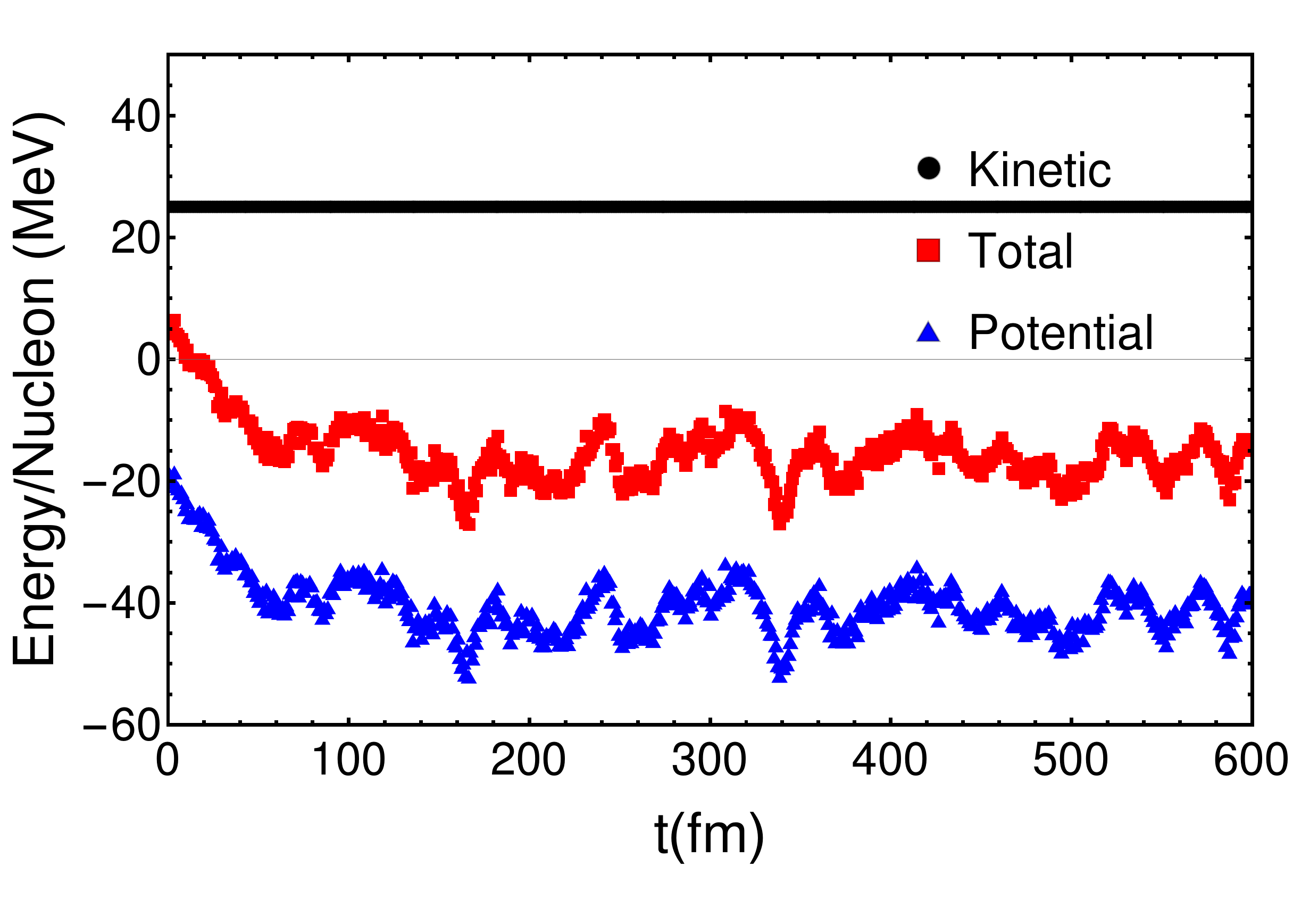}
\caption{Kinetic (black circles), total (red squares) and potential (blue triangles) energies per nucleon versus time in the infinite matter calculation. In addition to the Walecka $V_{A'}$ we implement the
localization potential in Eq.~(\ref{eq:locpot}) with $a=0.75$ fm$^3$.}
\label{fig:eloc}
\end{center}
\end{figure}

  The expected energy per nucleon at saturation $E/N=-16$ MeV provides the additional constraint that helps us to fix the remaining parameter of the simulation, the strength of the localization potential, 
to $a=0.75$ fm$^3$. The resulting energies versus time are given in Fig~\ref{fig:eloc}. After the equilibration time ($\sim$ 100 fm/$c$) we can measure the average total energy (binding energy) per nucleon.
We obtain $-16.6$ MeV, a fair value for our illustrative purposes. For dedicated computations a more precise value of $a$ can be extracted, using more nucleons in the simulation in order to reduce the 
statistical fluctuations of $E/N$ (going as $1/\sqrt{N}$).

\begin{figure}[ht]
\begin{center}
\includegraphics[width=6.5cm]{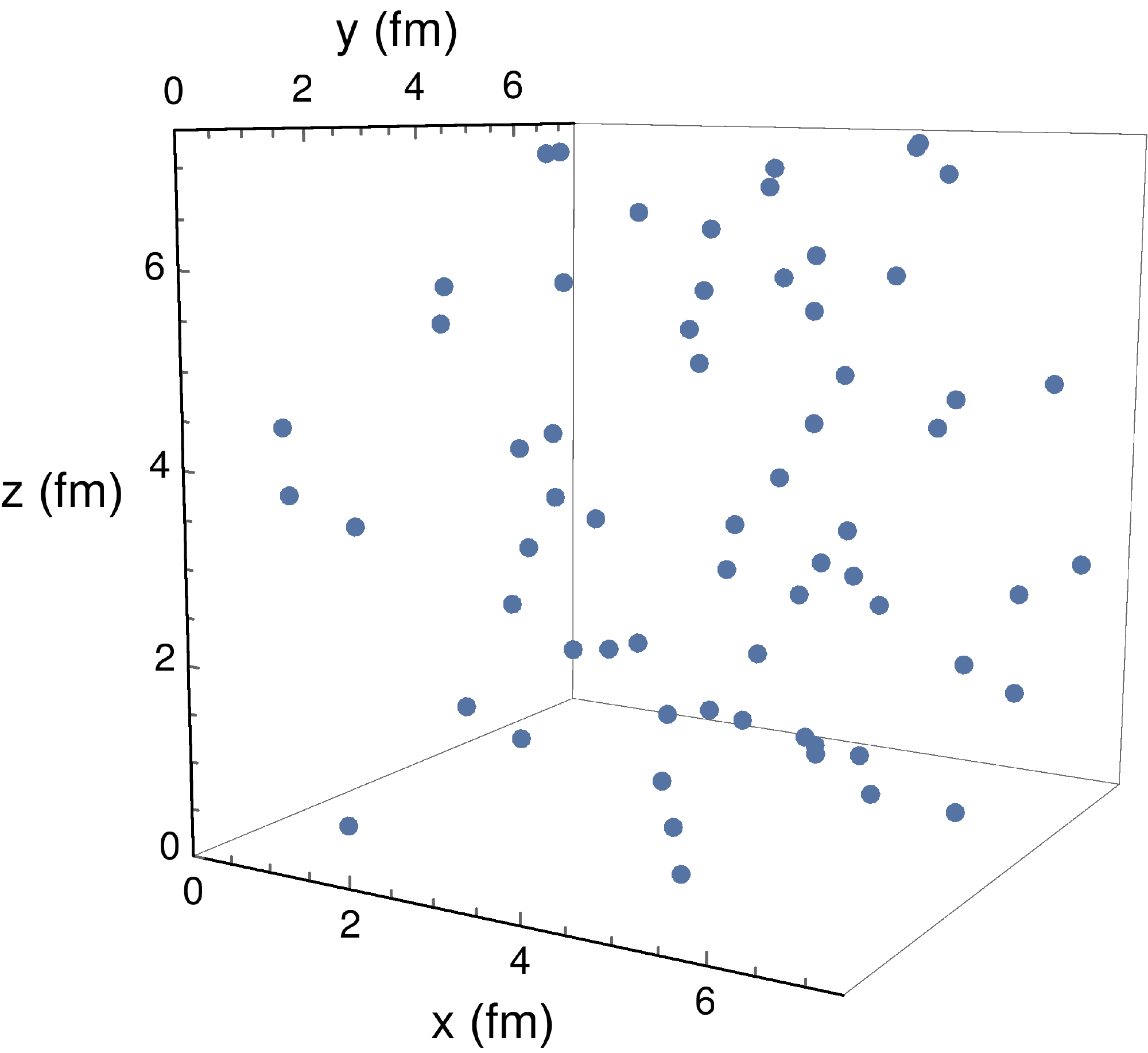}
\includegraphics[width=6.5cm]{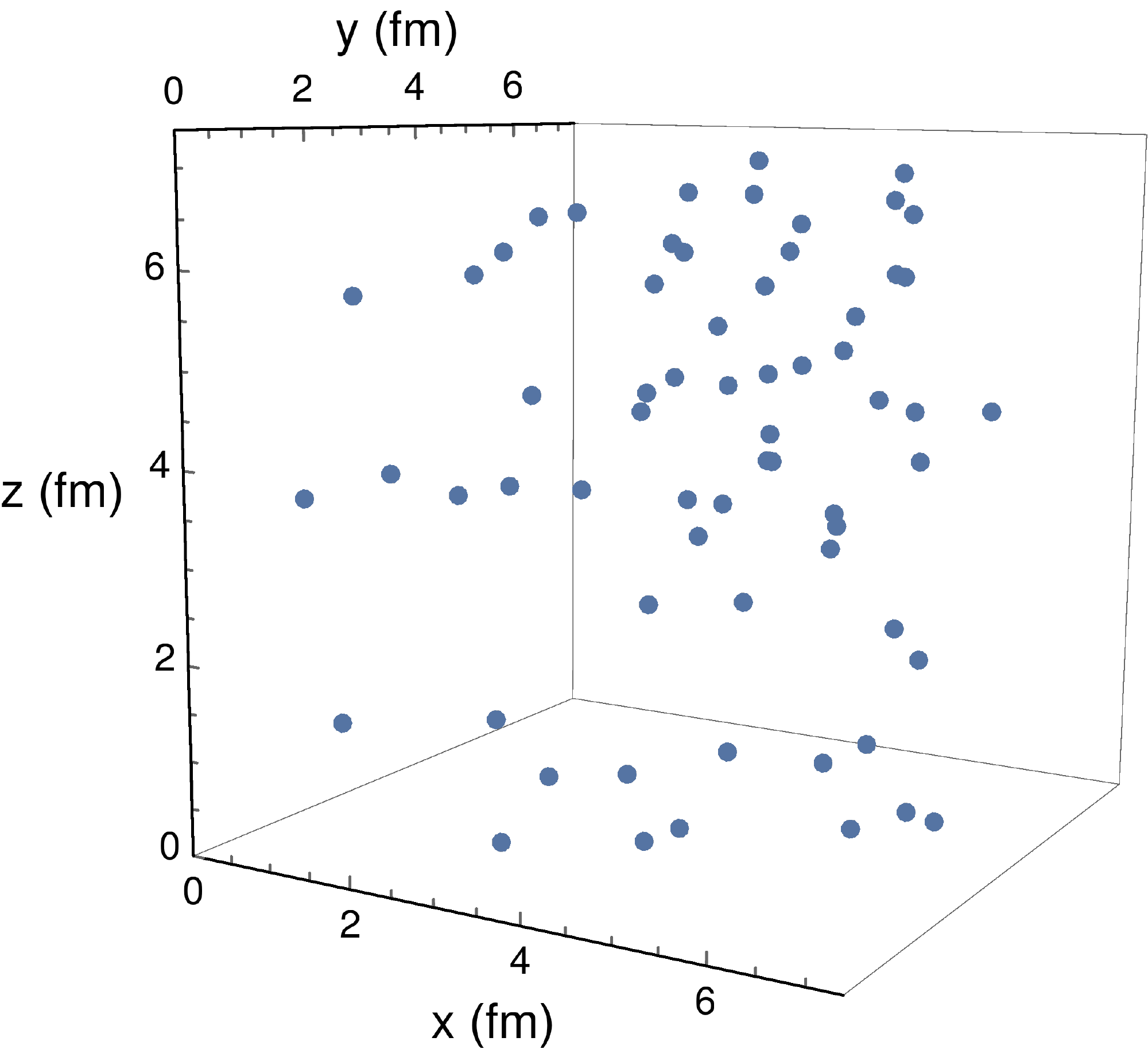}
\caption{Initial configuration (top) of nucleons in coordinate space and configuration at an arbitrary time after equilibration (bottom).}
\label{fig:distr}
\end{center}
\end{figure}

We find a rather homogeneous system at equilibrium with evidences of a slight grouping of nucleons. In the upper panel of Fig.~\ref{fig:distr} we show the initial configuration of nucleons at random positions in a volume
of (7.37 fm)$^3$. In the lower panel we show the spatial configuration of the nucleon for an arbitrary time well after the equilibration time.

 Quantum effects via localization potential---important for a $T \simeq 0$ calculation---will be absent around the freeze-out temperatures, where kinetic energy is expected to be dominated by thermal fluctuations.

\begin{acknowledgments}

We thank the STAR Collaboration (and Xiaofeng Luo) to provide the preliminary data of Ref.~\cite{Luo:2015ewa}. We acknowledge Ralf-Arno Tripolt for providing the $\sigma$ spectral functions in~\cite{Tripolt:2013jra,Tripolt:2015mtd}.
This work was supported in part by the U.S. Department of Energy under Contract No. DE-FG-88ER40388.

\end{acknowledgments}

\end{document}